\documentclass[12pt]{article}
\usepackage{endnotes}

\usepackage{amsmath}
\usepackage{amssymb}
\usepackage{amsthm}
\usepackage{lipsum}
\usepackage{forarray}
\usepackage{amstext}

\usepackage{amsfonts}

\usepackage{float}
\usepackage{cases}

\usepackage{longtable,lscape}
\usepackage{rotating,threeparttable}
\usepackage{dcolumn}
\usepackage{multirow}
\usepackage{graphicx}
\usepackage{pgf,tikz}
\usepackage{color}
\usepackage{caption}
\usepackage{graphicx}
\usepackage{amsfonts}
\usepackage{booktabs}
\usepackage{graphicx}
\usepackage{array}
\usepackage{ragged2e}
\usepackage{footnote}
\usepackage[colorlinks,citecolor=red]{hyperref}
\usepackage{epstopdf}
\usepackage[utf8x]{inputenc}

\usepackage{natbib}

\usepackage{lipsum}

\begin{document}
\baselineskip=.22in\parindent=30pt

\newtheorem{tm}{Theorem}
\newtheorem{dfn}{Definition}
\newtheorem{lma}{Lemma}
\newtheorem{assu}{Assumption}
\newtheorem{prop}{Proposition}
\newtheorem{cro}{Corollary}
\newcommand{\cor}{\begin{cro}}
\newcommand{\corr}{\end{cro}}
\newtheorem{exa}{Example}
\newcommand{\ex}{\begin{exa}}
\newcommand{\exx}{\end{exa}}
\newtheorem{remak}{Remark}
\newcommand{\rmk}{\begin{remak}}
\newcommand{\rmkk}{\end{remak}}
\newcommand{\thm}{\begin{tm}}
\newcommand{\nt}{\noindent}
\newcommand{\thmm}{\end{tm}}
\newcommand{\lm}{\begin{lma}}
\newcommand{\lmm}{\end{lma}}
\newcommand{\ass}{\begin{assu}}
\newcommand{\asss}{\end{assu}}
\newcommand{\df}{\begin{dfn} }
\newcommand{\dff}{\end{dfn}}
\newcommand{\prp}{\begin{prop}}
\newcommand{\prpp}{\end{prop}}
\newcommand{\bqu}{\sloppy \small \begin{quote}}
\newcommand{\equ}{\end{quote} \sloppy \large}
\newcommand\cites[1]{\citeauthor{#1}'s\ (\citeyear{#1})}

\newcommand{\eq}{\begin{equation}}
\newcommand{\eqq}{\end{equation}}
\newtheorem{claim}{Claim}
\newcommand{\cl}{\begin{claim}}
\newcommand{\cll}{\end{claim}}
\newcommand{\bit}{\begin{itemize}}
\newcommand{\eit}{\end{itemize}}
\newcommand{\ben}{\begin{enumerate}}
\newcommand{\een}{\end{enumerate}}
\newcommand{\bcen}{\begin{center}}
\newcommand{\ecen}{\end{center}}
\newcommand{\fn}{\footnote}
\def\qed{\hfill\vrule height4pt width4pt
depth0pt}
\def\reff #1\par{\noindent\hangindent =\parindent
\hangafter =1 #1\par}
\def\title #1{\begin{center}
{\Large {\bf #1}}
\end{center}}
\def\author #1{\begin{center} {\large #1}
\end{center}}
\def\date #1{\centerline {\large #1}}
\def\place #1{\begin{center}{\large #1}
\end{center}}

\def\date #1{\centerline {\large #1}}
\def\place #1{\begin{center}{\large #1}\end{center}}
\def\intr #1{\stackrel {\circ}{#1}}
\def\R{{\rm I\kern-1.7pt R}}
\def\N{{\rm I}\hskip-.13em{\rm N}}
\newcommand{\cprod}{\Pi_{i=1}^\ell}
\let\Large=\large
\let\large=\normalsize

\def\N{\mathbb{N}}

\begin{titlepage}
\def\thefootnote{\fnsymbol{footnote}}
\vspace*{1.1in}

\title{\large{On Sustainability and  Survivability in the Matchbox Two-Sector Model: 
 \vspace{4mm} \\A Complete Characterization of Optimal Extinction}\fn{A preliminary version of this work was presented to the Department of Economics at Cornell by the third author on October 26, 2021 -- he  thanks Tomasso Denti and  Marco Battaglini for this invitation, and in addition to the warm  hospitality of the department, acknowledges most stimulating conversations not only with  his hosts, but also with Arnab Basu, Kaushik Basu, Nancy Chau, David Easley, Chenyang Li, Suraj Malladi,  Francesca Molinari, J. Barkley Rosser Jr., Henry Wan and Si Zuo. The insights and comments of Bob Barbera, Larry Blume, Ralph Chami, Adel Guitoni  and Mukul Majumdar  call for especial gratitude.  Liuchun Deng acknowledges the support of the start-up grant from the Yale-NUS College.}}

\author{Liuchun Deng,\fn{Yale-NUS College, Social Sciences Division, Singapore.} Minako Fujio,\fn{Yokohama National University, Japan} and M. Ali Khan\fn{Department of Economics,
The Johns Hopkins University, Baltimore, MD 21218.}}

%\vskip 1.00em
\date{\today}

\bigskip

%\vskip 1.75em

\baselineskip=.18in

\noindent {\bf Abstract:} 
We provide a complete characterization of optimal extinction in a two-sector model of economic growth through three results, surprising in both their simplicity and intricacy. (i) When the discount factor is below a threshold identified by the well-known $\delta$-normality condition for the existence of a stationary optimal stock, the economy's capital becomes extinct in the long run. (ii) This extinction may be staggered if and only if the investment-good sector is capital-intensive. (iii) We uncover a sequence of thresholds of the discount factor, identified by a family of rational functions, that represent bifurcations for optimal postponements on the path to extinction. We also report various special cases of the model having to do with unsustainable technologies and equal capital-intensities that showcase long-term optimal growth, all of topical interest and all neglected in the antecedent literature.    \hfill(134 words)

\bigskip

\bigskip 
%\vskip 1.00em

%\vskip 1.75em

\noindent {\it Journal of Economic Literature} Classification Numbers:  C60, D90, O21

\medskip

%\vskip 1.00em

%\medskip
%\vskip 1.00em

\noindent {\it Key Words:} extinction, capital intensity, two-sector, $\delta$-normality, bifurcation

%\vskip 1.25em
\medskip

\noindent {\it Running Title:}  On Sustainability and  Survivability

\end{titlepage}

\tableofcontents

\pagebreak

%\bigskip

\setcounter{footnote}{0}

% To do:

% Split the document. The supplementary material needs its own table of content – look at Sannikov, for example.   
% 

\bqu  \textit{A formal presentation demands a precision in thinking and
encourages a search for the most direct route from a set of assumptions
to a conclusion. Despite its stark simplicity, a model may dramatically
confirm or reject an ``intuitive'' perception, and may  display highly complex, essentially unpredictable evolution, allowing for possibilities of extinction and indefinite sustainability. Even small
changes may set the stage for {\rm inevitable} rather than {\rm possible}  extinction and emergence of ``thresholds'' or ``tipping points'' that mark a change from growth to a stunning inevitability of extinction.}\fn{This epigraph is cobbled from several sentences: for the first two, see p. vii from the preface, and the third from pp. 25-26, all from \cite{ma20}.  Section 5.4 is directly relevant to this paper. More generally, this book addresses topical issues of the day, and merits a careful study.} \hfill{Majumdar (2020)}  \equ

\section{Introduction}

The notion of a stationary capital stock, also referred to as a stationary optimal program,  is central to the theory of aggregative and multi-sectoral descriptive and optimal growth, as it stems from the pioneering papers of \cite{ra28} and \cite{vn45}.\fn{It is now well understood that Ramsey's 1928 effort was rediscovered  by \cite{ca65} and \cite{ko65}, but the RCK label, common for the workhorse model of modern macroeconomics, does not acknowledge either \cite{sa65} and its earlier multi-sectoral extension by \cite{ss56}, or the independent analysis of  \cite{ma65}; see  \cite{sh67} for elaborations in  continuous time and the use of Pontryagin's principle;   also see \cite{sy14,sy15} for details.  \cite{ss56} concern themselves with multi-sectoral optimal growth theory, but closely follow Ramsey, while von Neumann's contribution sits astride descriptive and optimum growth theory in that it involves maximization but not that of a Ramseyian planner. The notion of a blanced growth  rate is, to be sure, directly connected to that of a stationary capital stock; see    \cite{ko64} and \cite{bu74}.  \label{fn:multi}}
 Adopting a primal approach, \cite{km86} obtain a sufficient condition concerning the discount factor and the technology, namely the $\delta$-normality condition, for the existence of a unique non-trivial stationary optimal stock for a large class of multi-sectoral optimal growth models.
  They also present an example in which the economy is not $\delta$-normal and the stationary optimal stock is trivial.\fn{The state of the art result is in  Section 7.5 of \cite{mck02} where the author refers to \cite{mck86} and to the work of Peleg-Ryder.  In his Handbook survey, he cites the work of Flynn, Khan-Mitra and Sutherland; the relevant result is Theorem 7.1 which uses Lemma 7.1 ascribed to the 1984 working paper version of  \cite{km86};   see the overview in the Handbook chapter of  \cite{mn06}.  Chapter 6 in \cite{ko85} also merits a careful study in this connection.   \label{fn:km} }  The literature has since largely presumed the existence of a non-trivial stationary optimal stock by explicitly or implicitly imposing this $\delta$-normality condition. Little of the existing work investigates the non-fulfilment of the condition and its resulting implications, especially in a disaggregated multi-sector economy.
This paper takes up this open question not merely to close a theoretical  lacuna, important though that is, but also to study the possibilities regarding   issues of survival and  optimal extinction of the capital stock that are opened up by the non-fulfilment of this condition.\fn{For the topicality, not to say immediacy of these issues, see,  in addition to  \cite{ma20},  \cite{ma15}, and their references. The latter is ostensibly phrased in the Asian context, yet testifies to the fact that the very nature of the problem spills beyond national boundaries; see for example the  chapter on environment and growth by \cite{ho15}.  \label{fn:top}} 

% Example 3 in Khan-Mitra 1986: economy not $\delta$-normal and the stationary optimal stock is trivial

  The question is best investigated in a model in which the existence of an optimal program is assured, but so is the non-existence of  a non-trivial stationary capital stock; a model  tractable enough for the question at hand, yet  with findings  whose robustness is   not called into question  in a fuller multi-sectoral setting.   The  canonical two-sector  Robinson-Srinivasan-Leontief (RSL) model of optimal growth, a special case of Morishima's matchbox two-sector  model,\fn{\cite{mo69} first introduces and analyzes a ``Walras-type model of matchbox size'', featuring Leontief production technologies in a two-sector setting. Lectures in  \cite{mo65c} are the natural precursor to the book.    \label{fn:morishima} }  fits this need well, and the results it furnishes are surprising both for their simplicity {\it and} their complication.  This  model  consists of a consumption-good and an investment-good sector, and with  Leontief production technologies  in both sectors. The (Ramseyian) social planner maximizes the discounted sum of future utilities by allocating capital and labor between the two sectors.
  
Under the aforementioned $\delta$-normality condition, more specifically, when the discount factor is above the inverse of the marginal rate of transformation (MRT) under full specialization in the investment-good sector, $(\delta > 1/\theta)$, this model has been employed as a workhorse to demonstrate how a wide array of dynamics,\fn{See the related literature on   two-sector RSL growth theory documented below.  \label{fn:rsl}}  ranging from monotone convergence to cycles and chaos, arise from a simple economic model.  The question then is  what happens when the $\delta$-normality condition is not fulfilled? And as befits any analysis of a two-sector model, we ask this question   under different   capital intensity conditions, a ``casual property of the technology'' being given prominence in Solow's rather immediate response to Uzawa's contribution: 

\bqu My second objective is to try to elucidate the role of the crucial capital-intensity condition in Uzawa's model. He finds that his model economy is always stable  if the
consumption-good sector is more capital-intensive than the investment-good sector. It seems paradoxical to me that such an important characteristic of the equilibrium path should depend on such a casual property of the technology. And since this stability
property is the one respect in which Uzawa's results seem qualitatively different from those of my 1956 paper on a one-sector model, I am anxious to track down the source of the difference.\fn{See \cite{so61}.  Solow specifies the notion of stability that is subscribing to: \lq\lq [The model economy is stable]   in the sense that full employment requires an approach to a state of balanced expansion."}
  \equ 

\nt We are anxious to see what happens to issues of survival and optimal extinction when there is no stationary capital stock and {\it a fortiori,} any convergence to it is precluded at the very outset.

In broad outline, the dispensation of the  $\delta$-normality condition in the RSL model furnishes  three results.  First, impatience leads to extinction.  If the discount factor is below $1/\theta$, capital stock will always converge to zero in the long run.  Second, investment on the optimal path to extinction hinges on the capital intensity condition, and  deferment of extinction by producing investment goods  arises only if the investment-good sector is more capital intensive; if less  intensive, the economy fully specializes in the consumption-good sector on the path to extinction. This asymmetry stems from the fact that production of consumption goods requires relatively less capital when the investment-good sector is capital intensive, and if the Ramseyian planner is not too impatient, it is optimal to trade off today's utility by diverting resources to investment for tomorrow's consumption gains. Third, perhaps most intriguingly, for the case of a capital-intensive investment-good sector,  we identify an {\it infinite} sequence of thresholds for the discount factor at which the optimal policy bifurcates. 
 As the discount factor rises, the economy will stay longer in the phase of diversification, with production resources fully utilized in consumption- and investment-good production, thus leading to a longer delay in extinction. Attainment of full utilization of resources along an optimal path is in itself a surprising result, since  there is (generically) excess capacity or unemployment for the case of a capital-intensive investment-good case when the $\delta$-normality condition is satisfied; see  \cite{fldk}.

Since the model no longer admits a a non-trivial stationary optimal stock when the $\delta$-normality condition fails, we exploit the full potential of the guess-and-verify approach.   A family of rational functions emerge in establishing the optimal policy for the case of a capital-intensive investment-good sector. The bulk of our characterization is to investigate the property of this family of rational functions and then use them to pin down the infinite sequence of bifurcation values for the discount factor.  This technical challenge that the guess-and-verify approach presents is new to us and may be of broader interest in the field of economic dynamics, and we rely on it to extend the characterization of optimal policy to three special cases. First, In the case of an unsustainable  RSL technology, that is,  any positive capital stock being technologically unsustainable in the long run,
the planner may still find it optimal to allocate resources to the investment-good sector along the path to extinction when the investment-good sector is capital intensive. Second, In the case when there is no difference in capital intensities, and the model reduces to the one-sector case, the optimal dynamics mirror the case of a capital-intensive consumption-good sector.
Third, and more to the point, in the  knife-edge case for the discount factor in which the optimal policy is no longer unique, and the optimal policy manifests itself as a correspondence, the door is opened to  a variety of long-term outcomes.

We now turn to the  relevant literature around which our model and results could possibly  be framed and evaluated. 
With regard to two-sector optimal growth theory, \cite{be92} and \cite{mmn00} still remain current as the go-to anthologies: in addition to neoclassical production functions,  they  include papers with Cobb-Douglas, CES and Leontieff technologies in one or both sectors, and emphasize the existence of cycles and chaos  even when intertemporal arbitrage opportunities are precluded by the assumption of an infinitely-lived Ramseyian planner.  They can be complemented by chapters in    \cite{dlmn}.  Where this literature needs updating is in regard to  a model which fell  in the crack between the one sector aggregate technology and the two-sector Uzawa one. This is the two-sector version of the so-called  
Robinson-Solow-Srinivasan (RSS) model that originated in the   development planning literature on the \lq\lq choice of technique," a technological specification in which labour is the only inter-sectorally mobile factor and exclusively used for the production of machines.\fn{See \cite{km05a} for references to the interesting exchange between Stiglitz and Robinson, and also \cite{in68} for the Hayekian setting where this assumption is relaxed. The  two-sector RSS model is acknowledged in  Handbook chapter of  \cite{mns} }   Since the revisiting of the model by    \cite{km05a,km05b}, 
  substantial work has accumulated, 
   and a general understanding has emerged that the new interesting results in the RSL setting can all be obtained in the simpler RSS setting; see for example \cite{fk06} and \cite{dkm20} for elaboration and references. In the context of this paper, what seems to have been missed however is that the $\delta$-normality condition holds in the RSS setting by default! There always exists a non-trivial stationary optimal stock.  As such, the possibilities  regarding extinction and survival explored here cannot arise. The two-sector RSL model is then a substantive alternative to the RSS model, and its direct generalization directly  relevant   for the problematic at issue.    
  
  The question then concerns the earlier results on the RSL model when the $\delta$-normality condition is satisfied.\footnote{It should be noted that there are two slightly different notions: $\delta$-normality and $\delta$-productivity. For the existence result, the more important is $\delta$-normality. For more detailed discussions, see \cite{mn06}. } 
  In a seminal paper, \cite{ny95} demonstrate in the RSL model with circulating capital that  optimal (ergodic) chaos can arise even for arbitrarily patient agents. \cite{fu05,fu08} characterizes the dynamic properties for the RSL model but without discounting. \cite{dfk19} and \cite{fldk} provide a partial characterization of the optimal policy for the case of a capital-intensive consumption-good sector and a complete characterization of that for the case of a capital-intensive investment-good sector. \cite{dfk20} depart from the optimal growth paradigm and obtain eventual periodicity in the RSL setting of equilibrium growth. The upshot of the existing work is that under the  $\delta$-normality condition,  the optimal policy for the RSL model is rich and complex for the case of a capital-intensive consumption-good sector and simple and uniform for the case of a capital-intensive investment-good sector.\fn{This is not to say that all is done when the $\delta$-normality is fulfilled: when 
  the discount factor lies  between the inverse of two MRTs ($1/\theta<\delta\leq1/\zeta),$ and the consumption-good sector is capital intensive,  the optimal policy has not been fully characterized, and this leads to the reasonable conjecture that chaos and complicated dynamics may arise.}     This paper demonstrates that it is this dichotomy that is reversed under the  non-fulfilment of the $\delta$-normality condition.

The rest of the paper is structured as follows. We introduce the model and preliminaries in the next section. In Section 3, we present the results on optimal extinction without investment. In Section 4, we explain the construction of thresholds for the discount factor, which are then used as bifurcation values in our characterization of optimal extinction with investment. Several special cases, including a numerical example, are discussed in Section 5. In keeping with the epigraph, the proofs of the results require  scrupulously detailed derivation, but we make do with geometry alone.\fn{All the proofs, lemmas, and additional characterization results are collected in the Appendix that constitutes the supplementary material to the work.}  We conclude this discussion of related work by pointing to two other streams of the literature that we shall take up in the concluding section of this essay: they concern the multi-sectoral and stochastic environments.

\section{The Model and Preliminaries}

\subsection{The Model}

We consider the two-sector RSL model of optimal growth with discounting.
There are two sectors: a consumption-good sector and an investment-good sector. The production technology is Leontief. It requires one unit of labor and $a_C>0$ units of capital to produce one unit of consumption good, and one unit of labor and $a_I\geq 0$ units of capital to produce $b>0$ units of investment good. If $a_C>a_I$, the consumption good sector is more capital intensive than the investment good sector, and if $a_C<a_I$, the investment good sector is more capital intensive than the consumption good sector. If $a_C=a_I$, the model boils down to its one-sector setting. Note that we assume $a_C>0$ because otherwise the planner would have no incentives to produce investment goods. However, we do not exclude the possibility of $a_I=0$ which corresponds to the two-sector RSS setting as in \cite{km05b}.

Labor supply is fixed and normalized to be one in each time period $t$.
Denote the capital stock in the current period by $x$, the capital stock in the next period by $x'$, and the depreciation rate of capital by 
$d\in(0,1]$. The {\it transition possibility 
set} is given by
$$\Omega=\{(x,x')\in \R_+\times \R_+: x'-(1-d)x\geq 0, x'-(1-d)x\leq
b\min\{1,x/a_{I}\}\},
$$ where $\R_+$ is the set of non-negative real numbers. 
Denote by $y$ the output of consumption good. 
For any $(x,x')\in\Omega$, we define a correspondence
\begin{eqnarray*}
\Lambda(x,x')=\left\{y\in \R_+: y\leq\frac{1}{a_C}\left(x-\frac{a_I}{b}(x'-(1-d)x)\right) \mbox{ and } y\leq 1-\frac{1}{b}(x'-(1-d)x)\right\}.
\end{eqnarray*}
A felicity function, $w: \R_+ \longrightarrow \R,$ is linear and given by $w(y)=y$.
The reduced form utility function, $u: \Omega
\longrightarrow \R_+,$ is defined as $$
u(x,x') = \max \{w(y): y \in \Lambda(x,x')\}. $$ The future utility
is discounted with a discount factor $\delta\in(0,1)$.
Define 
\begin{equation}\label{eq:zeta}
\zeta\equiv \frac{b}{a_C-a_I}-(1-d)
\end{equation} to be the MRT of capital between today and tomorrow under full utilization of both production factors.
Define 
\begin{equation}\label{eq:theta}
\theta \equiv \frac{b}{a_I}+(1-d)
\end{equation}
to be the MRT when the economy fully specializes in investment-good production with zero consumption good being produced for $x\leq a_I$.\fn{Later we will simply refer to $\theta$ as the MRT with zero consumption.} We then write explicitly the reduced-form utility function 
\begin{equation} \label{eq:reduced-form}
u( x,x') =\left\{\begin{array}{ll}
\frac{a_I\theta}{a_C b}x-\frac{a_I}{a_Cb}x', & \mbox{for } (a_C-a_I)x'\leq ((1-d)(a_C-a_I)-b) x+a_Cb \medskip \\

\frac{1-d}{b}x-\frac{1}{b}x'+1, & \mbox{for } (a_C-a_I)x'\geq ((1-d)(a_C-a_I)-b) x+a_Cb 
\end{array}\right.
\end{equation}
In the reduced-form utility function above, the first line stands for the case of full utilization of capital while the second line stands for the case of full employment of labor. 

An {\it economy} $E$ consists of a triplet $(\Omega, u, \delta)$. A {\it program from $x_0$} is
a sequence $\{x_t,y_t\}$ such that for all $t \in \N,\; (x_t,
x_{t+1}) \in \Omega$ and $y_t =\max \Lambda (x_t, x_{t+1}).$ A program $\{x_t,y_t\}$ is called {\it stationary}
if for all $t \in \N, (x_t,y_t) = (x_{t+1},y_{t+1}).$ For all $0 <
\delta < 1,$ a program $\{x^*_t,y^*_t\}$ from $x_0$ is said to be {\it
optimal} if
$$ \sum_{t=0}^\infty \delta^t[u(x_t, x_{t+1}) - u(x^*_t, x^*_{t+1})]
\leq 0 $$ for
every program $\{x_t, y_t\}$ from $x_0.$ A stationary optimal stock $x^*$ is said to be {\it non-trivial} if $u(x^*,x^*)>u(0,0).$

\begin{figure}[H]
\begin{center}
\includegraphics[width=10cm]{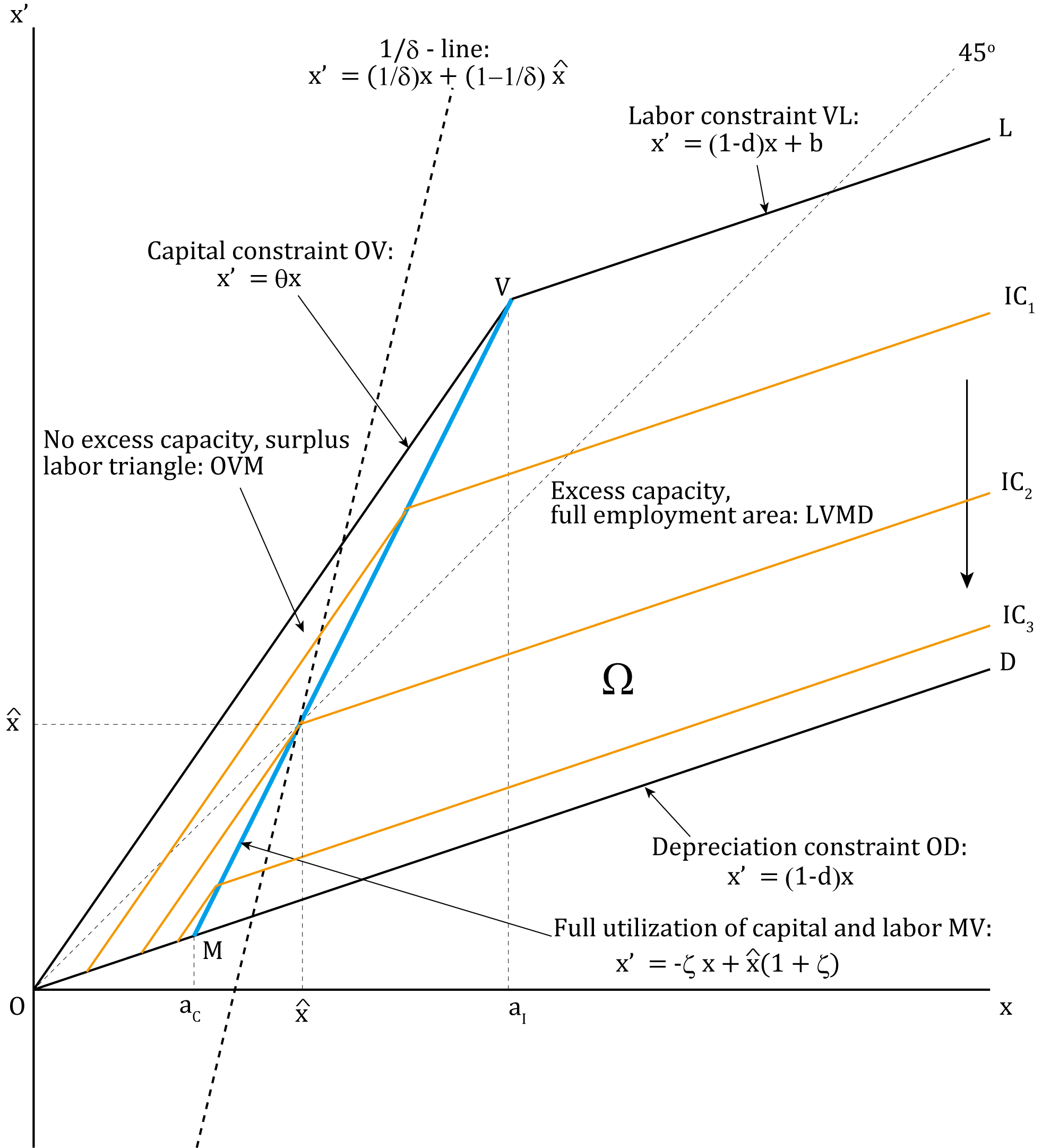}

%\vspace{-4.5cm}

\caption{The Basic Geometry for $a_C<a_I$ and $\delta<1/\theta$}\label{fig:basic-geometry}
\end{center}
\end{figure}

\subsection{Basic Geometry}

Before we turn to the formal discussion of the optimal policy, we describe in this subsection the basic geometry of the RSL model. Figure \ref{fig:basic-geometry} illustrates the transition possibility set for the case of a capital-intensive investment-good sector $(a_C<a_I).$ The $OD$ line corresponds to full specialization of the economy in the consumption-good sector. 
The $OVL$ line corresponds to full specialization of the economy in the investment-good sector. The slope of the $OV$ line is $\theta$. The $MV$ line corresponds to the case of full utilization of labor and capital. The slope of this line is $(-\zeta)$. When the investment-good sector is capital intensive as it is in Figure \ref{fig:basic-geometry}, if a production plan is above the $MV$ line, capital is fully utilized whereas there is surplus labor. If a production plan is below the $MV$ line, labor is fully employed whereas there is excess capacity. Moreover, $IC_1$, $IC_2$, and $IC_3$ in orange are the indifference curves for per-period utility. Lower indifference curves are associated with higher utility.

\subsection{Preliminaries}

We take the dynamic programming approach in our analysis.
Define the value function $V: \R_+\rightarrow\R$ as
$$ V(x)=\sum_{t=0}^{\infty}{\delta}^t u(x_t,x_{t+1})$$ 
where $\{x_t,y_t\}$ is an optimal program
starting from $x_0=x$. For each $x\in \R_+$, the Bellman
equation
$$ V(x)=\max_{x'\in \Gamma(x)}\{u(x,x')+\delta V(x')\}$$
holds where $\Gamma(x)=\{x':(x,x')\in\Omega\}$. For each $x\in \R_+$, define the
{\it optimal policy correspondence} 
$h(x)=\arg\max_{x'\in\Gamma(x)}\{u(x,x')+\delta V(x')\}.$ If $h(x)$ is a singleton for any $x\in \R_+$, then we define the {\it optimal policy function} $g: \R_+\rightarrow\R_+$ as $g(x)\in h(x)$ for any $x\in \R_+$.
A program $\{x_t,y_t\}$ from $x_0$ is optimal
if and only if it satisfies the equation:%\fn{This equivalence result is well known in the literature. We refer the interested reader to Footnote 22 in \cite{dfk19}.}
$$V(x_t)=u(x_t,x_{t+1})+\delta V(x_{t+1}) \mbox{ for }t\geq0.$$

The {\it modified golden rule} is formally defined as a pair $(\hat{x},\hat{p})\in\mathbb{R}_+^2$ such that
$(\hat{x},\hat{x})\in\Omega$ and
$$ u(\hat{x}, \hat{x}) + (\delta -1)
\hat{p} \hat{x} \geq u(x,x') + \hat{p} (\delta x' - x) \mbox{ for all
} (x,x') \in \Omega. $$
Or equivalently, the modified golden rule stock satisfies $u(\hat{x},\hat{x})\geq u(x,x')$ for all $(x,x')\in\Omega$ such that $x\leq (1-\delta)\hat{x}+\delta x'.$ Note that $x= (1-\delta)\hat{x}+\delta x'$ corresponds to the $1/\delta$-line in Figure \ref{fig:basic-geometry}.
An economy is said to be {\it $\delta$-normal} if there exists $(x,x')\in\Omega$ such that $x\leq \delta x'$ and $u(x,x')>u(0,0).$ The following lemma provides necessary and sufficient condition for $\delta$-normality in the RSL model. 

\lm\label{rho-normal}
The economy $E$ is $\delta$-normal if and only if $\delta>1/\theta.$ 
\lmm

We now state the main existence result from \cite{km86}.

\newtheorem*{tm2}{Theorem KM}
\begin{tm2}
For a class ${\mathcal E}$ of qualitatively-delineated economies, if the economy is $\delta$-normal, then there exists a modified golden-rule stock, which is also a non-trivial stationary optimal stock.
\end{tm2}

The RSL economy $E$ satisfies all the assumptions in \cite{km86} and thus is in ${\mathcal E}$. We apply Theorem KM to  obtain the following characterization of the modified golden rule which has been shown in \cite{dfk19} and \cite{fldk}.

% Proposition for the modified golden rule stock
\prp\label{golden-rule}
If $\delta>1/\theta$, then there exists a modified golden rule given by
$$(\hat{x},\hat{p})=\left(\frac{a_Cb}{b+d(a_C-a_I)},\frac{1}{(a_C-a_I)(1+\delta\zeta)}\right),$$
and the modified golden rule stock $\hat{x}$ is the unique non-trivial stationary optimal stock.
\prpp

The goal of our analysis in what follows is to characterize the optimal policy in the absence of $\delta$-normality. Without further explicit mention, from now on we will impose the following assumption on the discount factor, 
\begin{equation}
\delta\leq 1/\theta.
\end{equation}
This assumption stands in sharp contrast to the assumption of $\delta>1/\theta$ commonly imposed in the existing literature.
For $\delta\leq1/\theta$, the RSL model is no longer $\delta$-normal and thus Theorem KM no longer applies.
We will explore whether there still exists a stationary optimal stock and if not, how the economy evolves under the optimal policy. It should be noted that, in the two-sector RSS setting \cite{km05b}, the investment good sector is assumed to be infinitely productive ($a_I=0$) and as a result, this case of $\delta\leq 1/\theta$ is ruled out in the first place.

To facilitate exposition in the subsequent sections,
we give a formal definition of {\it extinction} in the long run. We distinguish the extinction phase {\it without} investment, in which the economy fully specializes in consumption-good production, from the extinction phase {\it with} investment, in which the economy may still allocate resources to the investment-good sector despite the gradual depletion of capital stock.

\df \label{extinction}
The economy is said to be in the extinction phase without investment if the optimal policy is given by $g(x)=(1-d)x$ for any $x>0$. The economy is said to be in the extinction phase with investment if the optimal policy yields the capital stock to converge to zero in the long run for any initial stock but there exists $x>0$ and $x'\in h(x)$ such that $x'>(1-d)x.$
\dff

According to this definition, the economy is in the extinction phase without investment if the transition path is entirely along the $OD$ line as in Figure \ref{fig:basic-geometry}, and the economy is in the extinction phase with investment if the optimal policy yields depletion of capital in the long run but the transition path is not entirely along the $OD$ line.

\section{Optimal Extinction without Investment}

We first examine the case of a capital-intensive consumption-good sector ($a_C > a_I$). It is known from the literature that, for $\delta>1/\theta$, the optimal policy for this case involves complicated bifurcation structures and a complete characterization has not been satisfactorily obtained even for the special case of $a_I=0$ \citep{km20}. However, for $\delta<1/\theta$, the optimal policy for the case of $a_C > a_I$ is surprisingly simple and uniform.

\thm\label{ac-greater-ai}
In the case of a capital-intensive consumption-good sector $(a_C>a_I)$, all rates of time preference $\delta$ less than the inverse of the MRT with zero consumption $(\delta<1/\theta)$ lead to an optimal policy under which the economy is in the extinction phase without investment.
\thmm

\begin{cro} \label{thm-ac-greater-ai}
In the case of a capital-intensive consumption-good sector $(a_C>a_I)$ and circulating capital $(d=1)$, if $\delta<1/\theta$, then the optimal policy yields immediate extinction: $g(x)=0$ for any $x>0.$
\end{cro}

\begin{figure}[H]
\begin{center}
\includegraphics[width=8cm]{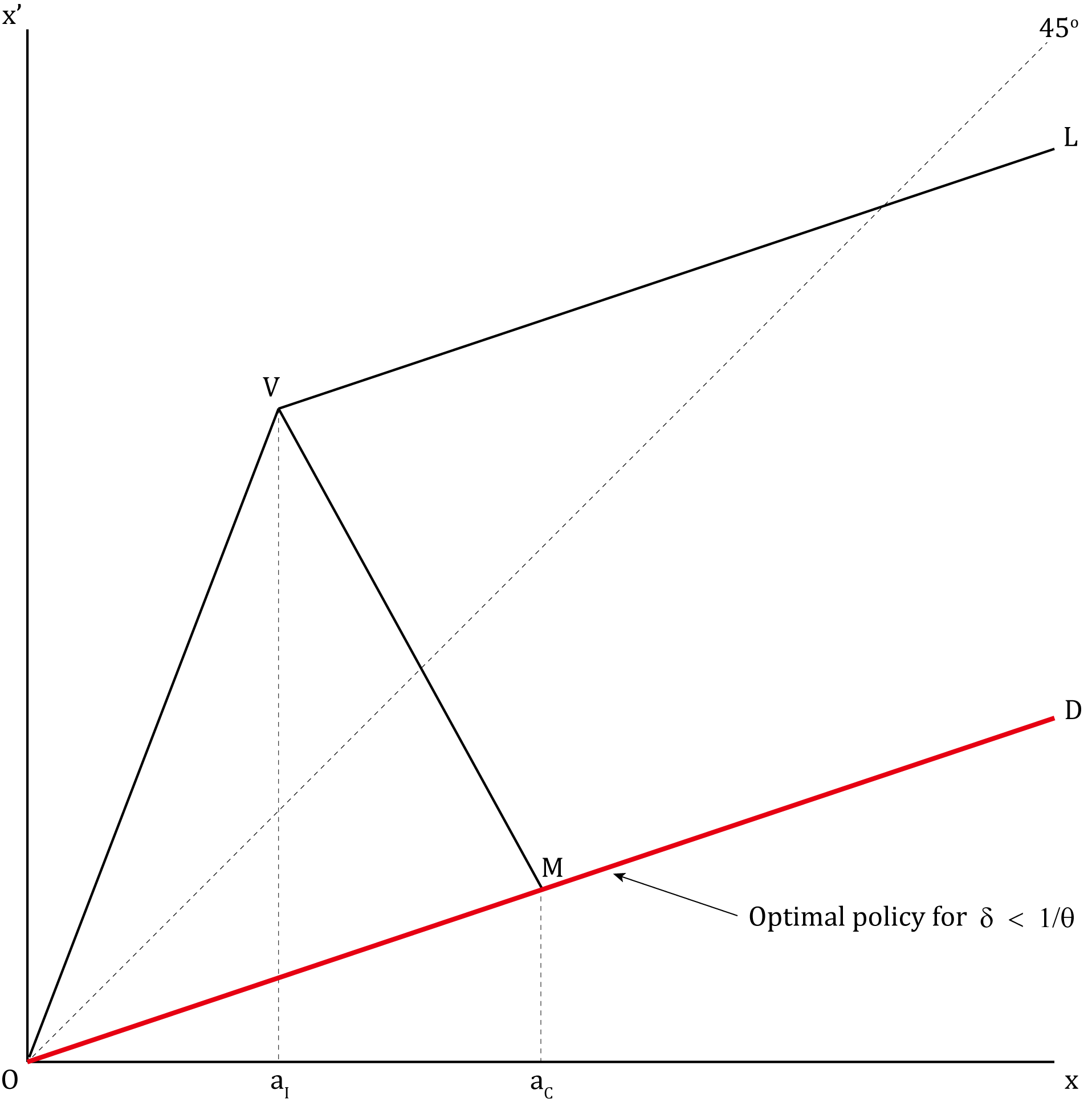}

\caption{The Optimal Policy for $a_C>a_I$ and $\delta< 1/\theta$}\label{fig:ac>ai}
\end{center}
\end{figure}

From Theorem \ref{ac-greater-ai}, there does not exist a non-trivial stationary optimal stock for $\delta<1/\theta$. As illustrated in Figure \ref{fig:ac>ai}, the optimal policy is represented by the $OD$ line: The economy converges monotonically to extinction ($x=0$) with no investment along the optimal path. Corollary \ref{thm-ac-greater-ai} further suggests that  if capital is circulating $(d=1)$, capital will be depleted just in one period.

We now turn to the case of a capital-intensive investment-good sector ($a_C<a_I$).
We first define
$$\mu_0\equiv \frac{1}{\frac{b}{a_C}+(1-d)}<\frac{1}{\frac{b}{a_I}+(1-d)}=\frac{1}{\theta},$$
where the inequality follows from $a_C<a_I.$ It is worth noting that from the formula above, there is a direct parallelism between $\mu_0$ and $1/\theta.$ % the proposal of relabeling them as theta_C and theta_I.

% what is the economic interpretation of \mu_0?

\thm\label{ac-less-ai-extinction}
In the case of a capital-intensive investment-good sector $(a_I>a_C)$, all rates of time preference $\delta$ less than a technological upper bound $(\delta<\mu_0)$ lead to an optimal policy under which the economy is in the extinction phase without investment.
\thmm

\begin{cro} \label{thm-ac-less-ai}
In the case of a capital-intensive investment-good sector $(a_I>a_C)$ and circulating capital $(d=1)$, if $\delta<\mu_0$, then the optimal policy yields immediate extinction: $g(x)=0$ for any $x>0.$
\end{cro}

Theorem \ref{ac-less-ai-extinction} says that if the discount factor is sufficiently low, the optimal policy for the case of $a_C<a_I$, represented by the $OD$ line in Figure \ref{fig:basic-geometry}, is the same as that for $a_C>a_I$. Theorems \ref{ac-greater-ai} and \ref{ac-less-ai-extinction} underscore that impatience leads to extinction: The economy fully specializes in the consumption-good sector if agents are sufficiently impatient.
Then, what remains open is for the discount factor between $\mu_0$ and $1/\theta$ in the case of a capital-intensive investment-good sector. This is what we turn to next.

\section{Optimal Extinction with Investment}

In the case of a capital-intensive investment-good sector $(a_I>a_C)$, we will show that if the discount factor $\delta$ is in $(\mu_0,1/\theta)$, the economy will converge to extinction in the long run but with positive investment along the transition path. 
The optimal policy bifurcates with respect to the discount factor in a rather intriguing manner. To characterize the optimal policy and its bifurcation structure, we first introduce a sequence of thresholds for the discount factor.

\subsection{Thresholds for the Discount Factor}

To define a sequence of thresholds for the discount factor  $\delta\in[\mu_0,1/\theta),$
we consider,
for any natural number $n$, the following rational function from $[0,1/\theta]$ to $\mathbb{R}$ as
\begin{equation} \label{eq:polynomial}
z_n(\delta)\equiv -\frac{1}{b}+\delta\left(-\frac{\sum_{i=0}^{n-1}(-\delta\zeta)^i}{a_I-a_C}+\frac{(-\delta\zeta)^n}{a_C(1-\delta(1-d))}\right).
\end{equation}
Further, we define 
\begin{equation} \label{eq:z0}
z_0(\delta)\equiv -\frac{1}{b}+\frac{\delta}{a_C(1-\delta(1-d))},
\end{equation}
which admits a unique root over the interval $[0,1/\theta]$ given by 
$\mu_0$ as defined in the last section. The function $z_n(\cdot)$ plays a central role in the establishment of the optimal policy. The following two lemmas state some useful properties of $z_n(\cdot)$.

\lm\label{property1}
Let $a_I>a_C.$
For any non-negative integer $n$, there exists $\mu_n\in(0,1/\theta)$ such that (i) $z_n(\mu_n)=0$; (ii) $z_n(\delta)<0$ for $\delta\in[0,\mu_n)$; (iii) $z_n(\delta)>0$ for $\delta\in(\mu_n,1/\theta].$
\lmm

\lm\label{property3}
Let $a_I>a_C.$
For any  $\delta\in(0,1/\theta)$ and any natural number $n$, $z_{n}(\delta)< z_{n-1}(\delta)$.
\lmm

\begin{figure}[H]
\begin{center}
\includegraphics[width=10cm]{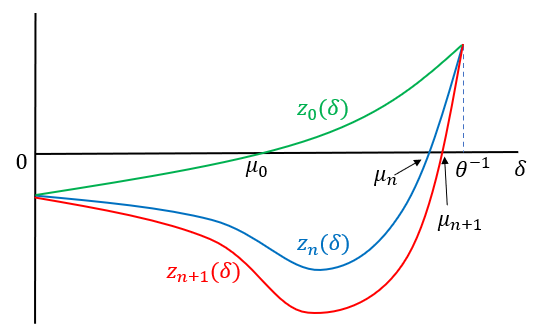}

\caption{The Single Crossing Property of $z_n(\cdot)$}\label{fig:zn-property}
\end{center}
\end{figure}

The qualitative features of $z_n(\cdot)$ are illustrated in Figure \ref{fig:zn-property}. As shown in Lemma \ref{property1}, $z_n(\cdot)$ has an important ``single-crossing'' property on $[0,1/\theta].$ The curve for $z_n(\cdot)$, starting from $z_n(0)<0$ and ending at $z_n(1/\theta)>0$, always cross the horizontal axis only once, which guarantees a unique root.
Moreover, according to Lemma \ref{property3}, for any non-negative integer $n$, the curve of $z_{n+1}(\cdot)$ always lies below that of $z_n(\cdot)$, which further suggests the monotonicity of the root associated with $z_n(\cdot)$ with respect to $n$.
Based on the properties of $z_n(\cdot)$ stated in Lemmas \ref{property1} and \ref{property3}, we can prove the following proposition.

% explain the proof idea of the two lemmas here

\prp\label{property-mu}
Let $a_I>a_C.$
For any $n\in\mathbb{N}$, there exists a unique root of $z_n(\delta)=0$ for $\delta\in(0,1/\theta)$, denoted by $\mu_n$. The sequence $\{\mu_n\}_{n=0}^\infty$ satisfies (i) $\mu_n>\mu_{n-1}$ for any $n\in\mathbb{N}$ and (ii) $\lim_{n\rightarrow\infty}{\mu}_n=1/\theta.$
\prpp

\noindent
According to Proposition \ref{property-mu}, there is a unique $\mu_n\in(0,1/\theta)$ such that $z_n(\mu_n)=0$. The family of rational functions $\{z_n(\cdot)\}_{n=0}^\infty$ then yield a well-defined sequence of technological parameters $\{\mu_n\}_{n=0}^\infty$. This sequence starts from $\mu_0$, is strictly increasing, and converges to $1/\theta.$ 
In what follows, we will demonstrate this sequence to be  the thresholds of the discount factor at which the optimal policy bifurcates.

\subsection{Optimal Delays in Extinction: Bifurcation Results}

In this subsection, we state the main theorem for extinction with investment for the case of $\theta\geq 1$, under which the economy can sustain a positive level of capital stock in the long run provided that a sufficient amount of recourse is allocated to the investment-good sector. 
The optimal policy for the (neglected) case of $\theta<1$, under which capital stock depletes in the long run even when the economy fully 
specializes in investment-good production, is qualitatively similar and will be discussed in the next section on the special cases of the model.

\begin{figure}[H]
\begin{center}
\includegraphics[width=8cm]{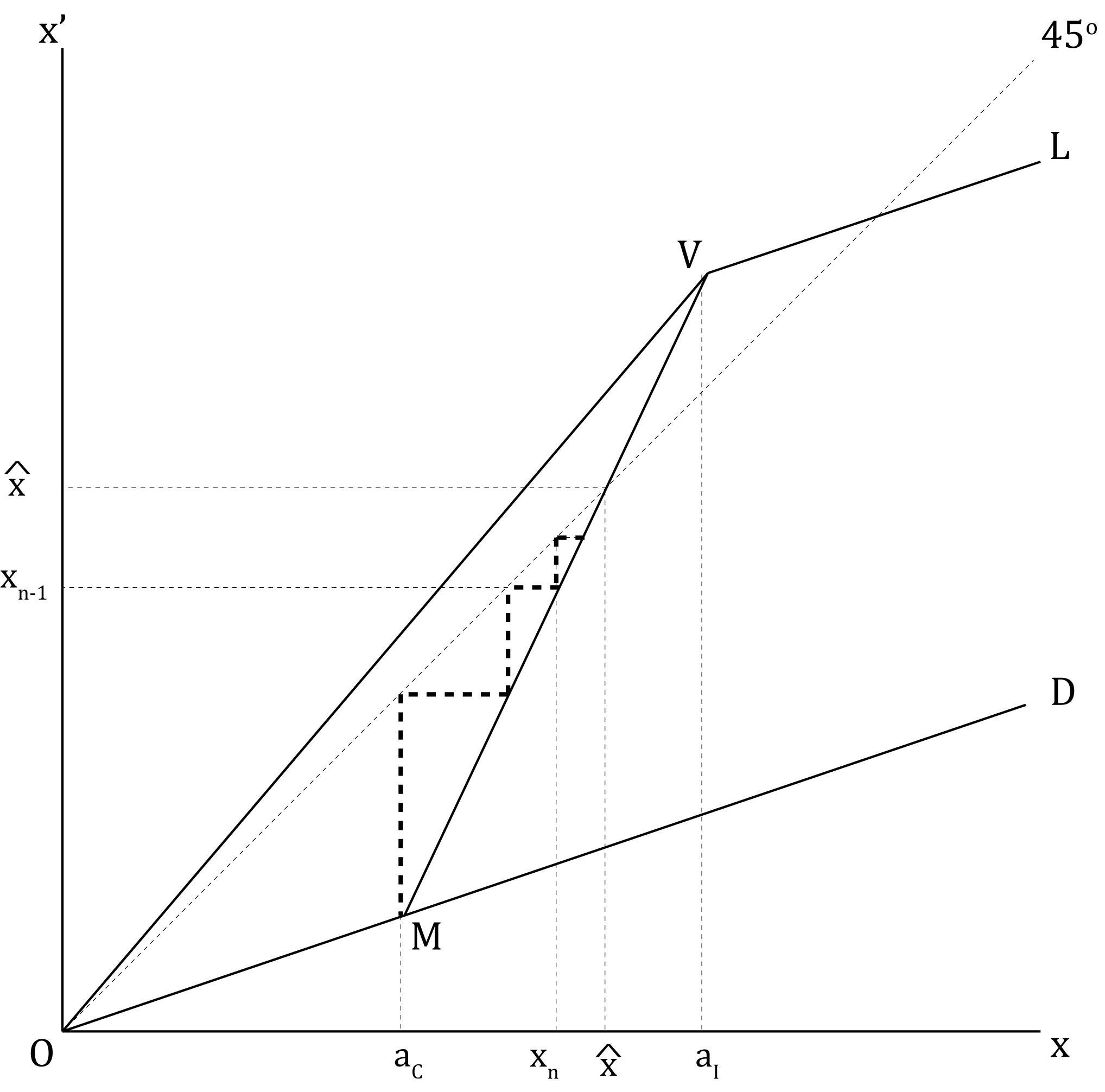}

\caption{The Construction of $x_n$ for $a_C<a_I$}\label{fig:xn}
\end{center}
\end{figure}

To ease the exposition of our characterization results, we define another sequence $\{x_n\}_{n=0}^\infty$ of thresholds for capital stock as follows: $x_0\equiv a_C$ and for any $n\in\mathbb{N}$,
\begin{equation}\label{eq:xn-def}
x_n=-\frac{1}{\zeta}\left(x_{n-1}-\frac{a_Cb}{a_C-a_I}\right). 
\end{equation}
We illustrate the construction of this sequence in Figure \ref{fig:xn}. The sequence starts from $x_0=a_C$. Given our construction, for any $n\in\mathbb{N}$, $(x_n,x_{n-1})$ is on the MV line where capital and labor are fully utilized. Geometrically, it is clear that this sequence converges to $\hat{x}.$

\lm \label{xn-property}
Let $a_C<a_I.$ The sequence $\{x_n\}_{n=0}^\infty$ is monotonically increasing: $x_n>x_{n-1}$ for any $n\in\mathbb{N}$. Further, $\lim_{n\rightarrow \infty} x_n=\hat{x}$ for $\theta>1$ and $\lim_{n\rightarrow \infty} x_n=a_I$ for $\theta=1.$
\lmm

Lemma \ref{xn-property} states formally the monotonicity and the limit of $\{x_n\}_{n=0}^\infty$ for $\theta\geq 1.$\fn{Recall $\hat{x}={a_Cb}/({b+d(a_C-a_I)})$, so the limit of $\{x_n\}_{n=0}^\infty$ can also be uniformly written as $\lim_{n\rightarrow \infty} x_n={a_Cb}/({b+d(a_C-a_I)})$ for both $\theta< 1$ and $\theta=1$.} With Lemma \ref{xn-property} and Proposition \ref{property-mu}, we are ready to present the main characterization result for extinction with investment.
The next proposition summarizes the bifurcation structure of the optimal policy with respect to the discount factor $\delta$ for $\delta\in(\mu_0,1/\theta).$ To bring out the most salient bifurcation pattern, we focus on the case of $\delta$ strictly between two consecutive thresholds. In the supplementary material, we present the additional characterization results for $\delta=\mu_n$ for which the optimal policy becomes a correspondence.

\prp\label{ac-less-ai-d<1-add}
Let $a_C< a_I$, $\theta\geq 1$, and $0<d<1$.
If ${\mu}_{n-1}<\delta<{\mu}_{n}$ for $n\in\mathbb{N}$,
then the optimal policy function is given by 
$$
g(x)=\left\{\begin{array}{ll}
(1-d)x & \mbox{for } x\in(0,a_C] \smallskip \\
-\zeta x+\frac{a_Cb}{a_C-a_I} & \mbox{for } x\in(a_C, x_n] \smallskip \\
x_{n-1} & \mbox{for } x\in(x_n,\frac{x_{n-1}}{1-d}] \smallskip \\
(1-d)x & \mbox{for } x\in(\frac{x_{n-1}}{1-d},\infty)
\end{array}\right..
$$
\prpp

\begin{figure}[H]
\begin{center}
\includegraphics[width=13cm]{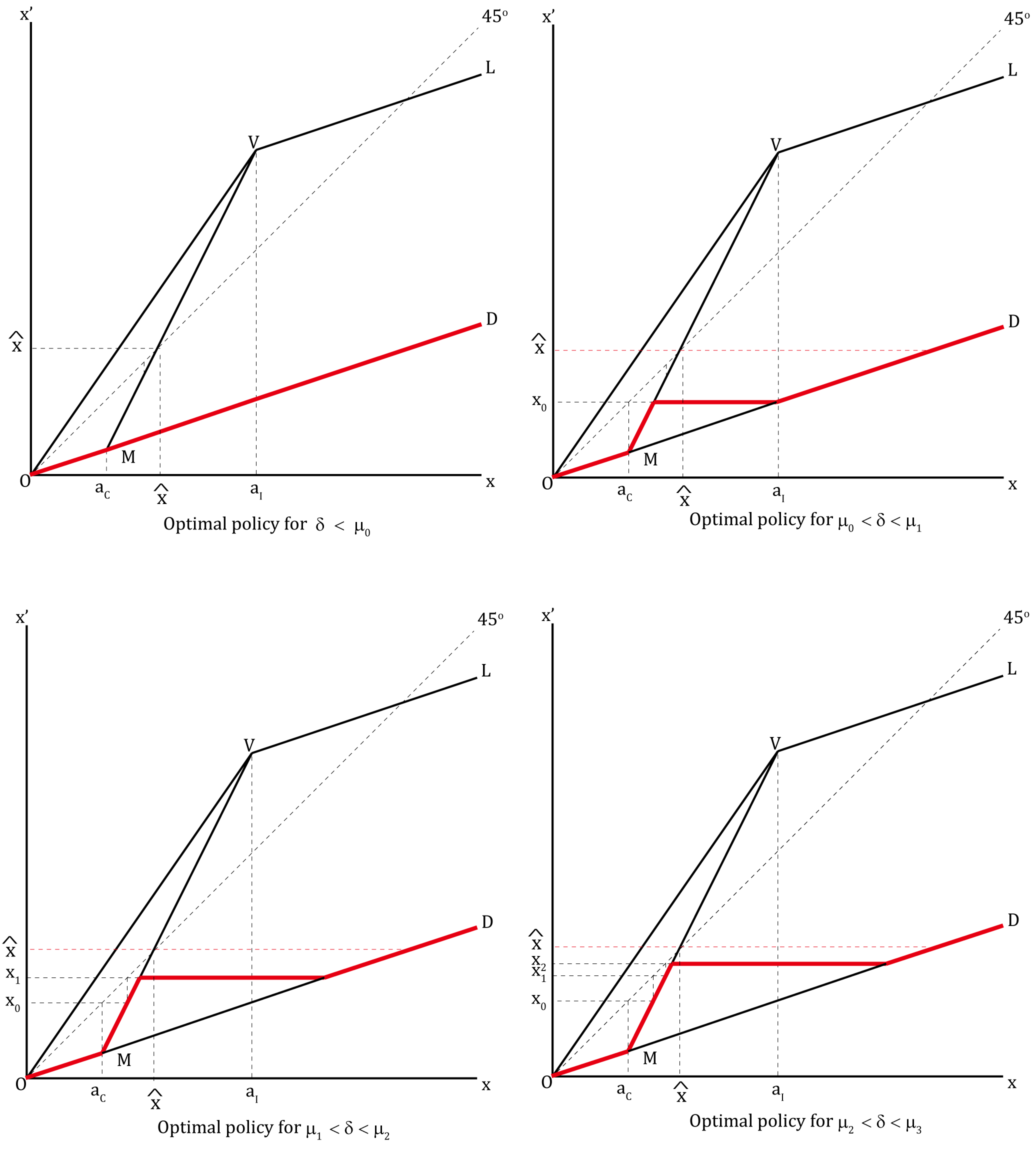}

\caption{The Optimal Policy for $a_C<a_I$, $\theta>1$ and $0<d<1$}\label{fig:ac<ai2}
\end{center}
\end{figure}

Figure \ref{fig:ac<ai2} shows how the optimal policy changes with the discount factor. The first panel corresponds to the case covered by Theorem \ref{ac-less-ai-extinction}. The second panel plots the optimal policy for $\delta\in(\mu_0,\mu_1)$. The policy deviates from the $OD$ line for $x\in(a_C,a_C/(1-d))$. For $x\in(a_C,x_1]$, the planner chooses to fully utilize the resources, and for $x\in(x_1,a_C/(1-d))$, the planner targets the investment-good production at a level such that capital stock tomorrow equals exactly $a_C$. Under this policy, for any initial stock above $a_C$, the economy deviates from the $OD$ line for exactly one period along its transition path.

The third and fourth panel of Figure \ref{fig:ac<ai2} illustrate the optimal policy for the discount factor in $(\mu_1,\mu_2)$ and that in $(\mu_2,\mu_3)$, respectively. The interval for capital stock at which the investment-good sector is activated enlarges as the discount factor increases, but the qualitative features of the transition dynamics remain the same: For $\delta\in(\mu_0,1/\theta),$ the optimal policy always consists of four segments, the middle two of which correspond to the case of positive investment. Moreover, for any positive integer $n$, if the discount factor is in $(\mu_{n-1},\mu_n)$, the economy will deviate from the $OD$ line by producing the investment goods for exactly $n$ periods. Since we know from Proposition \ref{property-mu} that the entire sequence $\{\mu_n\}_{n=0}^\infty$ is strictly increasing and converges to $1/\theta$, there are infinitely many bifurcations with respect to the discount factor. As the discount factor converges to $1/\theta$, the horizontal segment of the optimal policy will also approach the modified golden rule stock level $\hat{x}$, leading to more periods of delay in extinction.

\prp\label{ac-less-ai-d=1}
Let $a_C< a_I$, $\theta\geq 1$, and $d=1$.
If ${\mu}_{n-1}<\delta<{\mu}_{n}$ for $n\in\mathbb{N}$,
then the optimal policy function is given by 
$$
g(x)=\left\{\begin{array}{ll}
0 & \mbox{for } x\in(0,a_C] \smallskip \\
-\zeta x+\frac{a_Cb}{a_C-a_I} & \mbox{for } x\in(a_C, x_n] \smallskip \\
x_{n-1} & \mbox{for } x\in(x_n,\infty)
\end{array}\right..
$$
\prpp

\begin{figure}[H]
\begin{center}
\includegraphics[width=9cm]{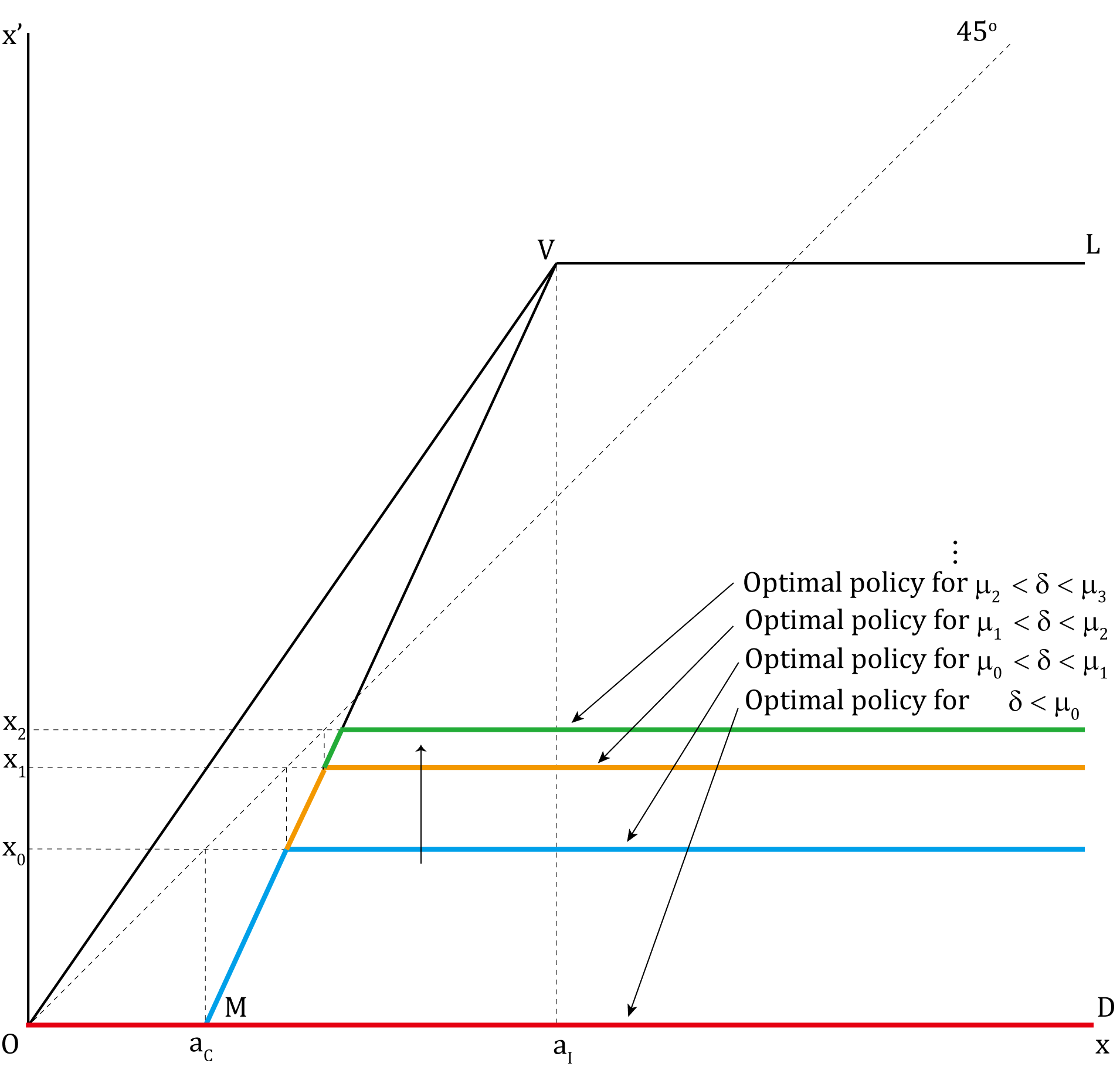}

\caption{The Optimal Policy for $a_C<a_I$, $\theta>1$, and $d=1$}\label{fig:ac<ai-d=1}
\end{center}
\end{figure}

To bring out optimal delays in extinction in its starkest form, we present the optimal policy for circulating capital ($d=1$) in Proposition \ref{ac-less-ai-d=1}. From Corollary \ref{thm-ac-less-ai}, we know the optimal policy yields immediate extinction for $\delta<\mu_0$. For $\delta\in(\mu_0,1/\theta)$, as shown in Figure \ref{fig:ac<ai-d=1}, the economy produces investment goods for any $x>a_C.$ The higher the discount factor is, the more periods the economy will sustain full utilization of resources (on the MV line) during the transition dynamics. In particular, for any initial stock above $x_n$ and any positive integer $n$, if the discount factor is in $(\mu_{n-1},\mu_n)$, the economy will produce $x_{n-1}$ units of investment goods in the first period, then stay on the phase of full utilization of production resources for $(n-1)$ periods, and reach the state of extinction after that.

From Proposition \ref{property-mu}, the interval $[\mu_0,1/\theta)$ can be partitioned into $\{[\mu_{n-1},\mu_n)\}_{n=1}^\infty$, so the following theorem follows immediately from the characterization results above.

\thm\label{ac-less-ai-no-extinction}
In the case of a capital-intensive investment-good sector $(a_I>a_C)$ and a positive capital stock being potentially sustainable $(\theta\geq1)$, all rates of time preference $\delta$ between the two technological bounds $(\mu_0\leq \delta<1/\theta)$ lead to an optimal policy under which the economy is in the extinction phase with investment.
\thmm

Theorem \ref{ac-less-ai-no-extinction} and the results in the previous section point to an important asymmetry: investment along the transition path to extinction can possibly occur {\it only}  in the case of a capital-intensive investment-good sector. To understand the source of this asymmetry, we consider the following intertemporal decision. Let capital stock today $x$ to be slightly above $a_C$ such that in the absence of any investment, capital stock tomorrow $x'=(1-d)x$ falls under $a_C$. Suppose the planner deviates from full specialization in consumption goods to allocate infinitesimal amount of resources to investment. Given $x>a_C$ and investment being infinitesimal, the economy is still in the region of excess capacity and thus the marginal cost of investment in terms of the consumption goods today is given by $1/b.$ We show that regardless of the capital intensity condition,
for $\delta<1/\theta,$ it is optimal for the economy to specialize in the consumption-good sector when capital stock is below $a_C$ and there is excess supply of labor. 
Thus, for $x'<a_C$, the economy enters the extinction phase without investment and the marginal return to investment   is given by $$\delta\left(\frac{1}{a_C}+\frac{\delta(1-d)}{a_C}+\frac{\delta^2(1-d)^2}{a_C}+\cdots\right)=\frac{\delta}{a_C(1-\delta(1-d))}.$$ When the consumption-good sector is capital intensive $(a_C>a_I)$, for $\delta<1/\theta$,
$$\frac{\delta}{a_C(1-\delta(1-d))}<\frac{1}{a_C(\theta-(1-d))}=\frac{a_I}{a_C}\cdot\frac{1}{b}<\frac{1}{b},$$
where the second inequality follows from $a_C>a_I$, which implies the marginal cost of investment exceeds the marginal return. In contrast, when the investment-good sector is capital intensive, for $\delta\in(\mu_0,1/\theta)$, 
$$\frac{\delta}{a_C(1-\delta(1-d))}>\frac{1}{b}.$$
Because it requires relatively less capital to produce consumption goods for $a_C<a_I$, the marginal return to investment can potentially exceed the marginal cost. As a result, optimal extinction with investment emerges in the case of a capital-intensive investment-good sector. Moreover, as the discount factor increases within the interval of $(\mu_0,1/\theta)$, the planner is more patient and thus has more incentives to invest, which translates into more periods of delay in extinction. 

\section{Optimal Policy: Some Special Cases}

\subsection{The Unsustainable Technology Case: $\theta<1$}

We now consider the case of $\theta<1$. In this case, regardless of the investment decision, it is technologically infeasible to sustain any positive capital stock in the long run and extinction is 
guaranteed for any discount factor.  Since the existing literature assumes the fulfillment of the $\delta$-normality condition with $\delta>1/\theta$, which requires $\theta>1$, this unsustainable technology case has largely been neglected. Since Theorem \ref{ac-greater-ai} applies to both $\theta\geq1$ and $\theta<1$, we focus on the case of a capital-intensive investment-good sector.
 Our next proposition establishes the possibility of deferred extinction for this neglected case.

\prp\label{ac-less-ai-d<1-theta<1-add}
Let $a_C< a_I$, $\theta< 1$, and $0<d<1$.

\noindent
(i) If $\mu_0\geq 1$,  the optimal policy function is given by $g(x)=(1-d)x$ for any $x.$

\noindent
(ii) If $\mu_0<1$,  there exists $n_0\in \mathbb{N}$ such that $\mu_{n_0-1}<1\leq \mu_{n_0}.$
For $\delta\leq \mu_{n_0-1}$,  characterization of the optimal policy follows the case of $\theta\geq 1$.
For $\mu_{n_0-1}<\delta<1$,  the optimal policy function is given by 
$$
g(x)=\left\{\begin{array}{ll}
(1-d)x & \mbox{for } x\in(0,a_C] \smallskip \\
-\zeta x+\frac{a_Cb}{a_C-a_I} & \mbox{for } x\in(a_C, x_{n_0}] \smallskip \\
x_{n_0-1} & \mbox{for } x\in(x_{n_0},\frac{x_{n_0-1}}{1-d}] \smallskip \\
(1-d)x & \mbox{for } x\in(\frac{x_{n_0-1}}{1-d},\infty)
\end{array}\right..
$$
\prpp

\begin{figure}[H]
\begin{center}
\includegraphics[width=9cm]{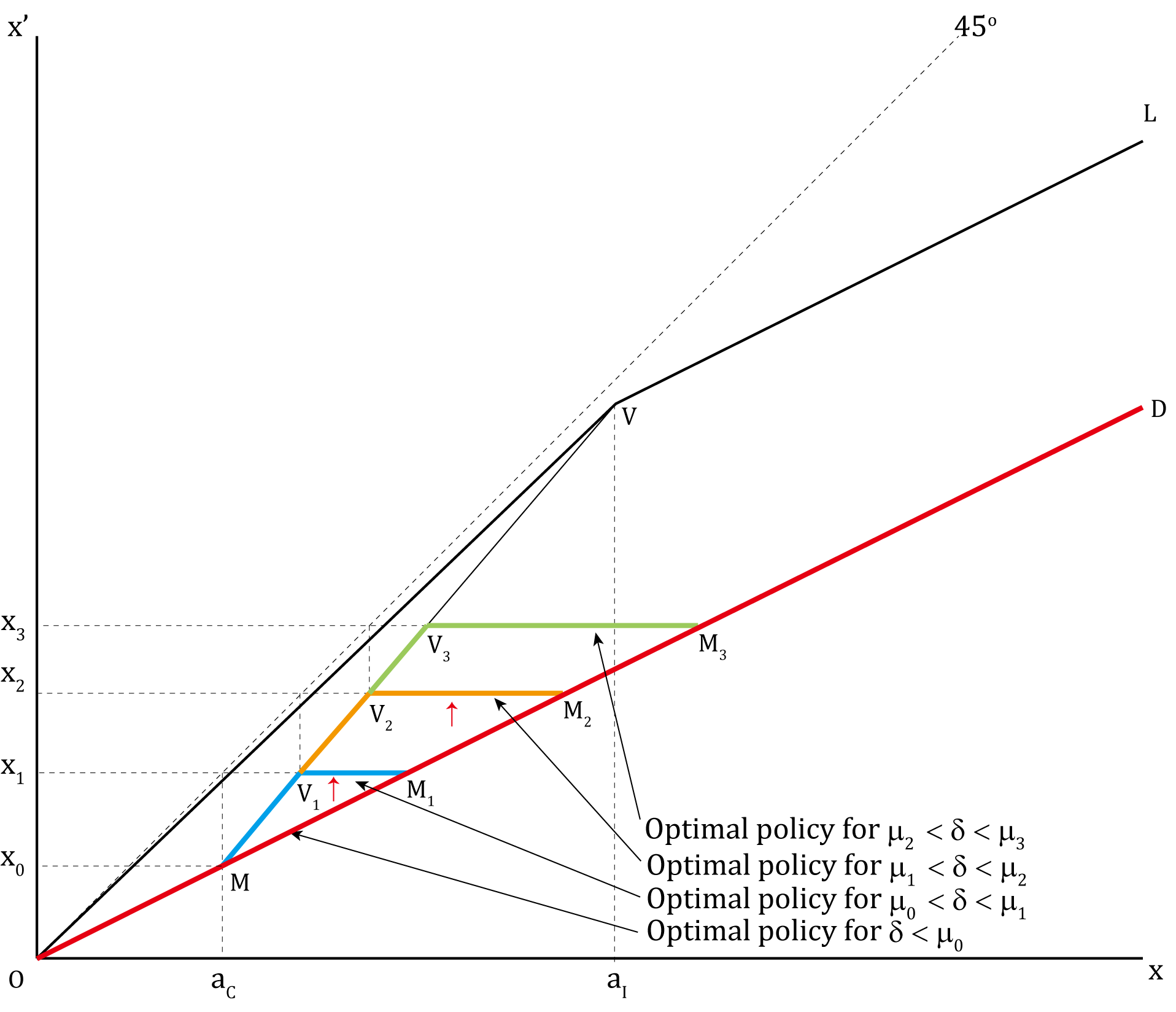}

\caption{The Optimal Policy for $a_C<a_I$ and $\theta<1$}\label{fig:ac<ai-theta<1}
\end{center}
\end{figure}

According to Proposition \ref{ac-less-ai-d<1-theta<1-add}, even the investment-good sector is highly unproductive, as long as the technological lower bound $\mu_0$ and the discount factor satisfy $\mu_0<\delta<1$, the social planner would still have the incentive to allocate resources to the investment-good sector along the transition path to extinction. Qualitatively, the main difference between this case and the benchmark case of $\theta\geq 1$ in the previous section is that there are only a {\it finite} number of bifurcations of the optimal policy with respect to the discount factor for $\theta<1$. 
Figure \ref{fig:ac<ai-theta<1} illustrates the bifurcation structure for the case of $n_0=3$. Since $\mu_3\geq 1>\mu_2$, there are three bifurcation values for the discount factor, $\mu_0$, $\mu_1$ and $\mu_2$. For $\delta>\mu_2$, the optimal policy is always represented by $OMV_3M_3D.$
In the next proposition, we extend the result above to the case of circulating capital.

\prp\label{ac-less-ai-d=1-add2}
Let $a_C< a_I$, $\theta< 1$, and $d=1$. 
If $\mu_0\geq 1$,  the optimal policy function is given by $g(x)=0$ for $x>0.$ If $\mu_0<1$,  there
exists $n_0\in \mathbb{N}$ such that $\mu_{n_0-1}<1\leq \mu_{n_0}.$
If $\delta\leq \mu_{n_0-1}$,  characterization of  the optimal policy  follows the case of $\theta\geq 1$. If $\mu_{n_0-1}<\delta<1$,  the optimal policy function is given by 
$$
g(x)=\left\{\begin{array}{ll}
0 & \mbox{for } x\in(0,a_C] \smallskip \\
-\zeta x+\frac{a_Cb}{a_C-a_I} & \mbox{for } x\in(a_C, x_{n_0}] \smallskip \\
x_{n_0-1} & \mbox{for } x\in(x_{n_0},\infty)
\end{array}\right..
$$
\prpp

\begin{figure}[H]
\begin{center}
\includegraphics[width=13cm]{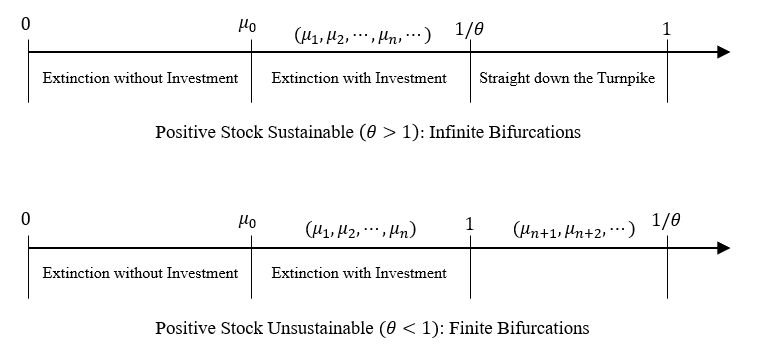}

\caption{The Sequence $\{\mu_n\}^\infty_{n=0}$ and $1/\theta$ for $a_C<a_I$}\label{fig:ordering}
\end{center}
\end{figure}

To summarize the bifurcation structure for the case of a capital-intensive investment-good sector,
Figure \ref{fig:ordering} illustrates the ordering of the thresholds for the discount factor with respect to $1/\theta$ and $1$. There are generically two possibilities. For $\theta>1$, the unit interval can be partitioned into three regions. The middle region  contains the sequence $\{\mu_n\}_{n=0}^\infty$, which gives rise to infinite bifurcations. For $\theta<1$, only a finite number of elements in the sequence will be in the unit interval, leading to finite bifurcations. It should be noted that the second panel of Figure \ref{fig:ordering} is based on the assumption of $\mu_0<1$. It is also possible to have $\mu_0\geq 1$, in which case there is  extinction without investment for any discount factor.

\subsection{The Knife-edge Case for the Discount Factor: $\delta=1/\theta$}

In this subsection, we present the results concerning an important bifurcation value for the discount factor, $\delta=1/\theta.$ For this knife-edge case, the optimal policy becomes a correspondence and there exists a continuum of non-trivial stationary optimal stocks.

% simulation results for \delta=1/\theta and a_C>a_I, see IJETConsumption
\prp\label{ac-greater-ai2}
Let $a_C>a_I$, $\theta>1$, and $\delta=1/\theta$. Then the optimal policy correspondence is given by 
$$
h(x)=\left\{\begin{array}{ll}
[(1-d)x,\min\{a_C,\theta x\}] & \mbox{for } x\in(0,a_I] \smallskip \\
\left[(1-d)x,\min\left\{a_C,-\zeta x+\frac{a_Cb}{a_C-a_I}\right\}\right] & \mbox{for } x\in(a_I, {a_C}] \smallskip \\
\{(1-d)x\} & \mbox{for } x\in({a_C},\infty)
\end{array}\right.
$$
\prpp

\prp\label{ac-less-ai-rho=1/theta}
Let $a_C< a_I$, $\theta> 1$, and $\delta=1/\theta$. The optimal policy correspondence is given by
$$
h(x)=\left\{\begin{array}{ll}
{[\max\{(1-d)x,-\zeta x+\frac{a_Cb}{a_C-a_I}\},\theta x]} & \mbox{for } x\in(0,\frac{\hat{x}}{\theta}] \smallskip \\
{[\max\{(1-d)x,-\zeta x+\frac{a_Cb}{a_C-a_I}\},\hat{x}]} & \mbox{for } x\in(\frac{\hat{x}}{\theta}, \hat{x}] \smallskip \\
\{\max\{\hat{x},(1-d)x\}\} & \mbox{for } x\in(\hat{x},\infty)
\end{array}\right..
$$
\prpp

\begin{figure}[H]
\begin{center}
\includegraphics[width=13cm]{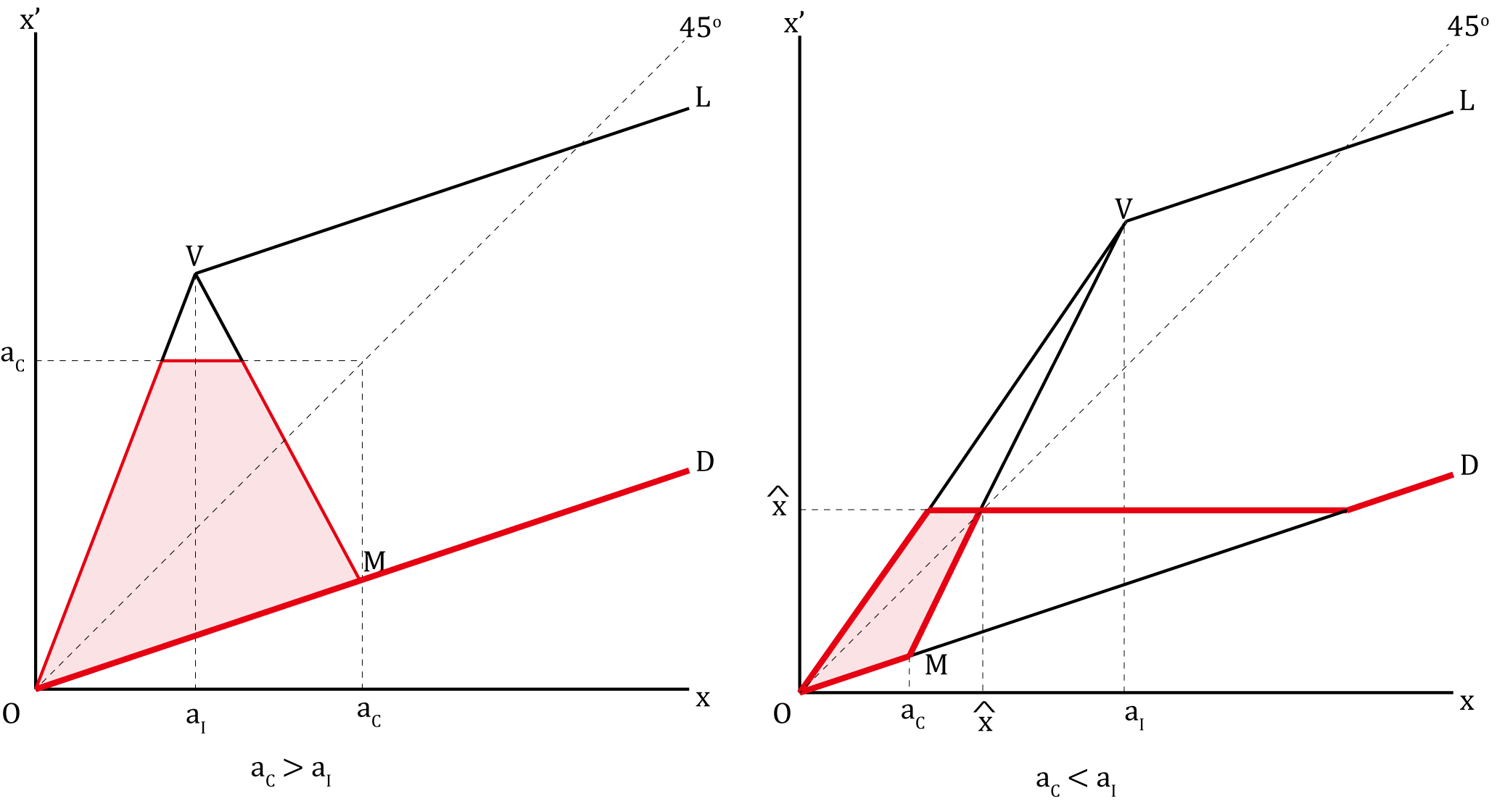}

\caption{The Optimal Policy for $\delta=1/\theta$}\label{fig:rho=1/theta}
\end{center}
\end{figure}

Propositions \ref{ac-greater-ai2} and \ref{ac-less-ai-rho=1/theta} present the optimal policy correspondence for $a_C>a_I$ and $a_C<a_I$, respectively. Figure \ref{fig:rho=1/theta} illustrates the optimal policy for both cases, in which the shaded area in red represents the optimal policy being non-unique. In particular, for both cases, any capital stock in $(0,\hat{x}]$ is a non-trivial stationary optimal stock, thus testifying that $\delta$-normality is not a necessary condition for the existence of a non-trivial stationary optimal stock. 

%\begin{figure}[H]
%\begin{center}
%\includegraphics[width=6cm]{figures/figure6.png}

%\caption{The Optimal Policy for $a_C=a_I$ and $\delta\leq1/\theta$}\label{fig:ac=ai}
%\end{center}
%\end{figure}

\subsection{The One-Sector Case: $a_C=a_I$}
We now consider the optimal policy for the case of two sectors having the same capital intensity ($a_C=a_I)$, which resembles a one-sector economy. The optimal policy for this case follows closely that for the case of a capital-intensive consumption-good sector ($a_C > a_I$), with a slight difference for $\delta=1/\theta$, as summarized in the following proposition.   

\prp\label{ac=ai}
Consider the one-sector case ($a_C=a_I$). If $\delta<1/\theta$, then the optimal policy function is given by $g(x)=(1-d) x$ for any $x>0$.
If $\delta=1/\theta$, then the optimal policy correspondence is given by 
$$
h(x)=\left\{\begin{array}{ll}
[(1-d)x,\min\{a_C,\theta x\}] & \mbox{for } x\in(0,a_C] \smallskip \\
{[(1-d)x,\max\{a_C,(1-d)x\}]} & \mbox{for } x\in(a_C,\infty)
\end{array}\right.
$$
\prpp

\subsection{A Numerical Example}

\begin{figure}[H]
\begin{center}
\includegraphics[width=6cm]{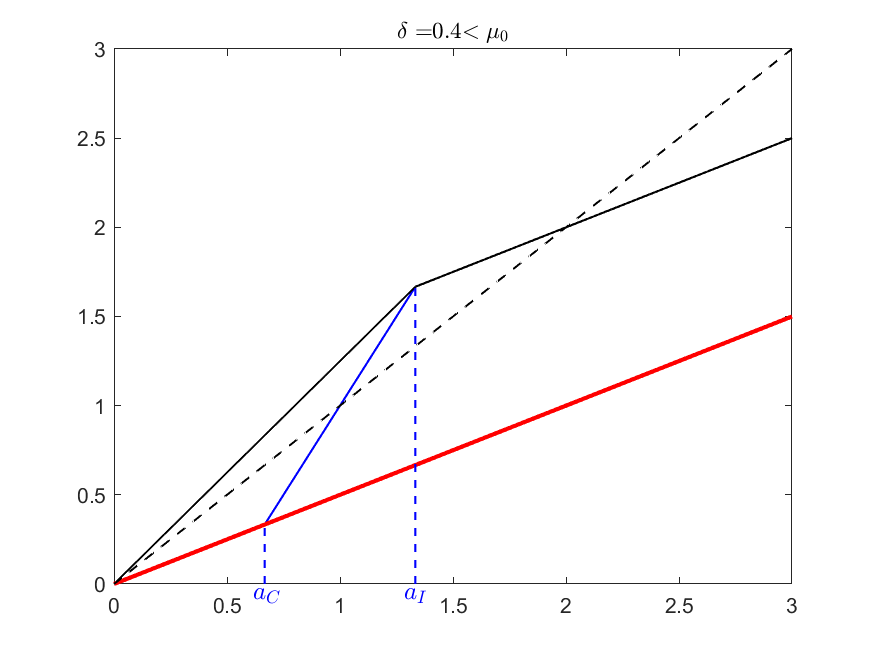} \includegraphics[width=6cm]{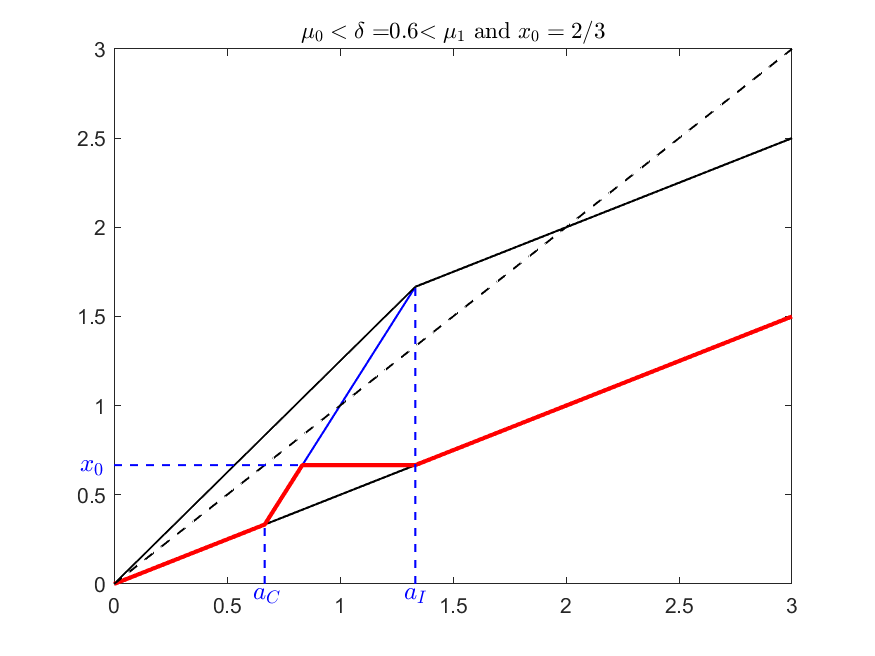}

\includegraphics[width=6cm]{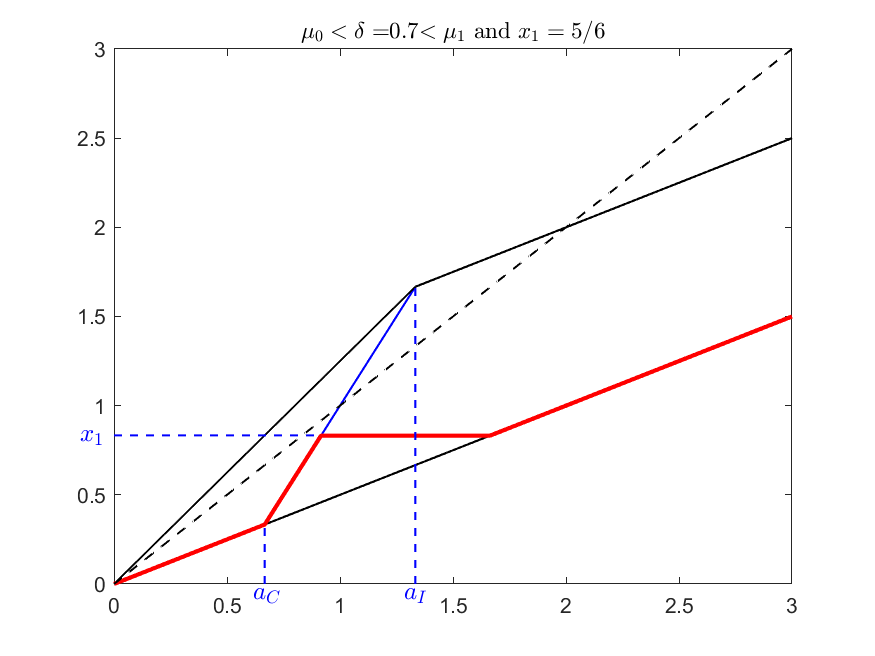}

\caption{A Numerical Example ($b=1$, $a_C=2/3$, $a_I=4/3,$ $d=1/2$)}\label{fig:numeric}
\end{center}
\end{figure}

We finally consider a numerical example of how the optimal policy bifurcates with respect to the discount factor in the case of a capital-intensive investment-good sector. Let $b=1$, $a_C=2/3$, $a_I=4/3$, and $d=1/2.$ From Equations (\ref{eq:zeta}) and (\ref{eq:theta}), we have $\theta=5/4$ and $\zeta=-2.$ Then, from Equations (\ref{eq:polynomial}) and (\ref{eq:z0}), we have
\begin{eqnarray*}
z_0(\delta)&=& -\frac{1}{b}+\delta\left(\frac{1}{a_C(1-\delta(1-d))}\right)=-1+\frac{3\delta}{2-\delta},\\
z_1(\delta)&=& -\frac{1}{b}+\delta\left(-\frac{1}{a_I-a_C}-\frac{\delta\zeta}{a_C(1-\delta(1-d))}\right)=-1+\delta\left(-\frac{3}{2}+\frac{6\delta}{2-\delta}\right),\\
z_2(\delta)&=& -\frac{1}{b}+\delta\left(-\frac{1-\delta\zeta}{a_I-a_C}+\frac{(\delta\zeta)^2}{a_C(1-\delta(1-d))}\right)=-1+\delta\left(-\frac{3+6\delta}{2}+\frac{12\delta^2}{2-\delta} \right),
\end{eqnarray*}
which yield $\mu_0=1/2$, $\mu_1=2/3$, $\mu_2\approx 0.73$, the first three bifurcation values for the discount factor, and we know $\lim_{n\rightarrow\infty}\mu_n=1/\theta=4/5.$ From Equation (\ref{eq:xn-def}), we obtain $x_0=a_C=2/3$, $x_1=5/6$, and $x_2=11/12.$ The optimal policy functions for $\delta=0.4\in(0,\mu_0)$, $\delta=0.6\in(\mu_0,\mu_1)$, and $\delta=0.7\in(\mu_1,\mu_2)$ are plotted in Figure \ref{fig:numeric}.

%Consider a numeric experiment: $a_C=1$, $a_I=3$, $b=1$, $d=0.5$, then $\zeta=-1$, $\theta=5/6<1.$ $$z_n(\delta)=\frac{-5\delta^{n+2}+6\delta^{n+1}-\delta^2+4\delta-4}{(2-\delta)(2-2\delta)},$$
%which yields $\mu_0=2/3$, $\mu_1=2/\sqrt{5}$, $\mu_2=1.$

%We consider a numerical experiment with the following parameter values: $b=1$, $a_C=1.2$, $a_I=1.5$, $d=0.5$, which implies $\theta\approx 1.166667$, $\zeta\approx-3.833333;$ $\hat{x}\approx 1.411765.$ The corresponding cutoffs for $\delta$ would be $\hat{\mu}_0=\frac{a_C}{b+a_C(1-d)}=0.75$, $\hat{\mu}_1\in(0.82,0.835)$, $\hat{\mu}_2\in(0.835,0.85)$, $\hat{\mu}_3\in(0.85,0.856)$; accordingly the ``height'' of the flat part is given by $c_0=a_C=1.2$, $c_1\approx1.3565$, $c_2\approx 1.3974$, $c_3\approx 1.4080.$

\section{Concluding Remarks}
In summary,  we provide a complete and comprehensive characterization of optimal policy for the two-sector RSL model in the absence of $\delta$-normality: a categorization of extinction when the discount factor is below the MRT with zero consumption ($\delta\leq 1/\theta$). For $\delta <1/\theta$, the optimal policy always yields extinction without investment along the transition path in the case of a capital-intensive consumption-good sector, whereas an intricate bifurcation structure emerges in the case of a capital-intensive investment-good sector. If the investment-good sector is capital intensive, and if the discount factor is between two technological bounds $(\mu_0<\delta<1/\theta)$, the planner needs to  allocate resources to the investment-good sector with resources sometimes being fully utilized so that extinction can be deferred.  The results are easy to state, but difficult to obtain.\fn{\cite{fk06} use as their epigraph  Amartya Sen's sentiments  that the two-sector model is not for the faint-hearted, and even though relegated to supplementary material, the detailed calculations may yield a mathematical insight that can be isolated as a lemma, and then offer in future work, methods that do rely quite as much on brute force in  \lq\lq guess and verify" analyses.}

 Not surprisingly,  our investigation leaves several questions open. For one thing, the results of this paper  lead us to pose the novel question as to the optimal policy for the unsustainable case $(\theta<1)$ in the RSL model without discounting, which is to say, when the discount factor is unity, its maximum value.  It is to us natural to pose survival and extinction issue, crucial as they are to environmental,  resource and ecological economics,  in an  
 undiscounted setting: if Ramsey's hesitations regarding discounting apply anywhere, they do so here. 
  Secondly, and more importantly,  from an abstract theoretical point of view, one could view both the RSS and RSL model   as exalted examples.\footnote{Even  Morishima's matchbox model has two activities in each of its two sectors; see \cite{mo65c}, and note the emphasis on von Neumann's paper in Parts II and III, two parts out of  four. Also see \cite{ko64, ko71} and Footnote~\ref{fn:morishima} above.   \label{fn:morishima1}} The  multi-sectoral extensions already available 
 admit  activities for intermediate goods and services, and are 
eminently suited to handle issues arising from  technological structures that pollute.\fn{We may single out \cite{ga56}, \cite{mck68, mck02},  \cite{in68} and   \cite{ko71}.}   But once one shifts from the RSL terrain to different formulations, it is far from clear as to the conditions under which extinction would obtain, and if it obtains; see for example \cite{ca91} for a condition under which sustainability obtains.    Furthermore,  the issues relate to the future and future-uncertainty, and surely demand that we move on from the stylized deterministic single-technique two-sector  setting of the RSL model to  a stochastic multi-sectoral setting.

Moving on to the stochastic environment in his last two chapters, \cite{ma20} quotes Frisch:

 \bqu   One way which I believe is particularly fruitful and promising is
to study what would become  the solution of a deterministic
dynamic system if it were exposed to a stream of erratic shocks
that constantly upsets its evolution.  \equ 

\nt     In an earlier analysis, \cite{ka06} offers  sufficient conditions for almost sure convergence to zero stock   in the context of the stochastic aggregative growth model. But this is just the tip of the iceberg:  even on limiting oneself to an aggregative stochastic environment, one has a rich literature to contend with.\fn{In a pioneering paper, \cite{st02} provides sufficient conditions for existence and stability of a positive steady state for a stochastic model of optimal growth with unbounded shock. \cite{ns05} apply an Euler equation technique to extend the stability result. For related discussions geared more towards the context of renewable resources management, see  \cite{or06}, \cite{mr06} and the references therein; also see \cite{ka07}, \cite{kr06,kr07}, \cite{ks14}, and  \cite{mr12,mr21}. In the context of stochastic equilibrium theory, see \cite{mh05} and their references.} 
In recent work devoted  to the treatment of the aggregative growth model,  \cite{kz21} present analysis of the  random two-sector RSS model.  The interest of this work lies in its showing   that the variety of cases in the deterministic setting of the  model get eliminated in a setting with uncertainty.    It is natural to ask whether the same would be true for the  RSL setting.

   In conclusion,  all  these questions  notwithstanding, one can hardly  forego  the larger overview of the issues of survival and extinction in economic theory:  we surely ought not to be  hamstrung  by the particular  exogenous growth setting studied here, and  allow ourselves to indulge in the big conceptual questions in alternative models, without showing any lack of respect for resolving small technical difficulties that may arise.  Two sets of models would be high on this aspirational research agenda: within economic dynamics and growth theory, models of endogenous growth,\fn{As such, the nod to the endogenous growth literature, and the stylized facts that it addresses, as for example in \cite{ghos}, \cite{jo05} and \cite{jr10}, is certainly not a mere strategic nod in a technical paper: it is very much the next step of the program. } and within resource economics, those having to do with exhaustible resources   as pioneered by Clark.\fn{Clarke writes, \lq\lq  Roughly speaking, conservation means saving for the future. The theory of resource conservation can therefore be addressed as a branch of the theory of capital and investment;" see the epigraph of \citeauthor{ma20} (2020, Chapter 5), and references to work on the S-shaped production function.  In addition to  \cite{cl10} and \cite{da82}, see  \cite{mi00} where among five examples  illustrating the intertemporal theory of resource allocation, Example 2.3 is Clarke's fishery model.}

\bibliographystyle{ecca} 
\bibliography{references}

\newpage

%%%%%%%%%%%%%%%%%%%%%%%%%%%
% Figures in the main text
%%%%%%%%%%%%%%%%%%%%%%%%%%%

\newpage

%%%%%%%%%%%%%%%%%%%%%%%%%%%%%%%%%
% Supplementary Material -- to be incorporated into the paper
%%%%%%%%%%%%%%%%%%%%%%%%%%%%%%%%%

\begin{appendix}

\clearpage
\pagenumbering{arabic}% resets `page` counter to 1
\renewcommand*{\thepage}{A\arabic{page}}

\newtheorem{propp}{Proposition}
\renewcommand{\thepropp}{A\arabic{propp}}
\setcounter{propp}{0} 

\newtheorem{lmaa}{Lemma}
\renewcommand{\thelmaa}{A\arabic{lmaa}}
\setcounter{lmaa}{0}

\numberwithin{equation}{section}
\setcounter{equation}{0}

\section{Supplementary Material}

We organize the supplementary material in three parts. We first present additional characterization results on the optimal policy correspondence. We then provide the proofs of all the main results presented in the paper. Last, we present and prove lemmas that are used in the proofs of the main results.

\subsection{Further Characterization Results}

We present two additional characterization results on the optimal policy when the discount factor is equal to a cutoff value $\mu_n$. Like the knife-edge case we identify in the paper for $\delta=1/\theta$, the optimal policy   becomes a correspondence. 
Proposition \ref{ac-less-ai-d<1-add2} concerns the case of durable capital $(0<d<1)$ and Proposition \ref{ac-less-ai-d=12} concerns the case of circulating capital $(d=1)$.

\begin{propp} \label{ac-less-ai-d<1-add2}
Let $a_C< a_I$, $\theta\geq 1$, and $0<d<1$.

\noindent
(i) If $\delta=\mu_0$, then the optimal policy correspondence is given by
$$
h(x)=\left\{\begin{array}{ll}
\{(1-d)x\} & \mbox{for } x\in(0,a_C] \smallskip \\
{[(1-d)x,-\zeta x+\frac{a_Cb}{a_C-a_I}]} & \mbox{for } x\in(a_C, x_1] \smallskip \\
{[(1-d)x,a_C]} & \mbox{for } x\in(x_1,\frac{a_C}{1-d}] \smallskip \\
\{(1-d)x\} & \mbox{for } x\in(\frac{a_C}{1-d},\infty)
\end{array}\right..
$$
\noindent
(ii) If $\delta=\mu_{n}$ for $n\in\mathbb{N}$, then the optimal policy correspondence is given by
$$
h(x)=\left\{\begin{array}{ll}
\{(1-d)x\} & \mbox{for } x\in(0,a_C] \smallskip \\
\{-\zeta x+\frac{a_Cb}{a_C-a_I}\} & \mbox{for } x\in(a_C, x_{n}] \smallskip \\
{[x_{n-1},\min\{-\zeta x+\frac{a_Cb}{a_C-a_I},x_n\}]} & \mbox{for } x\in(x_{n},\frac{x_{n-1}}{1-d}] \smallskip \\
{[(1-d)x,\min\{-\zeta x+\frac{a_Cb}{a_C-a_I},x_n\}]} & \mbox{for } x\in(\frac{x_{n-1}}{1-d},\frac{x_{n}}{1-d}] \smallskip \\
\{(1-d)x\} & \mbox{for } x\in(\frac{x_{n}}{1-d},\infty)
\end{array}\right..
$$
\end{propp}

\begin{propp} \label{ac-less-ai-d=12}
Let $a_C< a_I$, $\theta\geq 1$, and $d=1$.

\noindent
(i) If $\delta=\mu_0$, then the optimal policy correspondence is given by
$$
h(x)=\left\{\begin{array}{ll}
\{0\} & \mbox{for } x\in(0,a_C] \smallskip \\
{[0,-\zeta x+\frac{a_Cb}{a_C-a_I}]} & \mbox{for } x\in(a_C, x_1] \smallskip \\
{[0,a_C]} & \mbox{for } x\in(x_1,\infty)
\end{array}\right..
$$
\noindent
(ii) If $\delta=\mu_{n}$ for $n\in\mathbb{N}$, then the optimal policy correspondence is given by
$$
h(x)=\left\{\begin{array}{ll}
\{0\} & \mbox{for } x\in(0,a_C] \smallskip \\
\{-\zeta x+\frac{a_Cb}{a_C-a_I}\} & \mbox{for } x\in(a_C, x_{n}] \smallskip \\
{[x_{n-1},-\zeta x+\frac{a_Cb}{a_C-a_I}]} & \mbox{for } x\in(x_{n},x_{n+1}] \smallskip \\
{[x_{n-1},x_n]} & \mbox{for } x\in(x_{n+1},\infty)
\end{array}\right..
$$
\end{propp}

\subsection{Proofs}

%%%%%%%%%%%%%%%%%%%%%%%%%%%%%%%%%%%%%%%%%%%%%%
% Proof of LEMMA 1
%%%%%%%%%%%%%%%%%%%%%%%%%%%%%%%%%%%%%%%%%%%%%%

\noindent
{\it \textbf{Proof of Lemma \ref{rho-normal}}}: We first prove the ``if'' part.
Let $\delta>1/\theta$. Pick $\varepsilon>0$ such that $\varepsilon<a_I$ and $\varepsilon<a_C$. Since $\delta>1/\theta$, $\theta>1/\delta$ and $\theta>(\theta+1/\delta)/2>1/\delta$. Since $0<\varepsilon <a_C$, $0<\varepsilon<a_I,$ and $(\theta+1/\delta)/2<\theta$, $(\varepsilon, (\theta+1/\delta)\varepsilon/2)\in\Omega$
and $u(\varepsilon, (\theta+1/\delta)\varepsilon/2)>0=u(0,0).$ Moreover, since $\delta>1/\theta$, $\delta((\theta+1/\delta)\varepsilon/2)=(\delta\theta+1)\varepsilon/2>\varepsilon$. 
Thus, the economy is $\delta$-normal. Now we turn to the ``only if'' part. Let $\delta\leq 1/\theta$. For any $(x,x')\in\Omega$, $x'\leq \theta x\leq x/\delta$ which implies $x\geq \delta x'$. The equality holds only if $\delta=1/\theta$ and $x'=\theta x$. However, if $x'=\theta x$, $u(x,x')=0=u(0,0)$, so the economy is not $\delta$-normal. Then, we have obtained the desired conclusion. \qed

\bigskip

%%%%%%%%%%%%%%%%%%%%%%%%%%%%%%%%%%%%%%%%%%%%%%
% Proof of THEOREM 1
%%%%%%%%%%%%%%%%%%%%%%%%%%%%%%%%%%%%%%%%%%%%%%

% the proof does not depend on whether theta is >,=,or <1.
\noindent
{\it \textbf{Proof of Theorem \ref{ac-greater-ai}}}: We first consider $0<d<1.$ We adopt the standard ``guess-and-verify'' approach. Postulate a candidate value function based on the policy function $g(x)=(1-d)x$ for any $x$:
\begin{equation} \label{eq:value-function-thm1}
W(x)= \left\{\begin{array}{lll}

\frac{x}{a_C(1-\delta(1-d))} & \mbox{for } x\in [0,a_C] \medskip \\

\frac{1-\delta^n}{1-\delta}+\frac{\delta^n(1-d)^n x}{a_C(1-\delta(1-d))} & \mbox{for } x\in(\frac{a_C}{(1-d)^{n-1}},\frac{a_C}{(1-d)^{n}}]

\end{array} \right.,
\end{equation}
where $n=1,2,3...$ 
We now claim that for any $x$, $W(x)$ satisfies the Bellman equation $$W(x)=\max_{x'\in \Gamma(x)}\{u(x,x')+\delta W(x')\}.$$ To this end, we consider four cases: (i) $x\in(0,a_I);$ (ii) $x\in[a_I,a_C];$ (iii) $x\in(a_C,\frac{a_C}{1-d}];$ (iv) $x\in(\frac{a_C}{(1-d)^{n-1}},\frac{a_C}{(1-d)^{n}}]$ for $n=2,3,4...$

\medskip
\noindent {\bf Case (i):} For any $(x,x')\in\Omega$ such that $x<a_I$, we have $(a_C-a_I)x'< ((1-d)(a_C-a_I)-b) x+a_Cb$. Using the reduced-form utility function (\ref{eq:reduced-form}), for $x'\leq {a_C},$ we have
$$W_0(x,x')\equiv u(x,x')+\delta W(x')=\frac{a_I\theta}{a_Cb}x-\frac{a_I}{a_Cb}x'+\frac{\delta x'}{a_C(1-\delta(1-d))},$$
\begin{equation}\label{eq:case-i-thm1}
\frac{\partial W_0(x,x')}{\partial x'}=-\frac{a_I}{a_Cb}+\frac{\delta}{a_C(1-\delta(1-d))}=\frac{a_I(\delta\theta-1)}{a_Cb (1-\delta(1-d))} <0,
\end{equation}
where the inequality follows from $\delta\theta<1.$
For $x'>a_C$, there exists a natural number $n$ such that $x'\in(\frac{a_C}{(1-d)^{n-1}},\frac{a_C}{(1-d)^{n}}].$ Similarly, 
$$W_0(x,x')\equiv u(x,x')+\delta W(x')=\frac{a_I\theta}{a_Cb}x-\frac{a_I}{a_Cb}x'+\frac{\delta-\delta^{n+1}}{1-\delta}+\frac{\delta^{n+1}(1-d)^n x'}{a_C(1-\delta(1-d))},$$

\vspace{-5mm}

\begin{eqnarray*}
\frac{\partial W_0(x,x')}{\partial x'}&=&-\frac{a_I}{a_Cb}+\frac{\delta(\delta(1-d))^n}{a_C(1-\delta(1-d))}\\
&<&-\frac{a_I}{a_Cb}+\frac{\delta}{a_C(1-\delta(1-d))}=\frac{a_I(\delta\theta-1)}{a_Cb (1-\delta(1-d))} <0,
\end{eqnarray*}
where the first inequality follows from $\delta(1-d)<1$ and $n\geq 1$ and the second inequality follows from $\delta\theta<1.$
Then, for $x'>a_C$, $W_0(x,x')$ strictly decreases with $x'.$ Thus, for any $x'$, $W_0(x,x')$ strictly decreases with $x'$ and $W_0(x,x')$ attains its maximum when $x'$ attains its minimum: $x'=(1-d)x.$ Further, for $x\in (0,a_I),$
$$W(x)=\frac{x}{a_C(1-\delta(1-d))}=\frac{x}{a_C}+\frac{\delta (1-d)x}{a_C(1-\delta(1-d))}=u(x,(1-d)x)+\delta W((1-d)x).$$
Then the Bellman equation is satisfied for Case (i).

\medskip
\noindent {\bf Case (ii):} For $(x,x')\in\Omega$ such that $x\in[a_I,a_C]$, there are two subcases: (a) $(a_C-a_I)x'\geq ((1-d)(a_C-a_I)-b) x+a_Cb$ and (b) $(a_C-a_I)x'< ((1-d)(a_C-a_I)-b) x+a_Cb$. We first consider Subcase (a). Using the reduced-form utility function (\ref{eq:reduced-form}), for $x'\leq a_C$, 
$$W_0(x,x')\equiv u(x,x')+\delta W(x')=\frac{1-d}{b}x-\frac{1}{b}x'+1+\frac{\delta x'}{a_C(1-\delta(1-d))}.$$

\vspace{-5mm}

\begin{eqnarray}
\frac{\partial W_0(x,x')}{\partial x'}&=&-\frac{1}{b}+\frac{\delta}{a_C(1-\delta(1-d))} \nonumber \\
&=&\frac{\delta(b/a_C+(1-d))-1}{b (1-\delta(1-d))}< \frac{\delta\theta-1}{b (1-\delta(1-d))}<0, \label{eq:case-ii-thm1}
\end{eqnarray}
where the first inequality follows from $a_C> a_I$ and $\theta=b/a_I+(1-d)$ and the second inequality follows from $\delta\theta<1.$ So $W_0(x,x')$ strictly decreases with $x'.$ Similarly, we can show that $W_0(x,x')$ strictly decreases with $x'$ for $x'> a_C$. For Subcase (b), similar to Case (i), we can show that $W_0(x,x')$ strictly decreases with $x'$. In sum, $W_0(x,x')$ attains its maximum when $x'$ attains its minimum: $x'=(1-d)x.$ Then, similar to Case (i), we can show that the Bellman equation is satisfied for Case (ii).

\medskip
\noindent {\bf Case (iii):} For any $(x,x')\in\Omega$ such that $x\in(a_C,\frac{a_C}{1-d}]$, we have $(a_C-a_I)x'\geq ((1-d)(a_C-a_I)-b) x+a_Cb $. Similar to Subcase (a) of Case (ii), we can show that $(u(x,x')+\delta W(x'))$ attains its maximum when $x'$ attains its minimum: $x'=(1-d)x.$ 
Further, for $x\in (a_C,\frac{a_C}{1-d}],$
$$W(x)=1+\frac{\delta(1-d)x}{a_C(1-\delta(1-d))}=u(x,(1-d)x)+\delta W((1-d)x),$$
where the last equation follows from $u(x,(1-d)x)=1$ for $x\in(a_C,\frac{a_C}{1-d}]$ and $(1-d)x\leq a_C.$
Then, the Bellman equation is satisfied for Case (iii).

\medskip
\noindent {\bf Case (iv):} For any $(x,x')\in\Omega$ such that $x\in(\frac{a_C}{(1-d)^{n-1}},\frac{a_C}{(1-d)^{n}}]$ ($n=2,3,4...$), we have $(a_C-a_I)x'\geq ((1-d)(a_C-a_I)-b) x+a_Cb $, following again Subcase (a) of Case (ii), we can show that $(u(x,x')+\delta W(x'))$ attains its maximum when $x'$ attains its minimum: $x'=(1-d)x.$
Further, for $x\in(\frac{a_C}{(1-d)^{n-1}},\frac{a_C}{(1-d)^{n}}]$ and any positive integer $n\geq 2$,
\begin{eqnarray*}
W(x)=\frac{1-\delta^n}{1-\delta}+\frac{\delta^n(1-d)^n x}{a_C(1-\delta(1-d))} &=&1+\delta\left[\frac{1-\delta^{n-1}}{1-\delta}+ \frac{\delta^{n-1}(1-d)^{n-1} (1-d)x}{a_C(1-\delta(1-d))}\right]\\
& =&u(x,(1-d)x)+\delta W((1-d)x),
\end{eqnarray*}
where the last equation follows from $u(x,(1-d)x)=1$ for $x\in(\frac{a_C}{(1-d)^{n-1}},\frac{a_C}{(1-d)^{n}}]$ and $(1-d)x\in(\frac{a_C}{(1-d)^{n-2}},\frac{a_C}{(1-d)^{n-1}}].$
Then the Bellman equation is satisfied for Case (iv).

\medskip
In sum, we have verified that $W(x)$ is the value function satisfying the Bellman equation and the optimal policy is given by $g(x)=(1-d)x$ for any $x>0$ and $0<d<1.$ For the case of circulating capital ($d=1$), we can apply essentially the same argument as above to show that the optimal policy is given by $g(x)=(1-d)x=0$ for any $x>0$ with the value function $V(x)=x/a_C$ for $x\leq a_C$ and $V(x)=1$ for $x>a_C.$ 
Thus, we have obtained the desired conclusion.
\qed

\bigskip

%%%%%%%%%%%%%%%%%%%%%%%%%%%%%%%%%%%%%%%%%%%%%%
% Proof of THEOREM 2
%%%%%%%%%%%%%%%%%%%%%%%%%%%%%%%%%%%%%%%%%%%%%%

\noindent
{\it \textbf{Proof of Theorem \ref{ac-less-ai-extinction}}}: We first consider $0<d<1.$
Following the proof of Theorem \ref{ac-greater-ai}, we postulate a candidate value function $W(\cdot)$ to be the same as (\ref{eq:value-function-thm1}).
We now claim that if $\delta<\mu_0$, then $W(\cdot)$ satisfies the Bellman equation.
To this end, we consider two cases: (i) $x\leq a_C;$ (ii) $x>a_C$. For Case (i), the proof follows entirely Case (i) in the proof of Theorem \ref{ac-greater-ai}. For Case (ii), consider any $(x,x')\in\Omega$ such that $x>a_C$. There are two subcases: (a) $(a_C-a_I)x'\geq ((1-d)(a_C-a_I)-b) x+a_Cb$ and (b) $(a_C-a_I)x'< ((1-d)(a_C-a_I)-b) x+a_Cb$. We first consider Subcase (a). For $x'\leq a_C$, 
$$W_0(x,x')\equiv u(x,x')+\delta W(x')=\frac{1-d}{b}x-\frac{1}{b}x'+1+\frac{\delta x'}{a_C(1-\delta(1-d))}.$$
\begin{eqnarray}
\frac{\partial {W}_0(x,x')}{\partial x'}= -\frac{1}{b}+\frac{\delta}{a_C(1-\delta(1-d))}&=&\frac{(b+a_C(1-d))\delta-a_C}{a_Cb (1-\delta(1-d))}\nonumber \\ 
&=&\frac{b+a_C(1-d)}{a_Cb (1-\delta(1-d))}(\delta-\mu_0) <0, \label{eq:rho0}
\end{eqnarray}
where the inequality follows from $\delta<\mu_0.$ So $W_0(x,x')$ strictly decreases with $x'.$ For 
$x'\in(\frac{a_C}{(1-d)^{n-1}},\frac{a_C}{(1-d)^{n}}]$ with $n\in\mathbb{N}$, 
$$\frac{\partial { W}_0(x,x')}{\partial x'}= -\frac{1}{b}+\frac{\delta^{n+1}(1-d)^n}{a_C(1-\delta(1-d))}<-\frac{1}{b}+\frac{\delta}{a_C(1-\delta(1-d))} <0,$$
where the first inequality follows from $n\geq 1$ and $\delta(1-d)<1$ and the second inequality follows from (\ref{eq:rho0}).
This implies that ${W}_0(x,x')$ strictly decreases with $x'$ for $x'>a_C.$ Taken together, ${W}_0(x,x')$ strictly decreases with $x'$ for any $x'$ and it is maximized with $x'=(1-d)x.$ 
For Subcase (b), similar to Case (i), we can show that $W_0(x,x')$ attains its maximum for $x'=(1-d)x.$
Following the proof of Theorem \ref{ac-greater-ai}, we can further obtain ${W}(x)=u(x,(1-d)x)+\delta{W}((1-d)x)$ for any $x>0$. 
So we have verified that $W(\cdot)$ satisfies the Bellman equation. For the case of $d=1$, we can follow essentially the same argument to show that $g(x)=0$ for any $x>0.$ Thus, we have obtained the desired conclusion.
\qed

\bigskip

%%%%%%%%%%%%%%%%%%%%%%%%%%%%%%%%%%%%%%%%%%%%%%
% Proof of LEMMA 2
%%%%%%%%%%%%%%%%%%%%%%%%%%%%%%%%%%%%%%%%%%%%%%

\noindent
{\it \textbf{Proof of Lemma \ref{property1}}}:
For $n=0,$ since $z_0(\cdot)$ is strictly increasing on $[0,1/\theta]$ and by construction, $z_0(\mu_0)=0$, we have $z_0(\delta)<z_0(\mu_0)=0$ for $\delta\in[0,\mu_0)$ and $z_0(\delta)>z_0(\mu_0)=0$ for $\delta\in(\mu_0,1/\theta]$. 
Then, we just need to focus on $n\geq 1$ in this proof. 

From Lemma \ref{property1-tilde-zn}, we know for $\delta\neq-1/\zeta$, $z_n(\delta)=0$ if and only if $\tilde{z}_n(\delta)=0$, where $\tilde{z}_n(\cdot)$ is defined in Equation (\ref{eq:tilde-zn}). We first investigate the property of $\tilde{z}_n(\cdot)$.
From (c) of Lemma \ref{property1-tilde-zn}, there exists $\bar{\delta}$ such that $\tilde{z}_n'(\delta)>0$ for $\delta\in[0,\bar{\delta})$ and $\tilde{z}_n'(\delta)<0$ for $\delta\in(\bar{\delta},1/\theta],$ so $\tilde{z}_n(\cdot)$ is strictly increasing on $[0,\bar{\delta}]$ and strictly decreasing on $[\bar{\delta},1/\theta].$
 To better explain our proof, we illustrate the qualitative features of $\tilde{z}_n$  in Figure \ref{fig:proof-zn}.

 From (a.1)--(a.3) of Lemma \ref{property1-tilde-zn}, $\tilde{z}_n(-1/\zeta)=0$, $\tilde{z}_n(0)<0$, $\tilde{z}_n(1/\theta)<0$, and $\tilde{z}_n''(\delta)<0$ for $\delta\in[0,1/\theta)$.
There are three cases: (i) $\tilde{z}_n'(-1/\zeta)<0$; (ii) $\tilde{z}_n'(-1/\zeta)>0$; (iii) $\tilde{z}_n'(-1/\zeta)=0$. For Case (i), given the monotonicity property of $\tilde{z}_n$, we must have $-1/\zeta>\bar{\delta}.$ This case is illustrated in Panel (a) of Figure \ref{fig:proof-zn}. Since $-1/\zeta>\bar{\delta}$, $\tilde{z}_n'(\delta)<0$ for $\delta\in[-1/\zeta,1/\theta]$. From (a.1) of Lemma \ref{property1-tilde-zn}, $\tilde{z}_n(-1/\zeta)=0$, 
and since $\tilde{z}_n'(\delta)<0$ for $\delta\in[-1/\zeta,1/\theta]$,  $\tilde{z}_n(\delta)<0$ for $\delta\in(-1/\zeta,1/\theta].$
Since $\tilde{z}_n(\cdot)$ is strictly decreasing on $[\bar{\delta},-1/\zeta]$ and  $\tilde{z}_n(-1/\zeta)=0$, $\tilde{z}_n(\delta)>0$ for $\delta\in[\bar{\delta},-1/\zeta).$ In particular, $\tilde{z}_n(\bar{\delta})>0$. Since $\tilde{z}_n(\cdot)$ is strictly increasing on $[0,\bar{\delta}]$ and $\tilde{z}_n(0)<0$, by the continuity of $\tilde{z}_n(\cdot)$, there exists a unique root, denoted by $\mu_n$, of $\tilde{z}_n(\delta)=0$ on the interval $(0,\bar{\delta}),$ and $\tilde{z}_n(\delta)>0$ for $\delta\in(\mu_n,\bar{\delta}).$ In sum, if $\tilde{z}_n'(-1/\zeta)<0$, $\tilde{z}_n(\delta)=0$ admits two roots, $\mu_n$ and $(-1/\zeta)$, in $[0,1/\theta]$ such that $0<\mu_n<-1/\zeta<1/\theta$ and $\tilde{z}_n(\delta)>0$ for $\delta\in(\mu_n,-1/\zeta).$
For Case (ii), $\tilde{z}_n'(-1/\zeta)>0$, which is illustrated in Panel (b) of Figure \ref{fig:proof-zn}. Symmetrically, we can show that if $\tilde{z}_n'(-1/\zeta)>0$, $\tilde{z}_n(\delta)=0$ admits two roots, $\mu_n$ and $(-1/\zeta)$, in $[0,1/\theta]$ such that $0<-1/\zeta<\mu_n<1/\theta$ and $\tilde{z}_n(\delta)>0$ for $\delta\in(-1/\zeta,\mu_n).$
For Case (iii), $\tilde{z}_n'(-1/\zeta)=0,$ which is illustrated in Panel (c) of Figure \ref{fig:proof-zn}.
In this case, $\bar{\delta}=-1/\zeta$, and thus, $\tilde{z}_n(\delta)=0$ admits a unique root,  $(-1/\zeta)$, and we let $\mu_n\equiv -1/\zeta$ in this case.

From (b) of Lemma \ref{property1-tilde-zn},  $\tilde{z}_n'(-1/\zeta)=0$ if and only if $z_n(-1/\zeta)= 0.$ 
From Lemma \ref{property1-tilde-zn}, we also know that if $\delta\neq-1/\zeta$, $z_n(\delta)=0$ if and only if $\tilde{z}_n(\delta)=0$. Thus, if $\tilde{z}_n'(-1/\zeta)\neq 0,$ ${z}_n(-1/\zeta)\neq 0,$  so $\mu_n$ must be the unique root of $z_n(\delta)=0$ on $[0,1/\theta]$. On the other hand, if $\tilde{z}_n'(-1/\zeta)= 0,$ ${z}_n(-1/\zeta)= 0,$
and $-1/\zeta(=\mu_n)$   is the unique root of   $z_n(\delta)=0$ on $[0,1/\theta]$. Thus, $z_n(\delta)=0$ on $[0,1/\theta]$ always admits a unique root   $\mu_n.$

Next, we claim that $z_n(\delta)<0$ for $\delta\in[0,\mu_n)$. By construction, we know $z_n(0)=-1/b<0$. Suppose there exists $\hat{\delta}\in(0,\mu_n)$ such that $z_n(\hat{\delta})\geq 0.$ If $z_n(\hat{\delta})= 0$, then it contradicts to $\mu_n$ being the unique root, so $z_n(\hat{\delta})>0$. Since $z_n(0)=-1/b<0$, by the continuity of $z_n(\cdot)$ on $[0,1/\theta],$ there exists a root in $(0,\hat{\delta})$ of the equation $z_n(\delta)=0$. It again contradicts to $\mu_n$ being the unique root, thus establishing our claim.

Last, we claim that $z_n(\delta)>0$ for $\delta\in(\mu_n,1/\theta].$ There are two possible cases: (a) $\mu_n<-1/\zeta$ and (b) $\mu_n\geq-1/\zeta$. For (a), pick $\delta'$ in $(\mu_n,-1/\zeta)$. We have shown above that $\tilde{z}_n({\delta'})>0$ for $\delta'\in(\mu_n,-1/\zeta)$. Since $\delta'<-1/\zeta$ and $\zeta<0$ (for $a_I>a_C$), $(1+{\delta'}\zeta)>0$. Since $\tilde{z}_n({\delta'})>0$, $(1+{\delta'}\zeta)>0$, $a_I>a_C$, $\delta'\neq -1/\zeta$, from Lemmas \ref{property1-zn} and \ref{property1-tilde-zn}, we have 
$$z_n({\delta'})=\frac{\tilde{z}_n(\delta')}{a_C b(1-\delta(1-d))(a_I-a_C)(1+\delta'\zeta)}>0.$$
Suppose on the contrary, there exists $\delta''\in(\mu_n,1/\theta]$ such that $z_n(\delta'')\leq 0$. If $z_n(\delta'')=0$, it contradicts with $\mu_n$ being the unique root. If $z_n(\delta'')<0$, since $z_n({\delta'})>0$, by the continuity of $z_n$ on $[0,1/\theta],$
there exists another root in $(\mu_n,1/\theta)$,  leading to a contradiction. Thus, we must have 
$z_n(\delta)>0$ for $\delta\in(\mu_n,1/\theta].$ 
For (b), since $\mu_n\geq -1/\zeta$, from the discussion above we know $\tilde{z}_n(\delta)<0$ for any ${\delta}$ in $(\mu_n,1/\theta].$
For $\delta>\mu_n\geq-1/\zeta$, $(1+{\delta}\zeta)<0$ and since $\tilde{z}_n(\delta)<0$,  from Lemmas \ref{property1-zn} and \ref{property1-tilde-zn}, we have $z_n(\delta)>0$ for $\delta\in(\mu_n,1/\theta],$ thus establishing the claim. 

We have now obtained the desired conclusion. \qed

\bigskip

%%%%%%%%%%%%%%%%%%%%%%%%%%%%%%%%%%%%%%%%%%%%%%
% Proof of LEMMA 3
%%%%%%%%%%%%%%%%%%%%%%%%%%%%%%%%%%%%%%%%%%%%%%

\noindent
{\it \textbf{Proof of Lemma \ref{property3}}}:
From (\ref{eq:zn}) in Lemma \ref{property1-zn} stated below, if $\delta=-1/\zeta$, or equivalently, $-\delta\zeta=1,$ then 
$$z_n(\delta)=
\frac{-{nba_C(1-d)\delta}+nba_C-ba_I+2ba_C}{a_Cb(1-\delta(1-d))(a_I-a_C)\zeta}$$
Then, for $n> 1$, we have 
$$z_n(\delta)-z_{n-1}(\delta)=\frac{ba_C(1-(1-d)\delta)}{a_Cb(1-\delta(1-d))(a_I-a_C)\zeta}<0,$$
where the inequality follows from $\delta(1-d)<1$, $a_I>a_C$, and $\zeta<0.$
For $\delta\neq-1/\zeta$, from (\ref{eq:zn}), we can write $z_n(\delta)$ as 
$$z_n(\delta)=\frac{b a_I (-\zeta)^n (1-\theta\delta)\delta^{n+1}-a_C(a_I-a_C)(1-(1-d)\delta)^2}{a_Cb(1-\delta(1-d))(a_I-a_C)(1+\delta \zeta)}.$$
Then, for $n> 1$, we have
\begin{eqnarray*}
z_n(\delta)-z_{n-1}(\delta)&=&\frac{b a_I(-\zeta)^{n-1}(1-\theta \delta)\delta^n(-\delta\zeta-1)}{a_Cb(1-\delta(1-d))(a_I-a_C)(1+\delta \zeta)} \\
&=& - \frac{b a_I(-\zeta)^{n-1}(1-\theta \delta)\delta^n}{a_Cb(1-\delta(1-d))(a_I-a_C)}<0,
\end{eqnarray*}
where the inequality follows from $-\zeta>0$, $\delta<1/\theta$, $\delta(1-d)<1$, and $a_I>a_C.$
Thus, we have shown that $z_n(\delta)< z_{n-1}(\delta)$ for any $n>1$. Last, 
\begin{eqnarray*}
z_1(\delta)-z_0(\delta)&=&\delta\left(-\frac{1}{a_I-a_C}-\frac{\delta\zeta}{a_C(1-\delta(1-d))}\right)-\frac{\delta}{a_C(1-\delta(1-d))} \\
&=&-\delta\cdot\frac{a_C(1-\delta(1-d))+(1+\delta\zeta)(a_I-a_C)}{a_C(1-\delta(1-d))(a_I-a_C)}\\
&=&-\delta\cdot\frac{a_I-a_I\delta(1-d)-b\delta}{a_C(1-\delta(1-d))(a_I-a_C)}\\
&=&-\frac{\delta a_I(1-\delta \theta)}{a_C(1-\delta(1-d))(a_I-a_C)}<0,
\end{eqnarray*}
where the inequality follows from $\delta\theta<1$ and $a_I>a_C$. Thus, we have obtained the desired conclusion. \qed

\bigskip

%%%%%%%%%%%%%%%%%%%%%%%%%%%%%%%%%%%%%%%%%%%%%%
% Proof of PROPOSITION 2
%%%%%%%%%%%%%%%%%%%%%%%%%%%%%%%%%%%%%%%%%%%%%%

\noindent
{\it \textbf{Proof of Proposition \ref{property-mu}}}: 
From Lemma \ref{property1}, there is a unique root of $z_n(\delta)=0$ on the interval $(0,1/\theta)$. Denote this root by $\mu_n$. Consider the sequence $\{\mu_n\}_{n=0}^\infty$. We now want to establish the monotonicity and the limit of this sequence. In particular, we want to show that the sequence $\{\mu_n\}_{n=0}^\infty$ satisfies (i) $\mu_n>\mu_{n-1}$ for any $n\in\mathbb{N}$ and (ii) $\lim_{n\rightarrow\infty}{\mu}_n=1/\theta.$

To gain some intuition, we illustrate the determination of $\mu_n$ in Figure \ref{fig:rho-n}. Let $v_n(\delta)\equiv b a_I (-\zeta)^n (1-\theta\delta)\delta^{n+1}$ and $w(\delta)\equiv a_C(a_I-a_C)(1-(1-d)\delta)^2$. Then, $\tilde{z}_n(\cdot)$, as defined in (\ref{eq:tilde-zn}), can be written as $\tilde{z}_n(\delta)=v_n(\delta)-w(\delta),$ so $\tilde{z}(\delta)=0$ if and only if $v_n(\delta)=w(\delta)$. It is straightforward to show that $w(\cdot)$ is strictly decreasing and $v_n(\cdot)$ is first strictly increasing and then strictly decreasing on $[0,1/\theta].$ The two curves intersect with each other twice provided that $\mu_n\neq -1/\zeta.$ One of the points of intersection always corresponds to $\delta=-1/\zeta.$
The left panel shows how the curve of $v_n(\cdot)$ changes with $n$ for $\mu_n<-1/\zeta$ while the right panel illustrates the case of $\mu_n>-1/\zeta.$ As $n$ increases, the red curve shifts to the right, thus leading to $\mu_{n+1}>\mu_n.$

To establish the monotonicity formally, from Lemma \ref{property1}, we know for any $n\in\mathbb{N},$ $z_n(\delta)<0$ for $\delta\in[0,\mu_n)$ and $z_n(\delta)>0$ for $\delta\in(\mu_n,1/\theta].$ From Lemma \ref{property3}, $z_{n-1}(\delta)>z_n(\delta)$ for any $\delta\in(0,1/\theta)$. In particular,  $z_{n-1}(\delta)>z_n(\delta)$ for
 $\delta\in(\mu_n,1/\theta),$ and by the continuity of $z_n(\cdot)$ and $z_{n-1}(\cdot)$, we must also have  $z_{n-1}(1/\theta)\geq z_n(1/\theta)$. Then  for $\delta\in(\mu_n,1/\theta],$ $z_{n-1}(\delta)\geq z_n(\delta)>0$.
Further, $z_{n-1}(\mu_n)>z_{n}(\mu_n)=0$. Thus, $z_{n-1}(\delta)>0$ for $\delta\in[\mu_n,1/\theta],$ so $\mu_{n-1}$, defined as the unique root of the equation $z_{n-1}(\delta)=0$ for $\delta\in[0,1/\theta]$, has to be in $(0,\mu_n),$ which implies $\mu_{n-1}<\mu_n$.

We have now obtained the monotonic property of $\{\mu_n\}_{n=0}^\infty$. The next is to show that $\lim_{n\rightarrow\infty}{\mu}_n=1/\theta.$ We first note that, by construction, $\mu_n<1/\theta$ for any $n\in\mathbb{N}$. Thus, the sequence $\{\mu_n\}_{n=0}^\infty$ is bounded above by $1/\theta$ and monotonic, so it must have a limit and $\lim_{n\rightarrow \infty }\mu_n\leq 1/\theta.$
Since we have
\begin{eqnarray*}
z_n\left(-\frac{1}{\zeta}\right)&=& -\frac{1}{b}-\frac{1}{\zeta}\cdot\left(-\frac{\sum_{i=0}^{n-1}(-(-1/\zeta)\zeta)^i}{a_I-a_C}+\frac{(-(-1/\zeta)\zeta)^n}{a_C(1+(1-d)/\zeta)}\right) \\
&=&-\frac{1}{b}-\frac{1}{a_C(\zeta+1-d)}+\frac{n}{\zeta(a_I-a_C)},
\end{eqnarray*}
$z_n(-1/\zeta)$ is linear in $n$. Since $\zeta<0$ (for $a_I>a_C$), $z_n(-1/\zeta)$ is strictly decreasing in $n$. Thus, there exists $n_0\in \mathbb{N}$ such that for any $n>n_0$, $z_n(-1/\zeta)<0$, and by Lemma \ref{property1}, it implies $\mu_n>-1/\zeta.$
From Lemma \ref{property1-tilde-zn}, for any $n>n_0$,
we have
\begin{eqnarray*}
&&\tilde{z}_n(\mu_n)=b a_I (-\zeta)^n (1-\theta\mu_n)\mu_n^{n+1}-a_C(a_I-a_C)(1-(1-d)\mu_n)^2=0\\
&\Leftrightarrow& \lim_{n\rightarrow \infty } b a_I (-\zeta)^n (1-\theta\mu_n)\mu_n^{n+1}-a_C(a_I-a_C)(1-(1-d)\mu_n)^2=0 \\
&\Leftrightarrow& \lim_{n\rightarrow \infty } b a_I (-\zeta)^n (1-\theta\mu_n)\mu_n^{n+1}=\lim_{n\rightarrow \infty } a_C(a_I-a_C)(1-(1-d)\mu_n)^2<\infty,
\end{eqnarray*}
where the second line follows from $\tilde{z}_n(\mu_n)=0$ for any $n>n_0$ and the third line follows from the fact that $\{\mu_n\}_{n=0}^\infty$ has a limit and $\lim_{n\rightarrow \infty }\leq 1/\theta.$
Suppose $\lim_{n\rightarrow \infty }\mu_n< 1/\theta.$ Since $\mu_n>-1/\zeta$ for $n>n_0$ and $\{\mu_n\}_{n=0}^\infty$ is strictly increasing, 
then $$\lim_{n\rightarrow \infty } b a_I (-\zeta)^n (1-\theta\mu_n)\mu_n^{n+1}\geq \lim_{n\rightarrow \infty } b a_I (1-\theta \lim_{n\rightarrow \infty }\mu_n) \cdot (-\zeta)^n \mu_{n_0+1}^{n+1}=\infty,$$
where the last equality follows from $\lim_{n\rightarrow \infty }\mu_n< 1/\theta$ and $\mu_{n_0+1}>-1/\zeta$, leading to a contradiction. Thus, we must have $\lim_{n\rightarrow \infty }\mu_n=1/\theta.$

We have now obtained the desired conclusion. 
\qed

\bigskip

%%%%%%%%%%%%%%%%%%%%%%%%%%%%%%%%%%%%%%%%%%%%%%%%%
% Proof of Lemma 4
%%%%%%%%%%%%%%%%%%%%%%%%%%%%%%%%%%%%%%%%%%%%%%%%%

\noindent
{\it \textbf{Proof of Lemma \ref{xn-property}}}:
We first consider the case of $\zeta\neq -1$. Since $\zeta\neq-1$, $b/(a_C-a_I)+d\neq 0$ or equivalently, $b+(a_C-a_I)d \neq 0.$
Further, we have $$\frac{a_Cb}{b+d(a_C-a_I)}=\frac{a_C(\zeta+1-d)}{\zeta+1}\;\;\mbox{ and }\;\; \frac{a_Cb}{a_C-a_I}=a_C(\zeta+1-d).$$
Since by construction, $x_n=-(x_{n-1}-a_Cb/(a_C-a_I))/\zeta$, for any $n\in\mathbb{N}$, we have
\begin{eqnarray}
x_n- \frac{a_Cb}{b+d(a_C-a_I)}&=&-\frac{1}{\zeta}\left(x_{n-1}-\frac{a_Cb}{a_C-a_I}\right)- \frac{a_Cb}{b+d(a_C-a_I)}\nonumber \\
&=&-\frac{1}{\zeta}\left(x_{n-1}-a_C(\zeta+1-d)\right)-\frac{a_C(\zeta+1-d)}{\zeta+1}\nonumber \\
&=&-\frac{1}{\zeta}\left(x_{n-1}-\frac{a_C(\zeta+1-d)}{\zeta+1}\right) \nonumber \\
&=&-\frac{1}{\zeta}\left(x_{n-1}-\frac{a_Cb}{b+d(a_C-a_I)}\right) \nonumber \\
&=&\frac{1}{(-\zeta)^n}\left(x_0-\frac{a_Cb}{b+d(a_C-a_I)}\right) \nonumber \\
&=& -\frac{da_C(a_I-a_C)}{(b+d(a_C-a_I))(-\zeta)^n}, \label{eq:xn}
\end{eqnarray}
where the last equality follows from $x_0=a_C.$ Since $a_I>a_C,$ $\zeta<0$, and we consider $\zeta\neq-1$, so there are two cases: (i) $\zeta<-1$ and (ii) $0>\zeta>-1.$ For (i), since $\zeta<-1$, $(1+\zeta)<0$. Since $a_I>a_C$ and $\zeta<-1$, $b+d(a_C-a_I)>0$. Thus, $(1+\zeta)/(b+d(a_C-a_I))<0.$ For (ii), since $\zeta>-1$, $1+\zeta>0$. Since $\zeta>-1$ and $a_I>a_C$, $b+d(a_C-a_I)<0$. Again, we have $(1+\zeta)/(b+d(a_C-a_I))<0.$
For both cases, we then have
\begin{eqnarray*}
x_n-x_{n-1}&=&(x_n- \frac{a_Cb}{b+d(a_C-a_I)})-(x_{n-1}-\frac{a_Cb}{b+d(a_C-a_I)}) \\
&=&-\frac{da_C(a_I-a_C)}{(b+d(a_C-a_I))(-\zeta)^n}+\frac{da_C(a_I-a_C)}{(b+d(a_C-a_I))(-\zeta)^{n-1}} \\
&=&-\frac{da_C(a_I-a_C)(1+\zeta)}{(b+d(a_C-a_I))(-\zeta)^n}>0,
\end{eqnarray*}
where the inequality follows from $a_I>a_C$, $\zeta<0$ and $(1+\zeta)/(b+d(a_C-a_I))<0$ for both cases. Thus, $x_n>x_{n-1}$ for any $n\in\mathbb{N}$ and $\zeta\neq-1.$
For $\zeta=-1$, $x_n=x_{n-1}+a_Cb/(a_I-a_C)$. Since $a_I>a_C$, $x_n$ strictly increases with $n$.

Moreover, since $a_I>a_C>0$, $-\zeta>\theta$. Then, for $\theta\geq 1$, $\zeta<-\theta\leq -1,$ and from (\ref{eq:xn}), this implies 
$\lim_{n\rightarrow\infty} x_n=\frac{a_Cb}{b+d(a_C-a_I)}.$
For $\theta>1$, $\hat{x}=\frac{a_Cb}{b+d(a_C-a_I)}$, so we have $\lim_{n\rightarrow\infty} x_n=\hat{x}.$ For $\theta=1$, $b/a_I=d$, and thus, $\lim_{n\rightarrow\infty} x_n=a_I.$
We have then obtained the desired conclusion.
\qed

\bigskip

%%%%%%%%%%%%%%%%%%%%%%%%%%%%%%%%%%%%%%%%%%%%%%%%%
% Proof of Proposition 3
%%%%%%%%%%%%%%%%%%%%%%%%%%%%%%%%%%%%%%%%%%%%%%%%%

\noindent
{\it \textbf{Proof of Proposition \ref{ac-less-ai-d<1-add}}}: 
We adopt the standard guess-and-verify approach. 
Let ${\mu}_{n-1}<\delta<{\mu}_{n}$ for some $n\in\mathbb{N}$. Consider the following candidate policy function
$$
\bar{g}(x)=\left\{\begin{array}{ll}
(1-d)x & \mbox{for } x\in(0,a_C] \smallskip \\
-\zeta x+\frac{a_Cb}{a_C-a_I} & \mbox{for } x\in(a_C, x_n] \smallskip \\
x_{n-1} & \mbox{for } x\in(x_n,\frac{x_{n-1}}{1-d}] \smallskip \\
(1-d)x & \mbox{for } x\in(\frac{x_{n-1}}{1-d},\infty)
\end{array}\right.,
$$
where $x_0=a_C$ and from Lemma \ref{xn-property-remark}, $x_n= \hat{x}-{(\hat{x}-a_C)}/{(-\zeta)^n}$ (for $\theta=1$, let $\hat{x}=a_Cb/(b+d(a_C-a_I))=a_I$). 
To see that $\bar{g}(\cdot)$ is well defined, we need to verify (a) $a_C<x_n$, (b) $x_{n}<x_{n-1}/(1-d)$, and (c) $(x,\bar{g}(x))\in\Omega$ for any $x>0.$ 
Since $a_C<a_I$, from Lemma \ref{xn-property}, $x_n> x_{n-1}$. Since $x_0=a_C$, $x_{n}>x_0=a_C$, for any $n\in\mathbb{N}.$ Then, (a) is verified. 
Since $x_n>a_C$ and $a_C<a_I$, $(\zeta+1-d)x_n=bx_n/(a_C-a_I)<a_Cb/(a_C-a_I)$, or equivalently, $x_n<(a_Cb/(a_C-a_I)-\zeta x_n)/(1-d)=x_{n-1}/(1-d),$ where the equality follows from the construction of the sequence $\{x_n\}_{n=0}^\infty$. Then, (b) is verified.
For $x\in(0,a_C]\cup (x_{n-1}/(1-d),\infty)$, $(x,(1-d)x)\in\Omega.$ Since $a_C<a_I$ and $\theta\geq 1$, $\zeta<-\theta\leq -1$. Since $\zeta<-1$, $x_n= \hat{x}-{(\hat{x}-a_C)}/{(-\zeta)^n}$, and $\hat{x}>a_C$ (from $a_C<a_I$ and $\theta\geq 1$), we have $x_n<\hat{x}$. We have shown $x_n>a_C$, and from $a_C<a_I$ and $\theta\geq 1$, $\hat{x}\leq a_I$, so $a_C<x_n<\hat{x}\leq a_I.$ Since $a_C<x_n<a_I$, 
$(x,-\zeta x+a_Cb/(a_C-a_I))=(x,\hat{x}+\zeta(\hat{x}-x))\in\Omega$ for any $x\in(a_C,x_n].$
For $x\in(x_n,x_{n-1}/(1-d)],$ $x\leq x_{n-1}/(1-d)$. Then, $(1-d)x\leq x_{n-1}.$ Since $\theta\geq 1$ and $x_{n-1}<x_n$, $\theta x\geq x>x_n>x_{n-1}.$ Further, if $x>a_I$, $(1-d)x+b\geq (1-d)a_I+b=\theta a_I\geq a_I>x_n>x_{n-1}.$ Since $\theta x\leq (1-d)x+b$ if and only if $x\leq a_I$, we have shown that $(1-d)x\leq x_{n-1}<\min\{\theta x,(1-d)x+b\}$ for any $x\in(x_n,x_{n-1}/(1-d)].$ Then, (c) is verified. Based on the policy function $\bar{g}(\cdot)$, we postulate the following value function
$$W(x)= \left\{\begin{array}{lll}

\frac{x}{a_C(1-\delta(1-d))} & \mbox{for } x\in [0,a_C] \medskip \\

\left(-\frac{\sum_{i=0}^{m-1}(-\delta\zeta)^i}{a_I-a_C}+\frac{(-\delta\zeta)^m}{a_C(1-\delta(1-d))}\right)(x-\hat{x}) & \mbox{for } x\in(x_{m-1},x_m],\\
+\frac{(1-\delta^m)(a_I-\hat{x})}{(1-\delta)(a_I-a_C)}+\frac{\delta^m\hat{x}}{a_C(1-\delta(1-d))} & \;\;\;\;\; m=1,2,...,n \medskip \\

\frac{1-d}{b}x+\frac{b-x_{n-1}}{b} +\frac{\sum_{i=1}^{n}\delta^i(a_I-x_{n-i})}{a_I-a_C}+\frac{\delta^{n+1}(1-d)a_C}{1-\delta(1-d)} & \mbox{for } x\in (x_n,\frac{x_{n-1}}{1-d}] \medskip \\

\delta^\ell\left(-\frac{\sum_{i=0}^{n-1}(-\delta\zeta)^i}{a_I-a_C}+\frac{(-\delta\zeta)^n}{a_C(1-\delta(1-d))}\right)((1-d)^\ell x-\hat{x}) & \mbox{for } x\in\left(\frac{x_{n-1}}{(1-d)^\ell},\frac{x_n}{(1-d)^\ell}\right] \\
+\frac{1-\delta^\ell}{1-\delta}+\delta^\ell\left( \frac{(1-\delta^n)(a_I-\hat{x})}{(1-\delta)(a_I-a_C)}+\frac{\delta^n\hat{x}}{a_C(1-\delta(1-d))}\right) & \;\;\;\;\; \ell=1,2,... \medskip \\

\frac{\delta^{\ell}(1-d)^{\ell+1}}{b} x+\frac{1-\delta^\ell}{1-\delta}+\delta^\ell\left( \frac{b-x_{n-1}}{b} \right. & \mbox{for } x\in (\frac{x_n}{(1-d)^\ell},\frac{x_{n-1}}{(1-d)^{\ell+1}}] \\
\left.+\frac{\sum_{i=1}^{n}\delta^i(a_I-x_{n-i})}{a_I-a_C}+\frac{\delta^{n+1}(1-d)a_C}{1-\delta(1-d)}\right)& \;\;\;\;\; \ell=1,2,... 
\end{array} \right. $$ 

Before we verify that $W(\cdot)$ satisfies the value function, we first show how we obtain the postulated value function. For $x\in(0,a_C]$, 
$$W(x)=\sum_{i=0}^\infty \delta^i u((1-d)^ix, (1-d)^{i+1}x)=\sum_{i=0}^\infty \delta^i\frac{(1-d)^ix}{a_C} =\frac{x}{a_C(1-\delta(1-d))}.$$

For $x\in(x_{m-1},x_m]$ with $m\in\{1,2,...,n\}$, let $\bar{f}(x)=-\zeta x+a_C b/(a_C-a_I)=\zeta(\hat{x}-x)+\hat{x}.$ Then, we have
\begin{eqnarray*}
W(x)&=&\sum_{i=0}^{m-1}\delta^iu(\bar{f}^{i}(x),\bar{f}^{i+1}(x))+\delta^m W(\bar{f}^m(x)) \\
&=& \sum_{i=0}^{m-1} \delta^i\frac{a_I-\bar{f}^i(x)}{a_I-a_C}+\frac{\delta^m \bar{f}^m(x)}{a_C(1-\delta(1-d))}\\
&=& \sum_{i=0}^{m-1} \delta^i \frac{a_I-\hat{x}+(-\zeta)^i(\hat{x}-x)}{a_I-a_C}+\frac{\delta^m[\hat{x}-(-\zeta)^m(\hat{x}-x)]}{a_C(1-\delta(1-d))}\\
&=& \left(- \frac{\sum_{i=0}^{m-1}(-\delta\zeta)^i}{a_I-a_C}+\frac{(-\delta\zeta)^m}{a_C(1-\delta(1-d))}\right)(x-\hat{x})\\
&&+\frac{(1-\delta^m)(a_I-\hat{x})}{(1-\delta)(a_I-a_C)}+\frac{\delta^m\hat{x}}{a_C(1-\delta(1-d))}
\end{eqnarray*}
where the first equation follows from $\bar{g}^i(x)=\bar{f}^i(x)\in(x_{m-i-1},x_{m-i}]$ for $i=0,1,...,m-1$, the second equation follows from 
\begin{eqnarray*}
u(\bar{f}^{i}(x),\bar{f}^{i+1}(x))&=&u(\bar{f}^{i}(x),\zeta(\hat{x}-\bar{f}^{i}(x))+\hat{x})\\
&=&\frac{(1-d)}{b}\bar{f}^{i}(x)-\frac{\zeta(\hat{x}-\bar{f}^{i}(x))+\hat{x}}{b}+1=\frac{(a_I-\bar{f}^{i}(x))}{(a_I-a_C)},
\end{eqnarray*} 
and the third equation follows from $\bar{f}^i(x)=\hat{x}-(-\zeta)^i(\hat{x}-x)$. It should be noted that $-\zeta>\theta$, so we cannot rule out the possibility of $(-\delta\zeta)=1.$

For $x\in(x_n,x_{n-1}/(1-d)],$ since $(1-d)x_0=(1-d)a_C=-\zeta a_C+a_Cb/(a_C-a_I)=-\zeta x_0+a_Cb/(a_C-a_I)$, $u(a_C,(1-d)a_C)=u(\bar{f}^{n-1}(x_{n-1}),(1-d)\bar{f}^{n-1}(x_{n-1}))=u(\bar{f}^{n-1}(x_{n-1}),\bar{f}^{n}(x_{n-1})),$ and we have
\begin{eqnarray*}
W(x)&=& u(x,x_{n-1})+\sum_{i=1}^{n}\delta^iu(\bar{f}^{i-1}(x_{n-1}),\bar{f}^{i}(x_{n-1}))+\delta^{n+1} W(\bar{f}^{n} (x_{n-1}))\\
&=&\frac{1-d}{b}x-\frac{x_{n-1}}{b}+1+\sum_{i=1}^{n}\delta^i\frac{a_I-\bar{f}^{i-1}(x_{n-1})}{a_I-a_C}+\frac{\delta^{n+1} \bar{f}^n(x_{n-1})}{a_C(1-\delta(1-d))}\\
&=& \frac{1-d}{b}x+\frac{b-x_{n-1}}{b}+\frac{\sum_{i=1}^{n}\delta^i(a_I-x_{n-i})}{a_I-a_C}+\frac{\delta^{n+1} (1-d)a_C}{1-\delta(1-d)} 
\end{eqnarray*}
where the first equation follows from $\bar{g}^i(x)=\bar{f}^{i-1}(x_{n-1})=x_{n-i}$ for $i=1,...,n$; the second equation follows from $x_{n-1}= \zeta(\hat{x}-x_n)+\hat{x}< \zeta(\hat{x}-x)+\hat{x}$ for $x\in(x_n,x_{n-1}/(1-d)],$ $u(\bar{f}^{i-1}(x_{n-1}),\bar{f}^{i}(x_{n-1}))=u(\bar{f}^{i-1}(x_{n-1}),\bar{f}(\bar{f}^{i-1}(x_{n-1})))=(a_I-\bar{f}^{i-1}(x_{n-1}))/(a_I-a_C)$, and $\bar{f}^n(x_{n-1})=(1-d)a_C<a_C;$ the third equation follows from $\bar{f}^{i-1}(x_{n-1})=x_{n-i}$ and $\bar{f}^n(x_{n-1})=\bar{f}(x_0)=(1-d)a_C$.

For $x\in\left(\frac{x_{n-1}}{(1-d)^\ell},\frac{x_n}{(1-d)^\ell}\right]$ with $\ell\in\mathbb{N}$, we have
\begin{eqnarray*}
W(x)&=&\sum_{i=0}^{\ell-1} \delta^i+\delta^\ell W((1-d)^\ell x) \\
&=& \frac{1-\delta^\ell}{1-\delta}+\delta^\ell \left[\left(- \frac{\sum_{i=0}^{n-1}(-\delta\zeta)^i}{a_I-a_C}+\frac{(-\delta\zeta)^n}{a_C(1-\delta(1-d))}\right)((1-d)^\ell x-\hat{x})\right.\\
&&\left.+\frac{(1-\delta^n)(a_I-\hat{x})}{(1-\delta)(a_I-a_C)}+\frac{\delta^n\hat{x}}{a_C(1-\delta(1-d))}\right] \\
&=& \delta^\ell\left(-\frac{\sum_{i=0}^{n-1}(-\delta\zeta)^i}{a_I-a_C}+\frac{(-\delta\zeta)^n}{a_C(1-\delta(1-d))}\right)((1-d)^\ell x-\hat{x}) \\
&& +\frac{1-\delta^\ell}{1-\delta}+\delta^\ell\left( \frac{(1-\delta^n)(a_I-\hat{x})}{(1-\delta)(a_I-a_C)}+\frac{\delta^n\hat{x}}{a_C(1-\delta(1-d))}\right),
\end{eqnarray*}
where the first equation follows from $\bar{g}^i(x)>x_{n-1}/(1-d)$ for $i=0,1,...,\ell-1$ and the second equation follows from $(1-d)^\ell x\in(x_{n-1},x_n].$ 

For $x\in (\frac{x_n}{(1-d)^\ell},\frac{x_{n-1}}{(1-d)^{\ell+1}}]$ with $\ell\in\mathbb{N}$,
\begin{eqnarray*}
W(x)&=&\sum_{i=0}^{\ell-1} \delta^i+\delta^\ell W((1-d)^\ell x) \\
&=& \frac{1-\delta^\ell}{1-\delta}+\delta^\ell \left[\frac{1-d}{b}(1-d)^\ell x+\frac{b-x_{n-1}}{b}+\frac{\sum_{i=1}^{n}\delta^i(a_I-x_{n-i})}{a_I-a_C}+\frac{\delta^{n+1} (1-d)a_C}{1-\delta(1-d)} \right]\\
&=& \frac{\delta^{\ell}(1-d)^{\ell+1}}{b} x+\frac{1-\delta^\ell}{1-\delta}+\delta^\ell\left( \frac{b-x_{n-1}}{b}+\frac{\sum_{i=1}^{n}\delta^i(a_I-x_{n-i})}{a_I-a_C}+\frac{\delta^{n+1}(1-d)a_C}{1-\delta(1-d)}\right),
\end{eqnarray*}
where the first inequality follows from $\bar{g}^i(x)>x_{n-1}/(1-d)$ for $i=0,1,...,\ell-1$ and the second inequality follows from $(1-d)^\ell x\in(x_{n},x_{n-1}/(1-d)].$

We now turn to the verification of whether $W(\cdot)$ satisfies the Bellman equation. For $x\in(0,a_C]$, $(x,x')\in\Omega$ implies that $x'\geq \zeta(\hat{x}-x)+\hat{x}$. Using the reduced-form utility function (\ref{eq:reduced-form}), we have
$$W_0(x,x')\equiv u(x,x')+\delta W(x')=\frac{a_I\theta}{a_Cb}x-\frac{a_I}{a_Cb}x'+\delta W(x').$$
For $x'\in (0,a_C]$,
$$\frac{\partial W_0(x,x')}{\partial x'}= -\frac{a_I}{a_Cb}+\frac{\delta}{a_C(1-\delta(1-d))}=\frac{a_I(\delta\theta-1)}{a_Cb (1-\delta(1-d))} <0,$$
where the inequality follows from $\delta\theta<1.$ For $x'\in(x_{m-1},x_m]$ for some $m\in\{1,2,...,n\},$ 
\begin{eqnarray*}
\frac{\partial W_0(x,x')}{\partial x'}&=& -\frac{a_I}{a_Cb}+\delta\left(-\frac{\sum_{i=0}^{m-1}(-\delta\zeta)^i}{a_I-a_C}+\frac{(-\delta\zeta)^m}{a_C(1-\delta(1-d))}\right) =  -\frac{a_I}{a_Cb}+\delta\left(\frac{1}{b}+z_m(\delta)\right)\\
&\leq& -\frac{a_I}{a_Cb}+\delta\left(\frac{1}{b}+z_1(\delta)\right) = -\frac{a_I}{a_Cb}+\delta\left(-\frac{1}{a_I-a_C}-\frac{\zeta\delta}{a_C(1-\delta(1-d))}\right) \\
&<& -\frac{a_I}{a_Cb}+\frac{\delta}{a_C(1-\delta(1-d))}=\frac{a_I(\delta\theta-1)}{a_Cb (1-\delta(1-d))} <0,
\end{eqnarray*}
where the first inequality follows from Lemma \ref{property3}, the second inequality follows from $\delta\theta<1$ and Lemma \ref{add1}, and the third inequality follows from $\delta\theta<1.$ 
For $x'\in(x_n,x_{n-1}/(1-d)]$, we have
$$\frac{\partial W_0(x,x')}{\partial x'}= -\frac{a_I}{a_Cb}+\frac{(1-d)\delta}{b}<0,$$
where the inequality follows from $a_I>a_C$ and $(1-d)\delta<1.$ For $x'\in\left(\frac{x_{n-1}}{(1-d)^\ell},\frac{x_n}{(1-d)^\ell}\right]$ with $\ell\in\mathbb{N}$, then 
\begin{eqnarray*}
\frac{\partial W_0(x,x')}{\partial x'}&=& -\frac{a_I}{a_Cb}+\delta^{\ell+1}(1-d)^\ell\left(-\frac{\sum_{i=0}^{n-1}(-\delta\zeta)^i}{a_I-a_C}+\frac{(-\delta\zeta)^n}{a_C(1-\delta(1-d))}\right),\\
&\leq& -\frac{a_I}{a_Cb}+\delta^{\ell+1}(1-d)^\ell\left(-\frac{1}{a_I-a_C}-\frac{\zeta\delta}{a_C(1-\delta(1-d))}\right) \\
&\leq& -\frac{a_I}{a_Cb}+\delta\left(-\frac{1}{a_I-a_C}-\frac{\zeta\delta}{a_C(1-\delta(1-d))}\right) <0,
\end{eqnarray*}
where the first inequality follows from Lemma \ref{property3}, the second inequality follows from $\delta(1-d)<1$, and third inequality follows from $\delta\theta<1$ and Lemma \ref{add1}. If $x'\in (\frac{x_n}{(1-d)^\ell},\frac{x_{n-1}}{(1-d)^{\ell+1}}]$ with $\ell\in\mathbb{N}$, then 
$$\frac{\partial W_0(x,x')}{\partial x'}= -\frac{a_I}{a_Cb}+\frac{(1-d)^{\ell+1}\delta^{\ell+1}}{b}<0,$$
where the inequality follows from $a_I>a_C$ and $(1-d)\delta<1.$
Thus, for any $x'$, $W_0(x,x')$ strictly decreases with $x'$, so $W_0(x,x')$ attains its maximum when $x'$ attains its minimum: $x'=(1-d)x.$

In what follows, for any $(x,x')\in\Omega$ such that $x'\geq \zeta(\hat{x}-x)+\hat{x}$,  following the same argument as in the case of $x\leq a_C$ above, we can show that $W_0(x,x')$ strictly decreases with $x'$, so $W_0(x,x')$ attains its maximum only if $x'\leq \zeta(\hat{x}-x)+\hat{x}$. We thus focus on $x'\geq \zeta(\hat{x}-x)+\hat{x}$ for $x>a_C$. Using the reduced-form utility function, we have
$$W_0(x,x')\equiv u(x,x')+\delta W(x')=\frac{1-d}{b}x-\frac{1}{b}x'+1+\delta W(x').$$

Consider $x\in(x_{m-1},x_m]$ with $m\in\{1,2,...,n\}$. Since $\zeta<0$, $x'\leq \zeta(\hat{x}-x)+\hat{x}\leq \zeta(\hat{x}-x_{m})+\hat{x}=x_{m-1},$ where the last equality follows from the construction of $x_m$. If $x'<a_C$, then
$$\frac{\partial W_0(x,x')}{\partial x'}= -\frac{1}{b}+\frac{\delta}{a_C(1-\delta(1-d))}>0,$$
where the inequality follows from $(a_C(1-d)+b)\delta>a_C$, or equivalently, $\delta>\mu_0=1/(b/a_C+(1-d)).$
If $x'\in(x_{m'-1},x_{m'}]$ for some $m'\in\mathbb{N}$ and $m'\leq m-1$, then
$$\frac{\partial W_0(x,x')}{\partial x'}= -\frac{1}{b}+\delta\left(-\frac{\sum_{i=0}^{m'-1}(-\delta\zeta)^i}{a_I-a_C}+\frac{(-\delta\zeta)^{m'}}{a_C(1-\delta(1-d))}\right)=z_{m'}(\delta)>0,$$
where the inequality follows from $m'\leq m-1\leq n-1$, $\delta>\mu_{n-1}\geq \mu_{m'}$ (from Lemma \ref{property-mu}), and $z_{m'}(\delta)>0$ for $\delta>\mu_{m'}$ (from Lemma \ref{property1}).
Thus, for any $x'\leq \zeta(\hat{x}-x)+\hat{x}$, $W_0(x,x')$ strictly increases with $x'$ and for any $x'\geq \zeta(\hat{x}-x)+\hat{x}$, $W_0(x,x')$ strictly decreases with $x'$, so $W_0(x,x')$ attains its maximum when $x'=\zeta(\hat{x}-x)+\hat{x}.$

Consider $x\in(x_n,{x_{n-1}}/{(1-d)}]$. For $x'\in (0,a_C]$, following the argument for $x\in(x_{m-1},x_m]$ with $m\in\{1,2,...,n\}$, we can show that $W_0(x,x')$ strictly increases with $x'$. For $x'\in(x_{m-1},x_m]$ for some $m\in\{1,2,...,n\}$, 
\begin{equation}\label{eq:zm}
\frac{\partial W_0(x,x')}{\partial x'}= -\frac{1}{b}+\delta\left(-\frac{\sum_{i=0}^{m-1}(-\delta\zeta)^i}{a_I-a_C}+\frac{(-\delta\zeta)^{m}}{a_C(1-\delta(1-d))}\right)=z_{m}(\delta).
\end{equation}
Since $\mu_{n-1}<\delta<\mu_n$, from Lemma \ref{property-mu}, $\mu_m\leq \mu_{n-1}<\delta$ for $m=1,2,...,n-1,$
and $\mu_m=\mu_n>\delta$ for $m=n.$ From Lemma \ref{property1}, $z_m(\delta)>0$ for $m=1,2,...,n-1,$ and $z_m(\delta)<0$ for $m=n.$
Thus, $W_0(x,x')$ strictly increases with $x'$ for $x'\in(x_0, x_{n-1}]$ and strictly decreases with $x'$ for $x'\in(x_{n-1},x_n].$
For $x'\in(x_n,x_{n-1}/(1-d)]$, 
$$\frac{\partial W_0(x,x')}{\partial x'}= -\frac{1}{b}+\frac{\delta(1-d)}{b}<0,$$
which follows from $\delta(1-d)<1.$
For $x'\in\left(\frac{x_{n-1}}{(1-d)^\ell},\frac{x_n}{(1-d)^\ell}\right]$ for $\ell\in\mathbb{N}$, then 
\begin{eqnarray*}
\frac{\partial W_0(x,x')}{\partial x'}&=& -\frac{1}{b}+ \delta^{\ell+1}(1-d)^\ell\left(-\frac{\sum_{i=0}^{n-1}(-\delta\zeta)^i}{a_I-a_C}+\frac{(-\delta\zeta)^n}{a_C(1-\delta(1-d))}\right)\\
&=&-\frac{1}{b}+\delta^{\ell+1}(1-d)^\ell\left(z_n(\delta)+\frac{1}{b}\right)<-\frac{1}{b}+\frac{\delta^{\ell+1}(1-d)^\ell}{b}<0,
\end{eqnarray*}
where the first inequality follows from $\delta<\mu_n$ and $z_n(\delta)<0$ (from Lemma \ref{property1}) and the second inequality follows from $\delta(1-d)<1.$ For $x'\in (\frac{x_n}{(1-d)^\ell},\frac{x_{n-1}}{(1-d)^{\ell+1}}]$ for $\ell\in\mathbb{N}$, again, we have
$$\frac{\partial W_0(x,x')}{\partial x'}=-\frac{1}{b}+\frac{\delta^{\ell+1}(1-d)^\ell}{b}<0.$$
Thus, we have shown that $W_0(x,x')$ strictly increases with $x'$ for $x'<x_{n-1}$ and strictly decreases with $x'$ for $x'>x_{n-1}$, so $W_0(x,x')$ attains its maximum when $x'=x_{n-1}.$

Consider $x>{x_{n-1}}/{(1-d)}$ with $x'\leq \zeta(\hat{x}-x)+\hat{x}.$
Since $x'\geq (1-d)x> x_{n-1}$, following the argument for $x\in(x_n,\frac{x_{n-1}}{1-d}]$, we can show that $W_0(x,x')$ strictly decreases with $x'$ for $x'>x_{n-1}$. Thus, $W_0(x,x')$ attains its maximum when $x'=(1-d)x.$

We have now shown that for every $x>0$, $W_0(x,x')=u(x,x')+\delta W(x')$ is maximized for $x'=\bar{g}(x).$ Since $W(\cdot)$ is constructed from the policy function $\bar{g}(\cdot)$, we have $W(x)=u(x,\bar{g}(x))+\delta W(\bar{g}(x))$ for every $x>0$. So $W(\cdot)$ satisfies the Bellman equation and $\bar{g}(x)$ is the corresponding optimal policy. 
Thus, we have obtained the desired conclusion.
\qed
% continuity of W (x=x_n already checked)
% monotonicity of W
% concavity of W
% well-definedness of each interval

\bigskip

%%%%%%%%%%%%%%%%%%%%%%%%%%%%%%%%%%%%%%%%%%%%%%%%%
% Proof of Proposition 4
%%%%%%%%%%%%%%%%%%%%%%%%%%%%%%%%%%%%%%%%%%%%%%%%%

\noindent
{\it \textbf{Proof of Proposition \ref{ac-less-ai-d=1}}}: Let $d=1.$ 
Let ${\mu}_{n-1}<\delta<{\mu}_{n}$ for some $n\in\mathbb{N}$. We postulate the following value function 
$$W(x)= \left\{\begin{array}{lll}

\frac{x}{a_C} & \mbox{for } x\in [0,a_C] \medskip \\

\left(-\frac{\sum_{i=0}^{m-1}(-\delta\zeta)^i}{a_I-a_C}+\frac{(-\delta\zeta)^m}{a_C}\right)(x-\hat{x}) & \mbox{for } x\in(x_{m-1},x_m],\\
+\frac{(1-\delta^m)(a_I-\hat{x})}{(1-\delta)(a_I-a_C)}+\frac{\delta^m\hat{x}}{a_C} & \;\;\;\;\; m=1,2,...,n \medskip \\

\frac{b-x_{n-1}}{b} +\frac{\sum_{i=1}^{n}\delta^i(a_I-x_{n-i})}{a_I-a_C} & \mbox{for } x\in (x_n,\infty) \medskip 
\end{array} \right.. $$ 
The verification of $W(\cdot)$ satisfying the Bellman equation follows closely the proof of Proposition \ref{ac-less-ai-d<1-add}.
For $x\in(0,a_C]$, $(x,x')\in\Omega$ implies that $x'\geq \zeta(\hat{x}-x)+\hat{x}$. Using the reduced-form utility function, we have
$$W_0(x,x')\equiv u(x,x')+\delta W(x')=\frac{a_I\theta}{a_Cb}x-\frac{a_I}{a_Cb}x'+\delta W(x').$$
For $x'\in (0,a_C]$,
$$\frac{\partial W_0(x,x')}{\partial x'}= -\frac{a_I}{a_Cb}+\frac{\delta}{a_C}=\frac{a_I(\delta\theta-1)}{a_Cb} <0,$$
where the inequality follows from $\theta=b/a_I$ for $d=1$ and $\delta\theta<1.$ 
For $x'\in(x_{m-1},x_m]$ for some $m\in\{1,2,...,n\},$ 
\begin{eqnarray*}
\frac{\partial W_0(x,x')}{\partial x'}&=& -\frac{a_I}{a_Cb}+\delta\left(-\frac{\sum_{i=0}^{m-1}(-\delta\zeta)^i}{a_I-a_C}+\frac{(-\delta\zeta)^m}{a_C}\right)\\
&\leq& -\frac{a_I}{a_Cb}+\delta\left(-\frac{1}{a_I-a_C}-\frac{\zeta\delta}{a_C}\right) \\
&<& -\frac{a_I}{a_Cb}+\frac{\delta}{a_C}=\frac{a_I(\delta\theta-1)}{a_Cb } <0,
\end{eqnarray*}
where the first inequality follows from Lemma \ref{property3}, the second inequality follows from $\delta\theta<1$ and Lemma \ref{add1} with $d=1$, the third inequality follows from $\delta\theta<1.$ 
For $x'>x_n$, we have
${\partial W_0(x,x')}/{\partial x'}= -{a_I}/{(a_Cb)}<0.$
Thus, for any $x'$, $W_0(x,x')$ strictly decreases with $x'$, so $W_0(x,x')$ attains its maximum when $x'$ attains its minimum: $x'=(1-d)x=0$ for $x\in(0,a_C].$
In what follows, similar to the proof of Proposition \ref{ac-less-ai-d<1-add}, we  focus on $x'\geq \zeta(\hat{x}-x)+\hat{x}$ for $x>a_C$. Using the reduced-form utility function, we have
$$W_0(x,x')\equiv u(x,x')+\delta W(x')=-\frac{1}{b}x'+1+\delta W(x').$$

Consider $x\in(x_{m-1},x_m]$ with $m\in\{1,2,...,n\}$. Since $\zeta<0$, $x'\leq \zeta(\hat{x}-x)+\hat{x}\leq \zeta(\hat{x}-x_{m})+\hat{x}=x_{m-1},$ where the last equality follows from the construction of $x_m$. If $x'<a_C$, then,
$$\frac{\partial W_0(x,x')}{\partial x'}= -\frac{1}{b}+\frac{\delta}{a_C}>0,$$
where the inequality follows from $b\delta>a_C$, or equivalently, $\delta>\mu_0=1/(b/a_C)$ with $d=1.$
If $x'\in(x_{m'-1},x_{m'}]$ for some $m'\in\mathbb{N}$ and $m'\leq m-1$, then
$$\frac{\partial W_0(x,x')}{\partial x'}= -\frac{1}{b}+\delta\left(-\frac{\sum_{i=0}^{m'-1}(-\delta\zeta)^i}{a_I-a_C}+\frac{(-\delta\zeta)^{m'}}{a_C}\right)=z_{m'}(\delta)>0,$$
where the inequality follows from $m'\leq m-1\leq n-1$, $\delta>\mu_{n-1}\geq \mu_{m'}$ (from Lemma \ref{property-mu}), and $z_{m'}(\delta)>0$ for $\delta>\mu_{m'}$ (from Lemma \ref{property1}).
Thus, for any $x'\leq \zeta(\hat{x}-x)+\hat{x}$, $W_0(x,x')$ strictly increases with $x'$ and for any $x'\geq \zeta(\hat{x}-x)+\hat{x}$, $W_0(x,x')$ strictly decreases with $x'$, so $W_0(x,x')$ attains its maximum when $x'=\zeta(\hat{x}-x)+\hat{x}.$

Consider $x>x_n$. For $x'\in (0,a_C]$, following the argument for the previous case, we can show that $W_0(x,x')$ strictly increases with $x'$. For $x'\in(x_{m-1},x_m]$ for some $m\in\{1,2,...,n\}$, 
\begin{equation} 
\frac{\partial W_0(x,x')}{\partial x'}= -\frac{1}{b}+\delta\left(-\frac{\sum_{i=0}^{m-1}(-\delta\zeta)^i}{a_I-a_C}+\frac{(-\delta\zeta)^{m}}{a_C}\right)=z_{m}(\delta).
\end{equation}
Since $\mu_{n-1}<\delta<\mu_n$, from Lemma \ref{property-mu}, $\mu_m\leq \mu_{n-1}<\delta$ for $m=1,2,...,n-1,$
and $\mu_m=\mu_n>\delta$ for $m=n.$ From Lemma \ref{property1}, $z_m(\delta)>0$ for $m=1,2,...,n-1,$ and $z_m(\delta)<0$ for $m=n.$
Thus, $W_0(x,x')$ strictly increases with $x'$ for $x_0<x'\leq x_{n-1}$ and strictly decreases with $x'$ for $x'\in(x_{n-1},x_n].$
For $x'\in(x_n,\infty)$, 
${\partial W_0(x,x')}/{\partial x'}= -{1}/{b}<0.$
Thus, we have shown that $W_0(x,x')$ strictly increases with $x'$ for $x'<x_{n-1}$ and strictly decreases with $x'$ for $x'>x_{n-1}$, so $W_0(x,x')$ attains its maximum when $x'=x_{n-1}.$ Note that for $d=1$ and $x>x_n$, $(x,x_{n-1})$ is always in $\Omega.$

Last, we verify that the postulated value function is indeed consistent with the derived optimal policy function. For $x\in(0,a_C]$, 
$W(x)=x/a_C=u(x,0)+\delta W(0).$ For $x\in(x_0,x_1]=(a_C,x_1]$,
\begin{eqnarray*}
W(x)&=& \left(-\frac{1}{a_I-a_C}-\frac{\delta\zeta}{a_C}\right)(x-\hat{x}) +\frac{(1-\delta)(a_I-\hat{x})}{(1-\delta)(a_I-a_C)}+\frac{\delta\hat{x}}{a_C} \\
&=& \frac{a_I-x}{a_I-a_C} +\frac{\delta(\zeta(\hat{x}-x)+\hat{x})}{a_C} \\
&=& u(x,\zeta(\hat{x}-x)+\hat{x})+\delta W(\zeta(\hat{x}-x)+\hat{x}),
\end{eqnarray*}
where the last equality follows from the reduced-form utility function and $(\zeta(\hat{x}-x)+\hat{x})\in(0,a_C]$ for $x\in(x_0,x_1].$
For $x\in(x_{m-1},x_m]$ with $m\in\{2,3,...,n\}$, 
\begin{eqnarray*}
W(x)&=& \left(-\frac{\sum_{i=0}^{m-1}(-\delta\zeta)^i}{a_I-a_C}+\frac{(-\delta\zeta)^m}{a_C}\right)(x-\hat{x}) +\frac{(1-\delta^m)(a_I-\hat{x})}{(1-\delta)(a_I-a_C)}+\frac{\delta^m\hat{x}}{a_C} \\
&=& \left(-\frac{1+\sum_{i=1}^{m-1}(-\delta\zeta)^i}{a_I-a_C}+\frac{(-\delta\zeta)^m}{a_C}\right)(x-\hat{x}) +\frac{(1-\delta+\delta-\delta^m)(a_I-\hat{x})}{(1-\delta)(a_I-a_C)}+\frac{\delta^m\hat{x}}{a_C} \\
&=& \frac{a_I-x}{a_I-a_C}+ \left(-\frac{\sum_{i=1}^{m-1}(-\delta\zeta)^i}{a_I-a_C}+\frac{(-\delta\zeta)^m}{a_C}\right)(x-\hat{x}) +\frac{(\delta-\delta^m)(a_I-\hat{x})}{(1-\delta)(a_I-a_C)}+\frac{\delta^m\hat{x}}{a_C}\\
&=& \frac{a_I-x}{a_I-a_C}+\delta\left[ \left(-\frac{\sum_{i=0}^{m-2}(-\delta\zeta)^i}{a_I-a_C}+\frac{(-\delta\zeta)^{m-1}}{a_C}\right)(\zeta(\hat{x}-x)+\hat{x}-\hat{x})\right. \\
&&\left. +\frac{(1-\delta^{m-1})(a_I-\hat{x})}{(1-\delta)(a_I-a_C)}+\frac{\delta^{m-1}\hat{x}}{a_C} \right]\\
&=& u(x,\zeta(\hat{x}-x)+\hat{x})+\delta W(\zeta(\hat{x}-x)+\hat{x}),
\end{eqnarray*}
where the last equality follows from the $(\zeta(\hat{x}-x)+\hat{x})\in(x_{m-2},x_{m-1}]$ for $x\in(x_{m-1},x_m]$ with $m\in\{2,3,...,n\}$. 
For $x>x_n$ with $n=1$,  we have 
$$W(x)=\frac{b-x_{n-1}}{b} +\frac{\sum_{i=1}^{1}\delta^i(a_I-x_{n-i})}{a_I-a_C} =u(x,x_{n-1})+\delta=u(x,x_{n-1})+\delta W(x_{n-1}),$$
where the last equality follows from $x_{n-1}=x_0=a_C$ for $n=1$ and $W(a_C)=1.$ 
For $x>x_n$ with $n>1$, we have 
\begin{eqnarray*}
W(x)&=&\frac{b-x_{n-1}}{b} +\frac{\sum_{i=1}^{n}\delta^i(a_I-x_{n-i})}{a_I-a_C} \\
%&=&u(x,x_{n-1})+\delta\cdot \frac{\sum_{i=0}^{n-1}\delta^i(a_I-x_{n-i-1})}{a_I-a_C} \\
&=&u(x,x_{n-1})+\delta\left( \frac{\sum_{i=0}^{n-2}\delta^i(a_I-x_{n-i-1})}{a_I-a_C} +\delta^{n-1} \right) \\
&=& u(x,x_{n-1})+\delta\left( \frac{\sum_{i=0}^{n-2}\delta^i[a_I-\hat{x}+(-\zeta)^i(\hat{x}-x_{n-1})]}{a_I-a_C}+\frac{
\delta^{n-1}(a_C-\hat{x})}{a_C}+\frac{\delta^{n-1}\hat{x}}{a_C} \right) \\
%&=& u(x,x_{n-1})+\delta \left(-\frac{\sum_{i=0}^{n-2}(-\delta\zeta)^i}{a_I-a_C}(x_{n-1}-\hat{x})+\frac{(-\delta\zeta)^{n-1}}{a_C}(x_{n-1}-\hat{x})\right.\\
%&&\left.+\frac{\sum_{i=0}^{n-2}\delta^i(a_I-\hat{x})}{a_I-a_C}+\frac{\delta^{n-1}\hat{x}}{a_C} \right) \\
&=& u(x,x_{n-1})+\delta \left(\left(-\frac{\sum_{i=0}^{n-2}(-\delta\zeta)^i}{a_I-a_C}+\frac{(-\delta\zeta)^{n-1}}{a_C}\right)(x_{n-1}-\hat{x}) \right.\\
&& \left.+\frac{(1-\delta^{n-1})(a_I-\hat{x})}{(1-\delta)(a_I-a_C)}+\frac{\delta^{n-1}\hat{x}}{a_C} \right) \\
&=& u(x,x_{n-1})+\delta W(x_{n-1}).
\end{eqnarray*}
Thus, we have shown that $W(\cdot)$ satisfies the Bellman equation and the optimal policy is given by 
$$
g(x)=\left\{\begin{array}{ll}
0 & \mbox{for } x\in(0,a_C] \smallskip \\
-\zeta x+\frac{a_Cb}{a_C-a_I} & \mbox{for } x\in(a_C, x_n] \smallskip \\
x_{n-1} & \mbox{for } x\in(x_n,\infty)
\end{array}\right..
$$
\qed

\bigskip

%%%%%%%%%%%%%%%%%%%%%%%%%%%%%%%%%%%%%%%%%%%%%%%%%
% Proof of Propositions 5 and 6
%%%%%%%%%%%%%%%%%%%%%%%%%%%%%%%%%%%%%%%%%%%%%%%%%

\noindent
{\it \textbf{Proof of Propositions \ref{ac-less-ai-d<1-theta<1-add} and \ref{ac-less-ai-d=1-add2}}}: 
There are two possible cases: (i) $\mu_0\geq 1$ and (ii) $\mu_0<1.$ For (i), since $\mu_0>1$, we always have $\delta<1\leq \mu_0.$ Then Theorem \ref{ac-less-ai-extinction} applies. For (ii), from Lemma \ref{property-theta<1}, we know there exists a unique $n_0\in\mathbb{N}$ such that $\mu_{n_0-1}<1\leq \mu_{n_0}$ and 
$x_{n_0}<a_I.$ Since $x_{n_0}<a_I$, the optimal policy functions stated in the propositions are properly defined for $\delta\in(\mu_{n_0-1},1)\subset(\mu_{n_0-1},\mu_{n_0}).$ 
Then, following essentially the same argument as in the proofs of Propositions \ref{ac-less-ai-d<1-add} and \ref{ac-less-ai-d=1}, we can obtain the optimal policy.  \qed

\bigskip

%%%%%%%%%%%%%%%%%%%%%%%%%%%%%%%%%%%%%%%%%%%%%%
% Proof of PROPOSITION 7
%%%%%%%%%%%%%%%%%%%%%%%%%%%%%%%%%%%%%%%%%%%%%%

\noindent
{\it \textbf{Proof of Proposition \ref{ac-greater-ai2}}}:
For $0<d<1$, following the same argument as the proof of Theorem \ref{ac-greater-ai}, we can show the value function $V(\cdot)$ is again given by (\ref{eq:value-function-thm1}) for $\delta=1/\theta.$ Since $\delta=1/\theta$, the inequality in (\ref{eq:case-i-thm1}) for Case (i) in the proof of Theorem \ref{ac-greater-ai} becomes an equality. Then, we can establish that the optimal policy correspondence $h(x)=[(1-d)x,\min\{a_C,\theta x\}]$ for $x\in(0,a_I]$, $h(x)=[(1-d)x,\min\{a_C,-\zeta x+\frac{a_C}{a_C-a_I}\}]$ for $x\in(a_I,a_C],$ and $h(x)=\{(1-d)x\}$ for $x>a_C$. For $d=1$, a similar argument can be applied to obtain the optimal policy correspondence.
\qed

\bigskip

%%%%%%%%%%%%%%%%%%%%%%%%%%%%%%%%%%%%%%%%%%%%%%%%%
% Proof of Proposition 8
%%%%%%%%%%%%%%%%%%%%%%%%%%%%%%%%%%%%%%%%%%%%%%%%%

\noindent
{\it \textbf{Proof of Proposition \ref{ac-less-ai-rho=1/theta}}}:
Let $\delta=1/\theta$ with $\theta>1$ and $0<d<1.$ The proof follows closely the proof of Theorem 1 in \cite{fldk}. 
Postulate a candidate value function given by
$$W(x)= \left\{\begin{array}{lll}

\frac{a_I\theta}{a_Cb}\delta^n(\theta^nx-\hat{x})+\frac{\delta^n}{1-\delta}u(\hat{x},\hat{x}) & \mbox{for } x\in [\frac{\hat{x}}{\theta^{n+1}},\frac{\hat{x}}{\theta^n}) \smallskip \\
\frac{1-d}{b}\delta^n\left[(1-d)^nx-\hat{x}\right]+\frac{1-\delta^n+\delta^n u(\hat{x},\hat{x})}{1-\delta} & \mbox{for } x\in[\frac{\hat{x}}{(1-d)^n},\frac{\hat{x}}{(1-d)^{n+1}})

\end{array} \right. $$
where $n=0,1,2...$
We now verify if $W(x)$ satisfies the Bellman equation. We consider three cases: (i) $x\in(0,\hat{x});$ (ii) $x\in[\hat{x},\frac{\hat{x}}{1-d});$ (iii) $x\in[\frac{\hat{x}}{(1-d)^n},\frac{\hat{x}}{(1-d)^{n+1}})$ with $n\geq 1$.

For Case (i), there exists $n\in\mathbb{N}$ such that
$x\in[\hat{x}/{\theta}^{n},\hat{x}/\theta^{n-1}).$ Pick $x'$ such that $(x,x')\in\Omega.$ 
If $x' > \hat{x}\geq \zeta(\hat{x}-x)+\hat{x}$, then there exists $n_0\in\mathbb{N}$ such that $x'\in[\frac{\hat{x}}{(1-d)^{n_0-1}},\frac{\hat{x}}{(1-d)^{n_0}})$.
Since $x'\geq \zeta(\hat{x}-x)+\hat{x}$, we have
\begin{eqnarray*}
W_0(x,x')&\equiv&u(x,x')+\delta W(x')= \frac{a_I\theta}{a_Cb}x-\frac{a_I}{a_Cb}x'+\delta W(x'); \\
\frac{\partial W_0(x,x')}{\partial x'}&=& \frac{1}{{a_Cb}}\left[a_C\delta^{n_0}(1-d)^{n_0}-a_I\right]<0,
\end{eqnarray*}
where the inequality follows from
$a_C< a_I$ and $\delta(1-d)<1$. 
Consider $x'$ such that $\hat{x}> x'\geq \zeta(\hat{x}-x)+\hat{x}$. There exists $n_0\in\mathbb{N}$ such that $x'\in [\frac{\hat{x}}{\theta^{n_0}},\frac{\hat{x}}{\theta^{n_0-1}})$.
Since $x'\geq \zeta(\hat{x}-x)+\hat{x}$, 
\begin{eqnarray*}
W_0(x,x')&\equiv&u(x,x')+\delta W(x')= \frac{a_I\theta}{a_Cb}x-\frac{a_I}{a_Cb}x'+\delta W(x'); \\
\frac{\partial W_0(x,x')}{\partial x'}&=& \frac{a_I}{a_C{b}}\left[(\delta\theta)^{n_0}-1\right]=0,
\end{eqnarray*}
where the last equality follows from $\delta=1/\theta.$ For $x'<\zeta(\hat{x}-x)+\hat{x}<\hat{x}$, there exists $n_0\in\mathbb{N}$ such that $x'\in [\frac{\hat{x}}{\theta^{n_0}},\frac{\hat{x}}{\theta^{n_0-1}})$.
Since $x'< \zeta(\hat{x}-x)+\hat{x}$, 
\begin{eqnarray*}
W_0(x,x')&\equiv&u(x,x')+\delta W(x')= \frac{1-d}{b}x-\frac{1}{b}x'+1+\delta W(x'); \\
\frac{\partial W_0(x,x')}{\partial x'}&=& \frac{1}{{b}}\left[\frac{a_I}{a_C}(\delta\theta)^{n_0}-1\right]>0,
\end{eqnarray*}
where the inequality follows from $\delta\theta=1$ and $a_I>a_C.$ Taken together, we have shown that $W_0(x,x')$ strictly decreases with $x'$ for $x'>\hat{x}$, strictly increases with $x'$ for $x'< \zeta(\hat{x}-x)+\hat{x}$, and is constant with respect to $x'$ for $x'\in[\zeta(\hat{x}-x)+\hat{x},\hat{x}]$. Since $(x,x')\in\Omega$, $\theta x\geq x'\geq (1-d)x$. Thus, $W_0(x,x')$ is maximized for $x'\in{[\max\{(1-d)x,-\zeta x+\frac{a_Cb}{a_C-a_I}\},\min\{\theta x,\hat{x}\}]}.$

For Cases (ii) and (iii), following the proof of Theorem 1 in \cite{fldk}, we can show that for $x\in[\hat{x},\frac{\hat{x}}{1-d}),$
$W_0(x,x')$ is maximized with $x'=\hat{x}$, and for $x\in[\frac{\hat{x}}{(1-d)^n},\frac{\hat{x}}{(1-d)^{n+1}})$ with $n\in\mathbb{N}$, $W_0(x,x')$ is maximized with $x'=(1-d)x$.
Consider the policy correspondence
$$
\bar{h}(x)=\left\{\begin{array}{ll}
{[\max\{(1-d)x,-\zeta x+\frac{a_Cb}{a_C-a_I}\},\theta x]} & \mbox{for } x\in(0,\frac{\hat{x}}{\theta}] \smallskip \\
{[\max\{(1-d)x,-\zeta x+\frac{a_Cb}{a_C-a_I}\},\hat{x}]} & \mbox{for } x\in(\frac{\hat{x}}{\theta}, \hat{x}] \smallskip \\
\{\hat{x}\} & \mbox{for } x\in(\hat{x},\frac{\hat{x}}{1-d}] \smallskip \\
\{(1-d) x\} & \mbox{for } x\in(\frac{\hat{x}}{1-d},\infty)
\end{array}\right.,
$$
and the straight-down-the-turnpike policy
$$
\bar{g}(x)=\left\{\begin{array}{ll}
\theta x & \mbox{for } x\in(0,\frac{\hat{x}}{\theta}] \smallskip \\
\hat{x} & \mbox{for } x\in(\frac{\hat{x}}{\theta},\frac{\hat{x}}{1-d}] \smallskip \\
(1-d) x & \mbox{for } x\in(\frac{\hat{x}}{1-d},\infty)
\end{array}\right.,
$$
For any $x>0$, $\bar{g}(x)\in\bar{h}(x),$ and as shown above, $W_0(x,x')$ is maximized for any $x'\in \bar{h}(x)$ and in particular, for $x'=\bar{g}(x).$ Since $W(x)=u(x,\bar{g}(x))+\delta W(\bar{g}(x))$ for any $x$ (as in \cite{fldk}), $W(x)=u(x,x')+\delta W(x')$ for any $x'\in \bar{h}(x)$.
Thus, $W(\cdot)$ satisfies the Bellman equation and the optimal policy correspondence is given by $\bar{h}(\cdot).$
\qed

\bigskip

%%%%%%%%%%%%%%%%%%%%%%%%%%%%%%%%%%%%%%%%%%%%%%%%%
% Proof of Proposition 9
%%%%%%%%%%%%%%%%%%%%%%%%%%%%%%%%%%%%%%%%%%%%%%%%%

\noindent
{\it \textbf{Proof of Proposition \ref{ac=ai}}}:
For $\delta<1/\theta$, the proof follows the proof of Theorem \ref{ac-greater-ai}, so the optimal policy for $a_C=a_I$ is also given by $g(x)=(1-d)x$ for any $x>0$. For $\delta=1/\theta$, 
the proof of Proposition \ref{ac-greater-ai2} also carries over to the one-sector case ($a_C=a_I$) but with one modification: for $a_C=a_I$ and $\delta=1/\theta$, both inequalities in (\ref{eq:case-ii-thm1}) become equalities. This implies for $x>a_C$,
the optimal policy is given by $h(x)=[(1-d)x,\max\{a_C,(1-d)x\}]$. \qed

\bigskip

%%%%%%%%%%%%%%%%%%%%%%%%%%%%%%%%%%%%%%%%%%%%%%%%%
% Proof of Propositions 10 and 11
%%%%%%%%%%%%%%%%%%%%%%%%%%%%%%%%%%%%%%%%%%%%%%%%%

\noindent
{\it \textbf{Proof of Propositions \ref{ac-less-ai-d<1-add2} and \ref{ac-less-ai-d=12}}}:   We first consider the case of $0<d<1.$
For $\delta=\mu_0$, we can follow the proof of Theorem \ref{ac-less-ai-extinction} to establish the optimal policy correspondence. The only difference is that the inequality (\ref{eq:rho0}) becomes equality for $\delta=\mu_0$. This implies that for $x\in(a_C,\frac{a_C}{1-d}]$, ${\mathcal W}_0(x,x')$ is maximized for any $x'$ such that $x'\geq (1-d)x$, $x'\leq \zeta(\hat{x}-x)+\hat{x}$, and $x'\leq a_C.$ We thus establish Proposition  \ref{ac-less-ai-d<1-add2}(i).
For $\delta=\mu_n$ for some $n\in\mathbb{N}$, we can follow the proof Proposition \ref{ac-less-ai-d<1-add}. The only difference is that for (\ref{eq:zm}), $\frac{\partial W_0(x,x')}{\partial x'}=0$ for $m=n$ because $\delta=\mu_n$. Thus, for $x\in(x_n,\frac{x_{n-1}}{1-d}]$, $W_0(x,x')$ strictly increases with $x'$ for $x'\leq x_{n-1}$, is constant with respect to $x'$ for $x'\in(x_{n-1},x_n],$ and strictly decreases with $x'$ for $x'>x_n.$ Then, $W_0(x,x')$ is maximized for $x'\in(x_{n-1},x_n].$ Similarly, we can show that for $x\in(\frac{x_{n-1}}{1-d},\frac{x_{n}}{1-d}]$, $W_0(x,x')$ is maximized for $x'\in(x_{n-1},x_n].$ Using the fact that $(x,x')\in \Omega$, we then obtain the optimal policy correspondence as in Proposition \ref{ac-less-ai-d<1-add2}(ii). 
The argument is essentially the same for the case of $d=1.$
\qed

\subsection{Auxiliary Results}

%%%%%%%%%%%
% Lemma A1
%%%%%%%%%%%

\begin{lmaa} \label{property1-zn}
For any non-negative integer $n$, we can write $z_n(\delta)$ as 
\begin{equation}
z_n(\delta) = \left\{\begin{array}{ll} \frac{b a_I (-\zeta)^n (1-\theta\delta)\delta^{n+1}-a_C(a_I-a_C)(1-(1-d)\delta)^2}{a_Cb(1-\delta(1-d))(a_I-a_C)(1+\delta \zeta)} & \mbox{for } \delta\neq-\frac{1}{\zeta} \medskip \\ 
\frac{-{nba_C(1-d)\delta}+nba_C-ba_I+2ba_C}{a_Cb(1-\delta(1-d))(a_I-a_C)\zeta} & \mbox{for } \delta=-\frac{1}{\zeta} \label{eq:zn}
\end{array}\right..
\end{equation} 
\end{lmaa}

\begin{proof}
For $\delta\neq-1/\zeta$, $-\delta\zeta\neq1$ and we can simplify the geometric series in $z_n(\delta)$ as follows
\begin{eqnarray*}
z_n(\delta)&=& -\frac{1}{b}+\delta\left(-\frac{\sum_{i=0}^{n-1}(-\delta\zeta)^i}{a_I-a_C}+\frac{(-\delta\zeta)^n}{a_C(1-\delta(1-d))}\right) \\
&=&-\frac{1}{b}+\delta\left(-\frac{1-(-\delta\zeta)^n}{(1+\delta\zeta)(a_I-a_C)}+\frac{(-\delta\zeta)^n}{a_C(1-\delta(1-d))}\right) \\
&=& \frac{-(1+\delta\zeta)(a_I-a_C)a_C(1-\delta(1-d))-ba_C(1-\delta(1-d))\delta(1-(-\delta\zeta)^n))}{a_Cb(1-\delta(1-d))(a_I-a_C)(1+\delta \zeta)} \\
&& + \frac{\delta b(1+\delta\zeta)(a_I-a_C)(-\delta\zeta)^n}{a_Cb(1-\delta(1-d))(a_I-a_C)(1+\delta \zeta)} \\
&=& \frac{b a_I (-\zeta)^n (1-\theta\delta)\delta^{n+1}-a_C(a_I-a_C)(1-(1-d)\delta)^2}{a_Cb(1-\delta(1-d))(a_I-a_C)(1+\delta \zeta)},
\end{eqnarray*}
where we note that even though the first equation does not apply to $n=0$ because the summation is not property defined, the expressions in the second line onward apply for any non-negative integer $n$.

For $\delta=-1/\zeta$, $-\delta\zeta=1$, we have
\begin{eqnarray*}
z_n(\delta)&=& -\frac{1}{b}+\delta\left(-\frac{\sum_{i=0}^{n-1}(-\delta\zeta)^i}{a_I-a_C}+\frac{(-\delta\zeta)^n}{a_C(1-\delta(1-d))}\right) \\
&=& -\frac{1}{b}+\delta\left(-\frac{n}{a_I-a_C}+\frac{1}{a_C(1-\delta(1-d))}\right) \\
&=& \frac{-(a_I-a_C)a_C(1-\delta(1-d))-bn a_C(1-\delta(1-d))\delta+b(a_I-a_C)\delta}{a_Cb(1-\delta(1-d))(a_I-a_C)} \\
&=& \frac{-{nba_C(1-d)\delta}+nba_C-ba_I+2ba_C}{a_Cb(1-\delta(1-d))(a_I-a_C)\zeta},
\end{eqnarray*}
where again we note that even though the first equation does not apply to $n=0$, the expressions in the second line onward apply for any non-negative integer $n$.
We have thus obtained the desired conclusion.
\end{proof}

%%%%%%%%%%%
% Lemma A2
%%%%%%%%%%%

\begin{lmaa} \label{property1-tilde-zn}
For any positive integer $n$, define $\tilde{z}_n:[0,1/\theta]\to\mathbb{R}$ given by
\begin{equation}\label{eq:tilde-zn}
\tilde{z}_n(\delta)\equiv b a_I (-\zeta)^n (1-\theta\delta)\delta^{n+1}-a_C(a_I-a_C)(1-(1-d)\delta)^2.
\end{equation}
For $\delta\neq-1/\zeta$, $z_n(\delta)=0$ if and only if $\tilde{z}_n(\delta)=0$. Moreover, if $a_I>a_C$, then $\tilde{z}_n(\cdot)$ and its derivatives satisfy:

\medskip

(a.1) $\tilde{z}_n(-1/\zeta)=0$, (a.2) $\tilde{z}_n(0)<0$, (a.3) $\tilde{z}_n(1/\theta)<0$.

\medskip

(b) ${z}_n\left(-{1}/{\zeta}\right)\tilde{z}_n'\left(-{1}/{\zeta}\right)\leq 0$ 
and $\tilde{z}_n'(-1/\zeta)=0$ if and only if $z_n(-1/\zeta)=0.$

\medskip

(c) There exists $\bar{\delta}\in(0,1/\theta)$ such that $\tilde{z}_n'(\delta)>0$ for $\delta\in[0,\bar{\delta})$ and 
$\tilde{z}_n'(\delta)<0$ for $\delta\in(\bar{\delta},1/\theta].$
\end{lmaa}

\begin{proof}
From Lemma \ref{property1-zn}, for $\delta\neq-1/\zeta$, 
$$z_n(\delta)=\frac{b a_I (-\zeta)^n (1-\theta\delta)\delta^{n+1}-a_C(a_I-a_C)(1-(1-d)\delta)^2}{a_Cb(1-\delta(1-d))(a_I-a_C)(1+\delta \zeta)},$$
so $z_n(\delta)=0$ if the only if 
$$ b a_I (-\zeta)^n (1-\theta\delta)\delta^{n+1}-a_C(a_I-a_C)(1-(1-d)\delta)^2=\tilde{z}_n(\delta)=0.$$ 

Let $a_I>a_C.$ Since $\tilde{z}_n(\cdot)$ is continuous on $[0,1/\theta]$, we have
\begin{eqnarray*}
\tilde{z}_n(-1/\zeta)&=&\lim_{\delta\rightarrow-1/\zeta} \tilde{z}_n(\delta) \\
&=& \lim_{\delta\rightarrow-1/\zeta} a_Cb(1-\delta(1-d))(a_I-a_C)(1+\delta \zeta) {z}_n(\delta) \\
&=& \lim_{\delta\rightarrow-1/\zeta} a_Cb(1-\delta(1-d))(a_I-a_C)(1+\delta \zeta) {z}_n(-1/\zeta) =0,
\end{eqnarray*} % this continuity has been verified
where the second equality follows from the definition of $\tilde{z}_n(\cdot)$ and Lemma \ref{property1-zn}, the third equality follows from the continuity of $z_n(\cdot)$ on $[0,1/\theta]$, and the last equality follows from $(1+\delta\zeta)=0$ for $\delta=-1/\zeta$.
Since $a_I>a_C>0$, $\tilde{z}_n(0)=-a_C(a_I-a_C)<0$. Since $\theta>(1-d)$ and $a_I>a_C>0$, $\tilde{z}_n(1/\theta)=-a_C(a_I-a_C)(1-(1-d)/\theta)^2<0.$ Thus, we have established (a.1)--(a.3).

For (b), since 
\begin{equation}\label{eq:zn-tilde'}
\tilde{z}_n'(\delta)= b a_I(-\zeta)^n\left[ (n+1)(1-\theta \delta)\delta^n -\theta \delta^{n+1}\right]+2a_C(a_I-a_C)(1-d)(1-(1-d)\delta),
\end{equation}
we have  
\begin{eqnarray}
%\tilde{z}_n'\left(0\right) &=& 2a_C(a_I-a_C)(1-d)>0, \nonumber \\[5pt]
\tilde{z}_n'\left(-\frac{1}{\zeta}\right) &=& (n+1) b a_I\left(1+\frac{\theta}{\zeta}\right)+\frac{ba_I\theta}{\zeta} +2a_C(a_I-a_C)\left(1+\frac{1-d}{\zeta}\right)(1-d)\nonumber \\
&=& \frac{1}{\zeta}\left[\frac{(n+1)b^2a_C}{a_C-a_I}+b^2+b(1-d)a_I-2a_Cb(1-d)\right]\nonumber \\
&=&\frac{1}{\zeta}\left[ {nba_C}(\zeta+(1-d))+\frac{b^2 a_C}{a_C-a_I}+b^2+b(1-d)(a_I-2a_C)\right] \nonumber \\
&=& \frac{1}{\zeta}\left[ {nba_C}\zeta+(1-d)nba_C+\left(-\frac{b^2}{a_C-a_I}+b(1-d)\right)(a_I-2a_C)\right] \nonumber \\
&=& \frac{n(1-d)b a_C}{\zeta}+nb a_C-b(a_I-2a_C), \nonumber \\
&=& a_Cb(1+(1-d)/\zeta)(a_I-a_C)\zeta z_n\left(-\frac{1}{\zeta}\right), \nonumber
\end{eqnarray}
where the last equality follows from Lemma \ref{property1-zn} for $\delta=-1/\zeta$.
Since $\tilde{z}_n'(-1/\zeta)=a_Cb(1+(1-d)/\zeta)(a_I-a_C)\zeta z_n(-1/\zeta)$, $\tilde{z}_n'(-1/\zeta)=0$ if and only if $z_n(-1/\zeta)=0.$  Moreover, since $\zeta<0$ (for $a_I>a_C$) and $a_I>a_C,$
\begin{eqnarray*}
{z}_n\left(-\frac{1}{\zeta}\right)\tilde{z}_n'\left(-\frac{1}{\zeta}\right)&=&a_Cb(1-\delta(1-d))(a_I-a_C)\zeta z^2_n\left(-\frac{1}{\zeta}\right)\leq 0.
\end{eqnarray*}
Thus, we have established (b).

For (c), define $f(\delta)\equiv b a_I(-\zeta)^n\left[ (n+1)(1-\theta \delta)\delta^n -\theta \delta^{n+1}\right]$. Then, from Equation (\ref{eq:zn-tilde'}), we have
$$\tilde{z}_n'(\delta)= f(\delta) +2a_C(a_I-a_C)(1-d)(1-(1-d)\delta)>f(\delta),$$
where the inequality follows from $a_I>a_C$ and $\delta\in[0,1/\theta].$
Since $f'(\delta)=b a_I(-\zeta)^n (n+1)\delta^{n-1} [n-(n+2)\theta \delta]$,  $f'(\delta)\geq 0$ for $\delta\in[0, \frac{n}{(n+2)\theta}]$ and  $f'(\delta)< 0$ for $\delta> \frac{n}{(n+2)\theta}$.
Since $f(0)=0$ and $f'(\delta)\geq 0$ for $\delta\in[0, \frac{n}{(n+2)\theta}]$, $f(\delta)\geq 0$ for $\delta\in[0, \frac{n}{(n+2)\theta}].$ Then, $\tilde{z}_n'(\delta)>f(\delta)\geq 0$ for $\delta\in[0, \frac{n}{(n+2)\theta}].$ For $\delta> \frac{n}{(n+2)\theta},$
$\tilde{z}_n''(\delta)=f'(\delta)-2a_C(a_I-a_C)(1-d)^2<f'(\delta)<0,$ which implies that $\tilde{z}_n'(\cdot)$ is strictly decreasing on the interval $[\frac{n}{(n+2)\theta},\frac{1}{\theta}].$ Suppose $\tilde{z}_n'(1/\theta)\geq 0.$ By the monotonicity, we must have $\tilde{z}_n'(\delta)>0$ for $\delta\in( \frac{n}{(n+2)\theta},\frac{1}{\theta})$ and we have shown that $\tilde{z}_n'(\delta)>0$ for $\delta\in[0, \frac{n}{(n+2)\theta}],$ so $\tilde{z}_n(\cdot)$ is strictly increasing on $\delta\in[0,1/\theta].$ However, from (a.1) and (a.3), we know $\tilde{z}_n(-1/\zeta)=0>\tilde{z}_n(1/\theta)$ with $1/\theta>-1/\zeta$, contradicting to $\tilde{z}_n(\cdot)$ being strictly increasing. Thus, we  must have  $\tilde{z}_n'(1/\theta)< 0.$ Since 
$\tilde{z}_n'(\cdot)$ is strictly decreasing on the interval $[\frac{n}{(n+2)\theta},\frac{1}{\theta}]$ and $\tilde{z}_n'(\frac{n}{(n+2)\theta})>0,$ by the continuity of $\tilde{z}_n'$, there  exists $\bar{\delta}\in(\frac{n}{(n+2)\theta},\frac{1}{\theta})$ such that $\tilde{z}_n'(\bar{\delta})=0$, $\tilde{z}_n'({\delta})<0$ for $\delta>\bar{\delta}$ and  $\tilde{z}_n'({\delta})>0$ for $\delta\in[\frac{n}{(n+2)\theta},\bar{\delta}).$ Since we have already shown that $\tilde{z}_n'(\delta)>0$ for $\delta\in[0, \frac{n}{(n+2)\theta}],$ we have obtained the desired conclusion.
\end{proof}

%%%%%%%%%%%
% Lemma A3
%%%%%%%%%%%
\begin{lmaa} \label{xn-property-remark}
 For $\zeta\neq-1$, we can express $x_n$ more explicitly as
$$x_n= \frac{a_Cb}{b+d(a_C-a_I)}-\frac{da_C(a_I-a_C)}{(b+d(a_C-a_I))(-\zeta)^n}.$$
For $\theta>1$, we  further have $x_n= \hat{x}-{(\hat{x}-a_C)}/{(-\zeta)^n}.$
Moreover, for $\zeta=-1$, $x_n=a_C+{na_Cb}/{(a_I-a_C)}.$
\end{lmaa}
\begin{proof}
From Equation  (\ref{eq:xn}) in the proof of Lemma \ref{xn-property}, we directly obtain
$$x_n= \frac{a_Cb}{b+d(a_C-a_I)}-\frac{da_C(a_I-a_C)}{(b+d(a_C-a_I))(-\zeta)^n}.$$
For $\theta>1$, we have $\hat{x}=\frac{a_Cb}{b+d(a_C-a_I)}$, so we can further simplify the expression above to obtain
$$x_n= \hat{x}-{(\hat{x}-a_C)}/{(-\zeta)^n}.$$
For $\zeta=-1$, $x_n=x_{n-1}+{a_Cb}/{(a_I-a_C)}=x_0+{na_Cb}/{(a_I-a_C)}=a_C+{na_Cb}/{(a_I-a_C)},$ where the last equality follows from $x_0=a_C.$
\end{proof}

%%%%%%%%%%%
% Lemma A4
%%%%%%%%%%%

\begin{lmaa} \label{add1}
If $\delta\theta<1$ and $a_C<a_I$, then $-\frac{1}{a_I-a_C}-\frac{\zeta\delta}{a_C(1-\delta(1-d))}< \frac{1}{a_C(1-\delta(1-d))}.$
\end{lmaa}
\begin{proof}
Since $\delta\theta<1$ and $\theta=b/a_I+(1-d)$, we have
\begin{eqnarray*}
\delta\theta<1 &\Leftrightarrow& \delta(b+a_I(1-d))-a_I< 0 \\
&\Leftrightarrow& \delta(b-(a_C-a_I)(1-d))+\delta a_C(1-d)+(a_C-a_I)-a_C< 0 \\
&\Leftrightarrow& (a_C-a_I) \zeta \delta +(a_C-a_I)-a_C(1-\delta(1-d))<0 \\
&\Leftrightarrow& (-\zeta\delta-1)(a_I-a_C)-a_C(1-\delta(1-d))<0 \\
&\Leftrightarrow& \frac{-\zeta\delta-1}{a_C(1-\delta(1-d))}<\frac{1}{a_I-a_C} \\
&\Leftrightarrow& -\frac{1}{a_I-a_C}-\frac{\zeta\delta}{a_C(1-\delta(1-d))}< \frac{1}{a_C(1-\delta(1-d))},
\end{eqnarray*}
where the second to last line follows from $a_C<a_I$ and $\delta(1-d)<1.$
\end{proof}

%%%%%%%%%%%
% Lemma A5
%%%%%%%%%%%

\begin{lmaa}
\label{property-theta<1}
Let $a_C< a_I$, $\theta< 1$, and $\mu_0< 1$.
There exists a unique $n_0\in\mathbb{N}$ such that $\mu_{n_0-1}<1\leq \mu_{n_0}$ and 
$x_{n_0}<a_I.$
\end{lmaa}

\begin{proof}
Since $\theta<1$, from Lemma \ref{property-mu}, $\lim_{n\rightarrow\infty} \mu_n=1/\theta>1.$
We claim that
there exists a unique $n_0\in\mathbb{N}$ such that $\mu_{n_0-1}<1\leq \mu_{n_0}$. Suppose on the contrary, there does not exist a natural number $n_0$ such that $\mu_{n_0-1}<1\leq \mu_{n_0}$. 
Since the sequence $\left\{\mu_n\right\}_{n=0}^\infty$ is monotonically increasing and $\mu_0<1$, this implies that $\mu_n<1$ for any $n\in\mathbb{N}$. Since $\mu_n<1$ for any $n$, $\lim_{n\rightarrow\infty} \mu_n\leq 1$, leading to a contradiction. The strict monotonicity of $\left\{\mu_n\right\}_{n=0}^\infty$ further guarantees the uniqueness of $n_0.$ What remains to show is that
$x_{n_0}<a_I.$

Since $\mu_{n_0-1}<1\leq \mu_{n_0}$, from Lemma \ref{property1}, we have
$z_{n_0-1}(1)>0.$ 
For $\zeta=-1$, from Lemma \ref{property1-zn}, we have 
$$z_{n_0-1}(1)=\frac{(n_0-1)ba_Cd-ba_I+2ba_C}{a_Cbd(a_I-a_C)\zeta} > 0\;\Leftrightarrow\;(n_0-1)a_Cd< a_I-2a_C,$$ 
where the second inequality follows from $a_I>a_C$ and $\zeta<0.$
From Lemma \ref{xn-property-remark}, for $\zeta=-1$, 
$$x_{n_0}=a_C+\frac{n_0a_Cb}{a_I-a_C}=a_C+n_0a_Cd< a_C+a_Cd+a_I-2a_C=a_I-(1-d)a_C<a_I,$$ 
where the second equation follows from $\zeta=-1$ and the first inequality follows from $(n_0-1)a_Cd< a_I-2a_C.$ For $\zeta\neq -1$, from Lemma \ref{property1}, we have
\begin{eqnarray}
&&z_{n_0-1}(1)=\frac{b a_I (-\zeta)^{n_0-1} (1-\theta)-a_C(a_I-a_C)d^2}{a_Cbd(a_I-a_C)(1+ \zeta)}> 0 \nonumber \\
&\Leftrightarrow& \frac{a_C(a_I-a_C)d}{(-\zeta)^{n_0}(1+\zeta)}< \frac{b a_I(1-\theta)}{-\zeta d (1+\zeta)}\nonumber \\
&\Leftrightarrow& \frac{da_C(a_I-a_C)}{(b+d(a_C-a_I))(-\zeta)^{n_0}}>\frac{b a_I(1-\theta)}{-\zeta d(b+d(a_C-a_I))}, \label{eq:lemma1}
\end{eqnarray}
where the last inequality follows from $a_C<a_I.$
Further, from Lemma \ref{xn-property-remark}, we have
\begin{eqnarray*}
x_{n_0}&=&\frac{a_Cb}{b+d(a_C-a_I)}-\frac{da_C(a_I-a_C)}{(b+d(a_C-a_I))(-\zeta)^{n_0}}\\
&<&\frac{a_Cb}{b+d(a_C-a_I)}-\frac{ba_I(1-\theta)}{-\zeta d(b+d(a_C-a_I))}\\
&=&\frac{a_Cb}{b+d(a_C-a_I)}-\frac{b(da_I-b)}{-\zeta d(b+d(a_C-a_I))}\\
&=&\frac{a_Cb}{b+d(a_C-a_I)}-\frac{b(da_I-b)}{ d(b+d(a_C-a_I))}+\frac{b(da_I-b)}{ d(b+d(a_C-a_I))}-\frac{b(da_I-b)}{-\zeta d(b+d(a_C-a_I))}\\
&=& \frac{a_Cbd-a_Ibd+b^2}{(b+d(a_C-a_I))d}+\frac{(\zeta+1)b(da_I-b)}{ \zeta d(b+d(a_C-a_I))} \\
&=& \frac{b}{d} + \frac{b(da_I-b)}{\zeta d(a_C-a_I)}=a_I+\frac{b-a_Id}{d}+ \frac{b(da_I-b)}{\zeta d(a_C-a_I)}\\
&=&a_I+\frac{(b-a_Id)(\zeta (a_C-a_I)-b)}{\zeta d(a_C-a_I)}=a_I- \frac{(b-a_Id)(1-d)}{\zeta d}\leq a_I,
\end{eqnarray*}
where the first inequality follows from (\ref{eq:lemma1}) and the last inequality follows from $\zeta<0$ and $b<a_Id$ (from $\theta<1$). Then, we have obtained the desired conclusion. 
\end{proof}

\newpage

%%%%%%%%%%%%%%%%%%%%%%%%%%%
% Figures in the proofs
%%%%%%%%%%%%%%%%%%%%%%%%%%%

\renewcommand\thefigure{A.\arabic{figure}} 
\setcounter{figure}{0}

\begin{figure}
\begin{center}
\includegraphics[width=15cm]{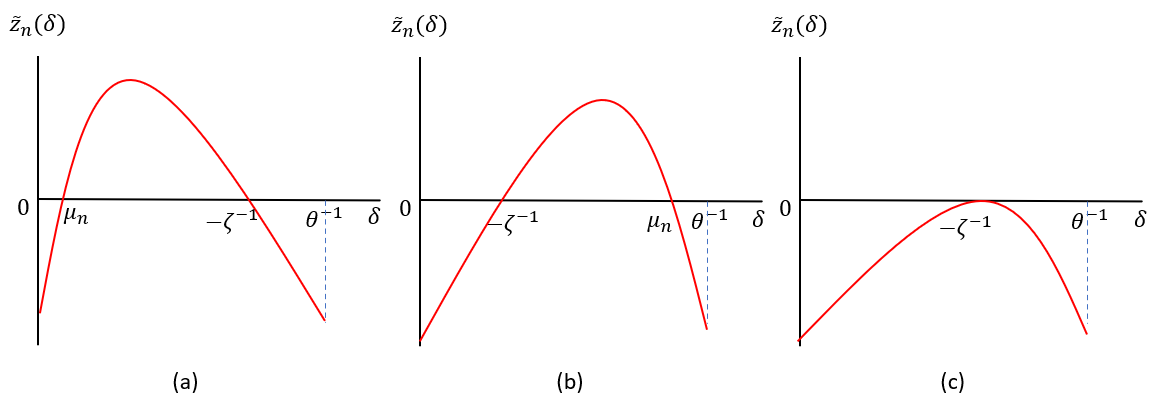}

\caption{Properties of $\tilde{z}_n$}\label{fig:proof-zn}
\end{center}
\end{figure}

\begin{figure}
\begin{center}
\includegraphics[width=15cm]{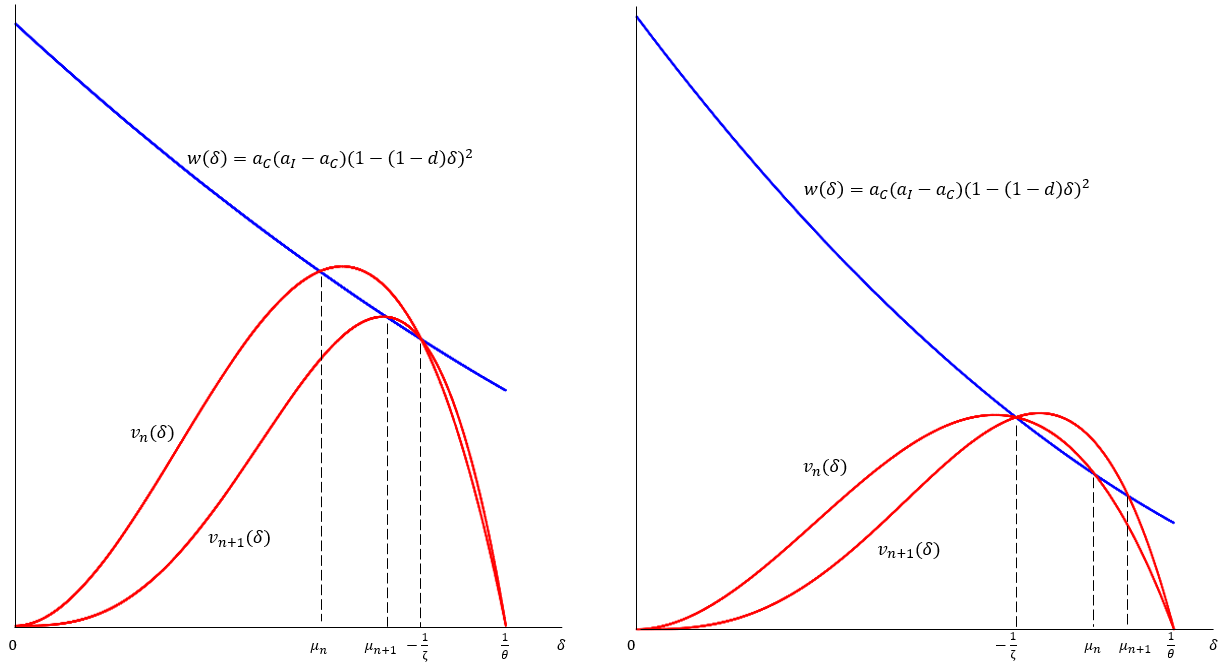}

\caption{Monotonicity of $\mu_n$}\label{fig:rho-n}
\end{center}
\end{figure}

\end{appendix}

\end{document}